\documentclass{article}

\usepackage{graphicx,amssymb,amsmath,ulem,url}
\title{From {\it Clarkia} to {\it Escherichia} and Janus: the physics of natural and synthetic active colloids}

\author{W. C. K. Poon\\School of Physics \& Astronomy, The University of Edinburgh\\James Clerk Maxwell Building, Kings Buildings,\\Mayfield Road, Edinburgh EH9 3JZ, United Kingdom.}

\begin{document}

\maketitle


\begin{abstract}
An active colloid is a suspension of particles that transduce free energy from their environment and use the energy to engage in {\it intrinsically non-equilibrium} activities such as growth, replication and self-propelled motility. An obvious example of active colloids is a suspension of bacteria such as {\it Escherichia coli}, their physical dimensions being almost invariably in the colloidal range. Synthetic self-propelled particles have also become available recently, such as two-faced, or Janus, particles propelled by differential chemical reactions on their surfaces driving a self-phoretic motion. In these lectures, I give a pedagogical introduction to the physics of single-particle and collective properties of active colloids, focussing on self propulsion. I will compare and contrast phenomena in suspensions of `swimmers' with the behaviour of suspensions of passive particles, where only Brownian motion (discovered by Robert Brown in granules from the pollen of the wild flower {\it Clarkia pulchella}) is relevant. I will pay particular attention to issues that pertain to performing experiments using these active particle suspensions, such as how to characterise the suspension's swimming speed distribution, and include an appendix to guide physicists wanting to start culturing motile bacteria. \\ \\
{\it Published as: Proceedings of the International School of Physics``Enrico Ferm'',
Course CLXXXIV ``Physics of Complex Colloid'', eds. C. Bechinger, F. Sciortino and P. Ziherl, IOS, Amsterdam: SIF, Bologna (2013),
pp. 317-386}\\ \\
Note: reference numbering is different in the published version!
\end{abstract}

\pagebreak


\pagebreak

\section{Introduction}

A colloid is a dispersion of non-density-matched particles in a liquid that remain suspended against gravity by virtue of their Brownian motion. This thermally-driven motion, which dominates the physics of colloids, was discovered by the Scottish botanist Robert Brown \cite{Brown,midworld}, who observed in 1827 the incessant movement of granules of approximately $1 \mu$m size extracted from the pollen of {\it Clarkia pulchella} \footnote{The type species of a genus of NW American wild flowers discovered in 1806. Named after one of its discoverers (William Clark; {\it pulchella} = `pretty' in Latin), it was brought to Britain in 1826 by, David Douglas (of Douglas fir fame). Brown was studying the germination its pollens.}.  Extensive experimentation showed Brown that this movement was {\it not} biological in origin; rather, it was a ubiquitous property of organic and inorganic particles suspended in liquids. 

Later, Albert Einstein \cite{EinsteinBM} predicted that in a dilute suspension, the number density of particles, $n$, as a function of height, $z$, in the earth's gravitational field (acceleration = $g$) should follow an exponential (or `barometric') distribution:
\begin{eqnarray}
n(z) & = & n(0)e^{-z/z_0}, \; \mbox{where} \label{eq:barometric}\\
z_0 & = & k_B T/\Delta m g \label{eq:gravheight}
\end{eqnarray}
is the sedimentation height; here $\Delta m$ is the buoyant mass of a particle, $k_B$ is Boltzmann's constant and $T$ is the absolute temperature. Einstein assumed that a suspended particle was in thermal equilibrium with the liquid molecule `heat bath', so that equipartition and therefore the Boltzmann distribution applied. Jean Perrin's experiments using  gum resin particles showed that this was indeed the case \cite{PerrinAtoms}. 

Einstein and Perrin laid the foundation for a physics of colloids. Indeed, Eq.~(\ref{eq:barometric}) defines a colloid: a suspension of particles for which $z_0 \gtrsim a$, where $a$ is the particle radius. For everyday densities, this criterion gives an upper bound to the `colloidal length scale' of $a \lesssim 10^0 \mu$m. Equation~(\ref{eq:barometric}) demonstrates that a colloid is fundamentally an {\it equilibrium} thermodynamic system. This insight has underpinned modern colloid physics \cite{Puseychapter}, the beginnings of which could perhaps be traced back to the classic demonstration by Pusey and van Megen in 1985 \cite{PvM} that concentrated suspensions of sterically-stabilised particles showed the equilibrium phase behaviour predicted for a collection of hard spheres \cite{Hoover}. 

Developments since then (see e.g. \cite{RusselBook,HenkBook,Wagner,Poonchapter} and elsewhere in this volume) have largely followed two paths. First, the complexity of the system can be increased from hard to soft and attractive interactions, and from spheres to a variety of anisotropic shapes (rods, etc.); mixtures bring further complexity. The statistical mechanics of Boltzmann and Gibbs provides the theoretical framework for understanding the equilibrium properties of these complex colloids. Secondly, external fields (electric, magnetic, shear, etc.) can be used to drive suspensions away from thermal equilibrium, and the processes of relaxation back to equilibrium (e.g. crystallisation), various long-lived metastable states (glasses and gels), and a plethora of transient phenomena and steady states can be studied. The statistical mechanics of driven colloids is less well developed, though progress is rapid.

An active colloid is a suspension in which the particles transduce free energy from their environment to engage in various {\it intrinsically non-equilibrium} processes. To date, attention has focussed on {\it self propelled} particles. The most obvious examples of such active colloids are provided by nature: various motile bacteria, of which {\it Escherichia coli} \footnote{First reported by the German doctor Theodor Escherich in his 1886 {\it habilitation} thesis. Escherich was studying the feces of healthy children to understand the relation between internal (enteric) bacteria and infant digestion. Called {\it Bacillus coli} in                                                                               earlier literature, this bacterium is well understood on the molecular genetic level, and is a model for the biophysics of motility.} is the best understood \cite{Berg}; but synthetic self-propelled colloids (colloidal `swimmers') have also been available for over a decade \cite{Howse,Sen}. For example, Janus \footnote{In Roman mythology, Janus is a two-faced god who simultaneously looks to the past and the future; his precise role in the pantheon is still not entirely certain.} polystyrene spheres half coated with platinum are motile in an aqueous solution of H$_2$O$_2$. Of course, bacteria also manifest their `active' status in many other ways, which invite mimesis from material scientists. Thus, these `natural active colloids' are capable of sensing their environment; the coupling of this ability to motility produces a class of active behaviour known as `taxis'. For example, a chemotactic bacterium moves up a concentration gradient of nutrients \cite{Bergchemo}. Some synthetic self-propelled particles appear capable of chemotaxis as well \cite{Hong}. Bacteria also grow and divide \cite{Koch}. Fully self-reproducing colloids have not yet been synthesised, though encouraging developments are already being reported \cite{ChaikinReproduce}.

While both driven and active colloids are non-equilibrium systems, there is a fundamental difference between them. A driven colloid is in an {\it extrinsic} non-equilibrium state due an external field. The individual particles themselves are passive. In contrast, an active colloid is intrinsically non-equilibrium: each particle is not in thermal equilibrium with its surroundings, so that even without external driving, a suspension of active colloids is already in a non-equilibrium (albeit perhaps steady) state. To bring out the contrast in another way, we can say that each active particle generates an `internal field', which affects its own state and the state of other particles. Thus, e.g., we shall see that a self-propelled particle generates a flow field around itself that typically has dipolar symmetry in the far field. Interestingly, the language of `internal fields' was indeed used as a defining feature when the concept of `active Brownian particles' first appeared  \cite{Malchow}. 

Active colloids are interesting for a number of reasons. Fundamentally, there is yet no general theory of the many-body physics of intrinsically non-equilibrium particles, in which detailed balanced, and therefore the Boltzmann distribution, do not apply. Experiments with well-characterised active colloids provide crucial data for mastering this next grand challenge in statistical mechanics. Studying bacteria as active colloids may also pay dividends for microbiology, e.g. elucidating how chemotaxis may be hindered by the structure of porous media \cite{Croze}. Active colloids will show novel forms of self assembly, both on their own and mixed with passive components. They also promise new strategies for delivering microscopic `cargos'. Indeed, the medical application of colloidal `nano-robots' has long featured in science fiction \footnote{The trail blazer was the 1966 movie {\it Fantastic Voyage}, in which a submarine carrying an American medical team was shrunk to $1 \mu$m and injected into the body of a Russian defector to destroy a blood clot in his brain. Its inspiration was likely Richard Feynman's 1959 lecture {\it There's plenty of room at the bottom} (see http://www.its.caltech.edu/~feynman/plenty.html), in which he credited Albert Hibbs with the notion of a nano-robot: `A friend of mine (Albert R. Hibbs) suggests \ldots a very wild idea \ldots You put the mechanical surgeon inside the blood vessel and it goes into the heart and ÒlooksÓ around \ldots finds out which valve is the faulty one and takes a little knife and slices it out.' Feynamn and Hibbs co-authored a famous text on path integrals.}.  

Active colloids is not the only kind of `active soft matter'. Other classes include active polymers, e.g. actin-myosin gels {\it in vitro} \cite{Bausch} and {\it in vivo} \cite{Rhoda}, and active emulsions, e.g. droplets undergoing Belousov-Zhabotinsky reactions \cite{Seth}. Research in these areas may lead to general principles for describing active materials, perhaps even living systems. 

A number of existing surveys of active colloids focus  on generic and theoretical aspects \cite{Sriram,Schweitzer,CatesReview2012,Romanczuk}. Below, I start from the more `particularist' perspective of the experimentalist, who first of all wants to know about actual active colloids available, of which there are two kinds: natural ones, bacteria, and synthetic ones, typically particles with heterogeneous surface chemistry. 
I review these two classes of active particles in Sections~\ref{sec:bacteria} and \ref{sec:janus}, exploring one propulsive mechanism in detail in each case. Next, I explain how experimentalists can characterise the activity of these systems, Section~\ref{sec:ddm}, before moving on to introduce aspects of the generic physics of active colloid, Section~\ref{sec:generic}. Concluding remarks in Section~\ref{sec:last} are followed by an appendix on `practical microbiology for  physicists'.

\section{Bacteria as active colloids} \label{sec:bacteria}

The bacterium is the simplest and smallest form of autonomous life on earth today, and the first living cells were probably bacteria-like. Interestingly, most bacteria have sizes in the range $10^{-1} - 10^{0} \mu$m ({\it Mycoplasma genitalium}, with the smallest known genome, is $\approx 0.2 \mu$m in diameter) and $\approx 10\%$ less dense than water; i.e., suspensions of bacteria are colloids. Before turning to the physics of bacterial self propulsion, I first pause to consider an intriguing question: Must bacteria be colloidal?

\subsection{Mu{\ss} es seine? Es mu{\ss} seine!\protect\footnote{`Must it be? It must be!' Words written in the score of the last movement of Beethoven's String Quartet No. 16, op. 135 in F major, the last large-scale work the composer completed.}}
That bacteria live in the colloidal domain may not be a biological accident; instead, a number of physical factors may dictate that the smallest units of life must be colloidal. 

First, the origin of life depended on the availability of micro-reactors: self-contained environments for the development of `individuals' with self-sustained chains of chemical reactions. Various possibilities have been suggested; one intriguing observation is that above the sea surface there is a population of aerosol droplets with a size distribution peaked at $1 - 2 \mu$m and a residence time of hours to days \cite{Porter}. In any case, it is likely that pre-biotic droplets suitable as micro-reactors  were colloidal for physical reasons.

Secondly, the first cells, like today's bacteria, have little internal structure compared to the eukaryotic cells in our bodies, so that the transport of small molecules relies entirely on three-dimensional (3D) diffusion throughout the cell volume. Well-rehearsed arguments \cite{Schulz} show that for 3D diffusion to sustain viable reaction rates, cells must be $\lesssim 10^0 \mu$m, i.e. they must be colloidal. Above this size, efficient intracellular transport depends on reduced dimensionality \cite{Delbruck}; hence the ubiquity of `rails' (e.g. actin filaments and microtubules) and membranes inside eukaryotes \footnote{In the $\lesssim 0.5$~ mm bacterium {\it Thiomagarita namibiensis} \cite{Schulz} each cell contains a large aqueous vacuo, and the cytoplasm, the seat of biochemistry, is confined to a $\sim 1 \mu$m layer.}.

Thirdly, there is an argument from information storage. While in most cases there is no obvious relationship between cell size and genome size, this is almost certainly not true for the smallest bacteria. The {\it M. genitalium} genome has 0.58~Mbp (mega base pairs) \cite{MycoplasmaGenome}. Approximating a double-stranded DNA molecule as a cylinder of diameter 2~nm, with each bp requiring a thickness 0.34~nm, we find that the {\it M. genitalium} genome occupies $\approx 0.0006 \mu$m$^3$. Since each cell is roughly a sphere of radius $0.2 \mu$m, the cell volume is $\approx 0.004 \mu$m$^3$. If most of the genome specifies proteins, and a few copies of each protein is constitutively expressed to give a 50\% protein solution, then taking the typical density of globular proteins, we find that the cell is full. In other words, if we could predict that it takes order 500 genes to specify a self-sustained chain of polymeric reactions, then the smallest life form utilising nucleic acids and proteins must be colloidal.

A final physical reason why the first cells should be colloidal (and not bigger) is that occupying this size range confers some motility without needing to wait for the evolution of specialised propulsive mechanisms. Particles in the colloidal length scale can remain suspended and diffuse significant distances by Brownian motion.

While each of these arguments on its own is more or less speculative, they combine with some force to suggest that the simplest cells must be colloidal for reasons of physics. 

\subsection{Life at low Reynolds numbers \label{subsec:Re} \protect\footnote{This is the title of a famous paper by Purcell \cite{PurcellAJP} on bacterial locomotion, but the foundations were laid earlier by G. J. Hancock, G. I. Taylor, J. Lighthill and others. The material in this section is covered more formally, but admirably clearly, in a recent review \cite{LaugaRev}.} }

An {\it E. coli} bacterium, with a $1 \mu\mbox{m} \times 2 \mu\mbox{m}$ cell body, swims typically at $\gtrsim 10 \mu$ms$^{-1}$ \footnote{See \cite{Dusenbery} for a fascinating study of velocity-size scaling from bacteria to mammals.}. To understand the significance of these `vital statistics' of {\it E. coli} motility, we need first to remind ourselves of the basics of fluid dynamics \footnote{For a basic introduction, see Chapter 5 of \cite{NelsonBook}. For a fuller treatment, see \cite{Guyon}, which is one of the best introductions to hydrodynamics as a branch of physics rather than `applied maths'.}

The forces per unit volume acting on an element of (Newtonian) fluid (density $\rho$) arise from the pressure ($p$) gradient, $\nabla p$, the viscous stress, $\eta \nabla^2 \mathbf{v}$ and external agencies,~$\mathbf{f}$. Newton's law of motion for a fluid element (per unit volume), taking into account its advection by the flow field, gives the Navier-Stokes equation
\begin{equation}
\rho \left(\frac{\partial \mathbf{v}}{\partial t} + \mathbf{v} \cdot \nabla \mathbf{v} \right) = -\nabla p + \eta \nabla^2 \mathbf{v} + \mathbf{f}. \label{eq:navier}
\end{equation}
The magnitude of the inertial term, $ \rho \mathbf{v} \cdot \nabla \mathbf{v}$, scales as $\rho U^2/L$, where $U$ and $L$ set the velocity and length scales of the problem respectively, while the magnitude of the viscous term, $\eta \nabla^2 \mathbf{v}$, scales as $\eta U/L^2$. Their ratio gives the Reynolds number, so that when Re~$= \rho U L/\eta \ll 1$, we can neglect the inertial term. For {\it E. coli} swimming in water ($\rho \simeq 10^3$kgm$^{-3}$, $\eta \simeq 10^{-3}$Pa.s), Re~$\sim 10^{-5}$. The remoteness of this regime from everyday fluid dynamics can be appreciated by estimating stopping distances at the cessation of propulsion. Typical inertial and viscous drag forces scale as $(\rho U^2/L) \times L^3$ and $(\eta U/L^2)\times L^3$ respectively, and the swimmer mass (density $\rho_s$) scales as $\rho_s L^3$, so the typical decelerations are $(\rho/\rho_s)(U^2/L)$ and $(\eta/\rho_s)(U/L^2)$ respectively, from which we obtain the typical stopping distances in units of $L$ to be $\rho_s/\rho$ and $(\rho_s/\rho)\times \mbox{Re}$ for high and low Re regimes. Thus, humans coast for $\sim 10^0$m when we stop swimming, but a bacterium coasts for $\lesssim 1$\AA$\,$.

At low Re we can neglect the non-linear inertial term in Eq.~(\ref{eq:navier}). Moreover, time-dependent forces scale as $\rho U/\tau$, where $\tau$ is the timescale over which velocities change. If $L^2/\tau \nu \ll 1$, where $\nu = \eta/\rho$ \footnote{The kinematic viscosity $\nu$ is the diffusion coefficient for the transport of vorticity, $\nabla \times \mathbf{v}$.}, such as in the kind of quasi-stationary flows we are interested in (where $\tau \rightarrow \infty$), we can neglect the time derivative as well. What is left is the Stokes equation governing flow at low Reynolds numbers (or `creeping flow') \footnote{Chapter 8 of \cite{Guyon} gives an insightful introduction to creeping flow.}:
\begin{equation}
-\nabla p + \eta \nabla^2 \mathbf{v} + \mathbf{f}= 0 , \label{eq:stokes}
\end{equation}
to be solved subject to the incompressibility of the fluid, $\nabla \cdot \mathbf{v} = 0$, and appropriate boundary conditions (such as no slip on all solid surfaces).

Since there are no time derivatives, the flow field responds instantaneously to applied forces and boundary conditions. This means that Stokes flow is not so much  a problem in fluid dynamics as a problem in `fluid statics' --- the forces and torques on each fluid element balance at every instant, because time lag effects from inertia are absent. 

We now consider two important consequences of the form of Eq.~(\ref{eq:stokes}). 

\subsubsection{Linearity and superposition} Since Eq.~(\ref{eq:stokes}) is linear, one way to solve particular problems is to superpose various `singularity solutions' \cite{RusselBook} to satisfy given boundary conditions, and then appeal to the uniqueness of the solutions to this equation. The most well known `singularity solution' is the flow field corresponding to a point force, i.e. Eq~(\ref{eq:stokes}) with $\mathbf{f} = \mathbf{F}\delta(\mathbf{r})$. For $\mathbf{F} = (F, 0, 0)$ (i.e. a point force acting along $x$) \footnote{A concise derivation is given by Lighthill \cite{LighthillBook}; the book is now printed on demand by CUP.}: 
\begin{equation}
\mathbf{v} = \frac{F}{8\pi\eta} \left( \frac{x^2 + r^2}{r^3} , \frac{xy}{r^3}, \frac{xz}{r^3} \right), \;p = \frac{Fx}{4\pi r^3}. \label{eq:oseen}
\end{equation}
Note that this velocity field, known as the Oseen tensor or a `stokeslet', is long range: it falls off as the inverse of the distance to the point force, $v \sim r^{-1}$ \footnote{Throughout, `$A \sim B$' means `$A$ scales as $B$' (so that $A$, $B$ may have different units) and `$A \simeq B$' means $A$ is equal to $B$ up to a numerical multiplier (so that $A$, $B$ have the same units).}. A second salient feature is that $v \sim F$, which is a direct consequence of the linearity of Eq.~(\ref{eq:stokes}). Other singularity solutions, e.g. the `stresslet' corresponding to a point force dipole at the origin, also show proportionality between `response' and `stimulus'. 

To understand the essence of the approach to solving particular problems by superposing singularity solutions, consider (schematically) how to obtain the Stokes drag on a sphere. It can be shown that superposing the stokeslet, Eq.~(\ref{eq:oseen}), with the flow field of another singularity solution, that of a `potential doublet' of strength $G = Fa^2/6\eta$ at the origin \footnote{This gives rise to a velocity field $v = \nabla \phi$ with $\phi = Gx/4\pi r^3$. See Lighthill \cite{LighthillBook} for details.} gives rise to a uniform velocity $\mathbf{U} = \mathbf{F}/6\pi\eta a$ on the surface of a sphere of radius $a$ centred on the origin. The force on such a sphere, obtained by integrating the pressure and viscous stresses over its surface, must be equal in magnitude and opposite in direction to the force on the fluid, which we know to be $\mathbf{F}$ from the stokeslet (the flow due to the potential doublet exerts no net force on the fluid), from which we obtain directly the well known result that the force applied by the fluid of viscosity $\eta$ on a sphere of radius $a$ moving through it at velocity $\mathbf{U}$ is $\mathbb{F} = - 6\pi\eta a \mathbf{U}$ \footnote{Throughout, the font $\mathbb{F}$ denotes {\it drag}.}.

This proportionality between force and velocity can be generalised to any body, the motion of which is specified by its centre-of-mass velocity, $\mathbf{U}$, and its angular velocity, $\mathbf{\Omega}$, so that the boundary condition to be satisfied is that the fluid velocity on the surface of the body is $\mathbf{v} = \mathbf{U} + \mathbf{\Omega} \times \mathbf{r}$. The linearity of Eq.~(\ref{eq:stokes}) then ensures that the force and torque applied to the body by the fluid, $\mathbb{F}$ and $\mathbb{N}$, are related linearly to $\mathbf{U}$ and $\mathbf{\Omega}$:
\begin{equation}
\left( \begin{array}{cc}
\mathbb{F} \\
\mathbb{N}
\end{array}
\right) = 
-\left( \begin{array}{cc}
{\rm \uuline A} & {\rm \uuline B} \\
{\rm \uuline C} & {\rm \uuline D} \\
\end{array}
\right)
\left( \begin{array}{cc}
\mathbf{U} \\
\mathbf{\Omega}
\end{array}
\right). \label{eq:linear}
\end{equation}
For a body of arbitrary shape, each of \uuline{A}, \uuline{B}, \uuline{C}, \uuline{D} is a $3\times3$ matrix. The full matrix of coefficients is called the resistance matrix in the fluid dynamics literature, but christened the `propulsion matrix' by Purcell \cite{PurcellPNAS}. It can be shown quite generally that, in the absence of hydrodynamic interactions, $\uuline{\mbox{A}}$ and $\uuline{\mbox{B}}$ are symmetric, and $B_{ij} = C_{ji}$ \cite{Guyon}. Dimensional analysis tells us that  $A_{ij} \sim \eta L$, $B_{ij} \sim \eta L^2$ and $D_{ij} \sim \eta L^3$, where $L$ is a typical length scale in the problem. Since \uuline{A} and \uuline{D} are symmetric, there exist a set of orthogonal axes under which they can be simultaneously diagonalised. For a sphere of radius $a$, \uuline{B} = \uuline{C} = \uuline{0}, and \uuline{A} and \uuline{D} reduce to \uuline{A}~$ = 6\pi \eta a$\uuline{I} and \uuline{D}~$ = 8\pi \eta a^3$\uuline{I} (where \uuline{I} is the identity matrix), with the expected $\eta$ and $a$ scalings.

\begin{figure}
\begin{center}
\includegraphics[height=3.5cm]{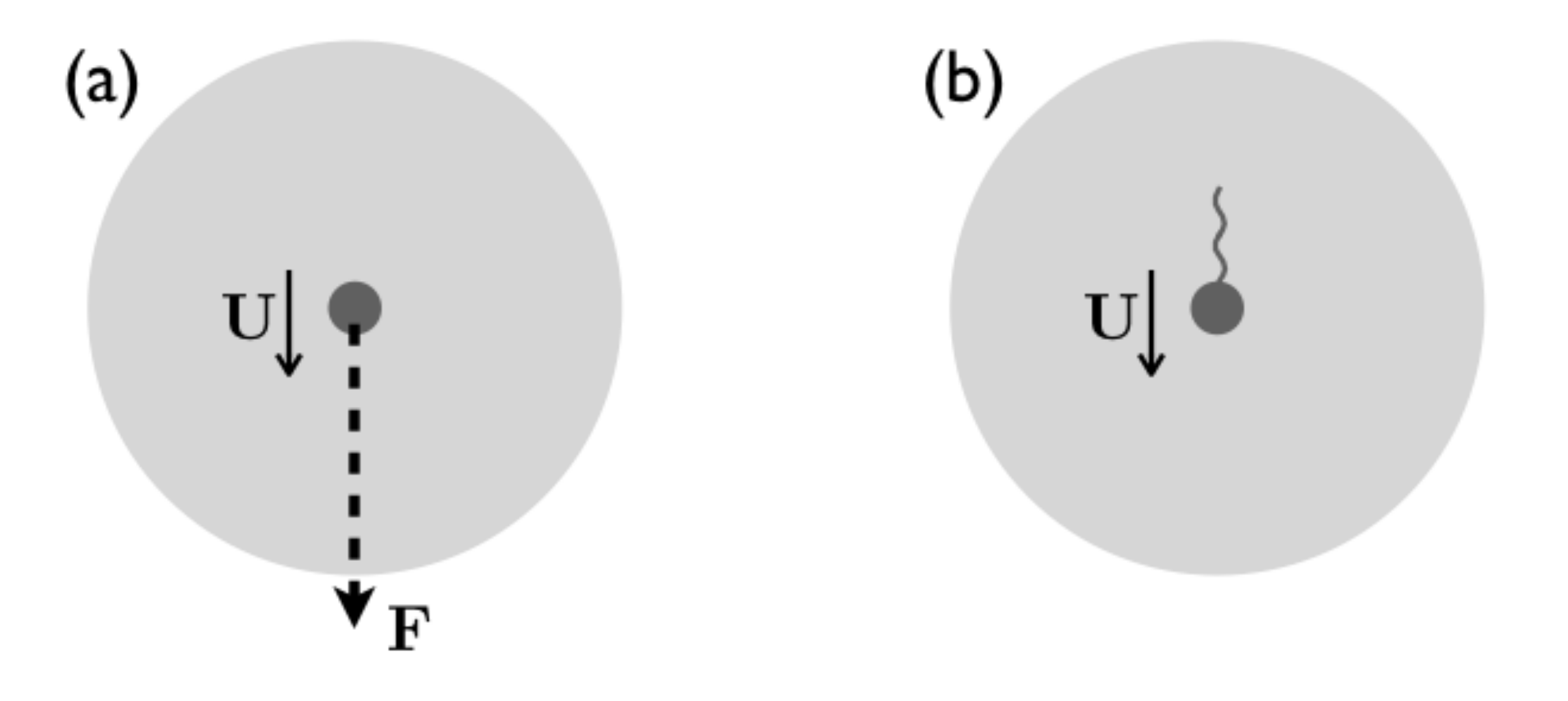}  
\end{center}  
\caption{Schematic illustration of an essential difference between a `dragged' particle and a self-propelled particle, each contained in an isolated liquid droplet. (a) The particle moves under an external force $\mathbf{F}$, so that the whole system (droplet + particle) is under the same external force, and its centre of mass accelerates in the direction of $\mathbf{F}$. (b) The particle is self propelled. The whole system (droplet + particle) is force free, and its centre of mass remains stationary.} \label{fig:noforce}
\end{figure}

We have seen that a fluid subject to a point force displays a flow field that decays as $r^{-1}$ away from the point of application, Eq.~(\ref{eq:oseen}). This long-range decay describes the far-field flow generated by any solid body translating at low Reynolds number in a fluid under an external force, e.g. a particle of arbitrary shape sedimenting under gravity. In these situations, a net external force is exerted on the fluid. Note that the case of any self-propelled body, whatever the mechanism of propulsion, is very different. In this case, there is {\it no external force} acting on the particle, and therefore on the fluid, Fig.~\ref{fig:noforce}. The far-field flow therefore must decay not as a stokeslet, but as the superposition of  higher-order multipoles, the lowest-order of which is the `stresslet', the singularity solution due to a dipole of strength $Fb$ at the origin pointing along  $x$:
\begin{equation}
\mathbf{v} = \frac{Fb}{8\pi\eta}\left( \frac{1}{r^3} + \frac{3x^2}{r^5} \right) \mathbf{r},
\end{equation}
which decays $\sim r^{-2}$. Hydrodynamic interactions between self-propelled particles are therefore weaker than those between particles translated by external forces. 

\subsubsection{The scallop theorem} Another consequence of the form of Eq.~(\ref{eq:stokes}) is reversibility. If $\mathbf{v}(\mathbf{r})$ is a solution of Eq.~(\ref{eq:stokes}) with associated pressure field $p(\mathbf{r})$, then $-\mathbf{v}(\mathbf{r})$ is also a solution with all the forces reversed and the reverse pressure gradient. A dramatic demonstration of this was  given by G. I. Taylor in a well known movie, in which a vertical streak of ink injected into a viscous fluid between double cylinders was smeared by rotating the inner cylinder, and then perfectly `de-smeared' by exactly counter-rotating the inner cylinder back to its starting point \footnote{See \url{http://web.mit.edu/hml/ncfmf.html} under `Low Reynolds Number Flow'.}. One consequence, Purcell's `scallop theorem', is that propulsion cannot be obtained by reciprocating movement: a perfectly symmetric scallop opening and shutting at Re~$\ll 1$ is not going anywhere!

Purcell proposed an artificial `three-linked swimmer' with a non-reciprocating motion cycle \cite{PurcellAJP}. This device, now realised macroscopically as part of the thesis work of of a graduate student at MIT \footnote{Thesis available at \url{http://web.mit.edu/chosetec/Public/thesis}; a movie is currently available from \url{http://www.youtube.com/watch?v=f-sIaYrH45U}.}, harbours much more complexity than Purcell originally envisaged \footnote{Purcell left the direction of motion of his swimmer as an exercise for the reader; it turns out that both directions are realisable in different parts of parameter space \cite{Stone}.}. A version suitable for implementation with colloids has been proposed \cite{GolestanianSwimmer}, which, although not yet been realised as a swimmer, has been implemented using laser tweezers as a pumping device \cite{Cicuta}. In either case, the key point is that non-reciprocating motion leads to relative motion between the colloids and the surrounding liquid. 

Micro-organisms generate a variety of non-reciprocating motion for propulsion, often involving some form of flagellum \footnote{Reviewed in Lighthill's 1975 John von Neumann Lecture \cite{LighthillSIAM}. Though now dated in some aspects, this older work (including the hand-drawn Fig. 1) is still a {\it tour de force} in its scope.}. I now review the case of certain flagellated bacteria in more detail, which serves to illustrate many points of general relevance.

\begin{figure}
\begin{center}
\includegraphics[height=2.3cm]{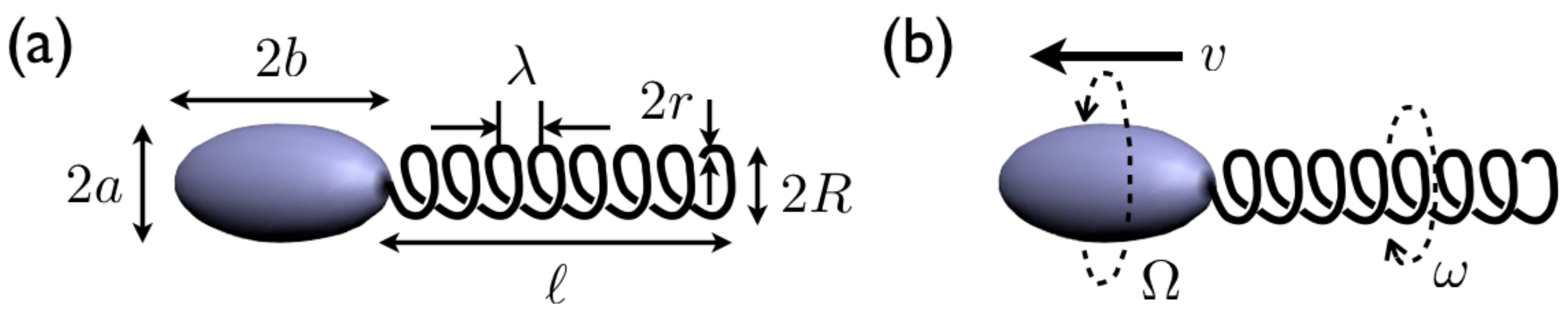}  
\end{center}  
\caption{Model of a bacterium with an ellipsoidal cell body and a single helical flagellum attached to the trailing cell pole. (a) The relevant geometric parameters defining the model. (b) The three dynamical quantities characterising the locomotion of such a model bacterium.  } \label{fig:coli}
\end{figure}

\subsection{{\textit E. coli} in motion} \protect\footnote{Howard Berg's book of this title \cite{Berg} contains extensive references.} The cell body of the {\it E. coli} bacterium is approximately a spherocylinder. Immediately after cell division, this spherocylinder has end-to-end length $L \gtrsim 2 \mu$m and diameter $D \lesssim 1 \mu$m, so that the aspect ratio is $L/D \gtrsim 2$. The cell body, often modelled as a prolate ellipsoid with the same aspect ratio, is propelled by a number (typically $\sim 6-10$) of helical flagella, each of which is powered by an intricate rotary motor driven by proton (H$^+$) currents flowing from outside the cell under a proton-motive force (pmf) of $\approx 150$mV during `normal operation'. Each flagellum is a left-handed helix, Fig.~\ref{fig:coli}(a), made up of a coiled filament of diameter $r \approx 20$nm and pitch $\lambda \approx 1.5 \mu$m. The diameter and length of the helix are $R \approx 0.2 \mu$m and $\ell \approx 6-10 \mu$m respectively. When all the flagella are turning counterclockwise (CCW) viewed from behind, they bundle  and propel the cell forward at a speed of $v \gtrsim 10 \mu$ms$^{-1}$.  

\subsubsection{Propulsive mechanism} \label{subsubsec:propulsion} 
\begin{figure}
\begin{center}
\includegraphics[height=3cm]{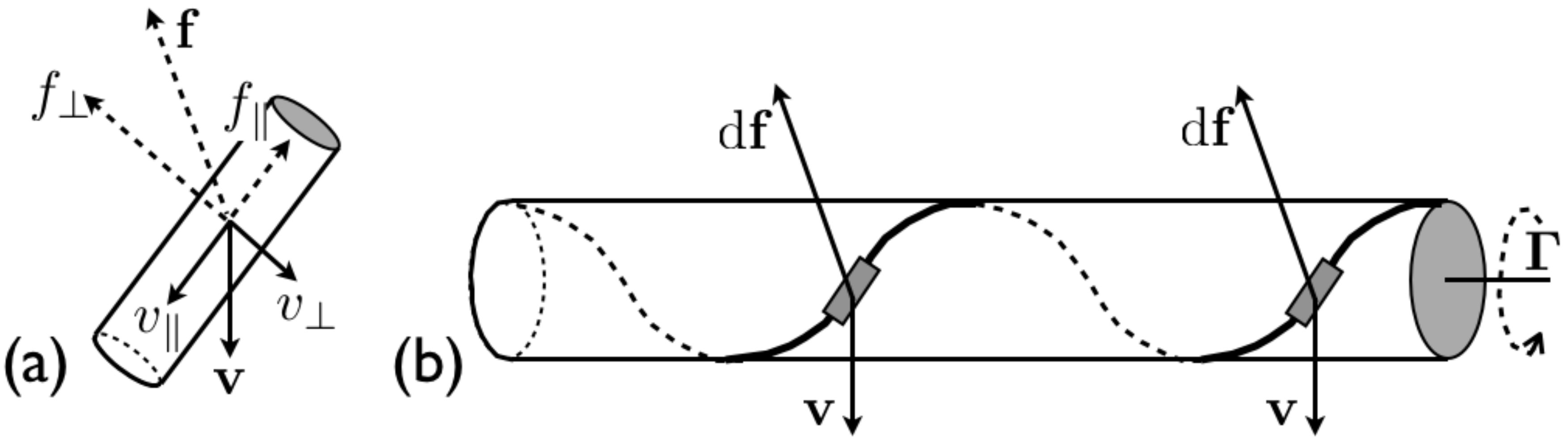}  
\end{center}  
\caption{Rotational-translational coupling in a helix. (a) A small cylindrical element. The drag parallel (perpendicular) to the axis is $f_\parallel = \xi_\parallel v_\parallel$  ($f_\perp = \xi_\perp v_\perp$). Since $\xi_\perp > \xi_\parallel$, the net drag force $\mathbf{f}$ is {\it not} anti-parallel to $\mathbf{v}$. (b) A left-handed helix (wrapped onto a cylinder for visual clarity) with a torque $\boldsymbol{\Gamma}$ applied in the direction indicated, showing the velocity and force on two differential cylindrical elements on successive turns of the helix. Summing ${\rm d}\mathbf{f}$ along the helix contour, we see that the fluid exerts a net force to the left on the helix. Redrawn after \cite{NelsonBook}.} \label{fig:rottrans}
\end{figure}

Partly for simplicity, and partly because the physics of flagella bundling is still far from understood, the real swimming {\it E. coli} with a flagella bundle is almost invariably modelled as a cell body being propelled by a single `effective' helical flagellum from behind, although real bacteria propelled by a single trailing polar flagellum do exist (e.g. {\it Pseudomonas aeruginosa}). Until as late as the mid-1970s, it was not clear whether the filamentous flagellum (whether genuinely a single filament or an `effective' bundle) propels by continuously sending helical waves down its length, or by virtual of being a rotating helix with more or less rigid conformation (see e.g. the discussion in \cite{LighthillSIAM}). It is now known that the latter is the case for bacteria like {\it E. coli} and {\it P. aeruginosa}. Interestingly, flagella propulsion by either mechanism ultimately relies on the same basic physics: that it is easier to drag a cylinder in a fluid along its length than perpendicular to it, Fig.~\ref{fig:rottrans}.

Consider first the drag on the cylinder shown in  Fig.~\ref{fig:rottrans}(a). The drag per unit length parallel and perpendicular to the axis are given by $\xi_\parallel v_\parallel$ and $\xi_\perp v_\perp$ respectively (linearity of Stokes flow). Since, $\xi_\perp > \xi_\parallel$, the net drag is slanted to the left of the direction of $-\mathbf{v}$. In  Fig.~\ref{fig:rottrans}(b), we divide a (left-handed) helix into cylindrical elements, and subject each element to the same analysis as in Fig.~\ref{fig:rottrans}(a). An external torque is applied to the helix with the sense shown. Summing the differential force elements along the contour of the helix, we find that the fluid exerts a net force to the left on the helix. Thus, in the notation of Eq.~(\ref{eq:linear}), $\uuline{\rm B}, \uuline{\rm C} \neq \uuline{0}$ for the helix, because $\xi_\perp/\xi_\parallel \neq 1$ for cylinders.

Formally, a general argument can be given \cite{Stone} why a putatively self-propelled, and therefore force-free (remember Fig.~\ref{fig:noforce}(b)), filament of constant length $L$ with {\it isotropic} local drag cannot in fact change the position of its average centre of mass, denoted by $\bar{\mathbf{r}}$:
\begin{equation}
\frac{d \bar{\mathbf{r}}}{dt} = \frac{1}{L} \frac{d}{dt} \int_0^L \mathbf{r}(s,t)ds = \frac{1}{L} \int_0^L \frac{\partial \mathbf{r}}{\partial t} ds = \frac{1}{L} \int_0^L \mathbf{U} ds \propto - \int_0^L \mathbf{f} ds = 0 . \label{eq:stone}
\end{equation}
Here $\mathbf{r}(s,t)$ is the position of the differential element $ds$ along the contour length $s$ of the filament. We have used the key assumption that the local drag and velocity are strictly anti-parallel, $\mathbf{U} \propto -\mathbf{f}$, so that the situation in Fig.~\ref{fig:rottrans}(a) does not occur. The last equality follows from the force-free nature of the motion. Thus, propulsion by filamentous appendages depends essentially on the drag anisotropy of slender bodies. 

Three kinematic quantities describe the locomotion of our model bacterium, Fig.~\ref{fig:coli}(b): the translation velocity of the organism $\mathbf{v}$ and the the angular velocities of the flagellum, $\boldsymbol{\omega}$, and the body, $\boldsymbol{\Omega}$; the two angular velocities must be in opposite directions so that the whole organism is torque free. The angular velocity of the motor in the stationary frame of the body is  $\boldsymbol{\omega}_m = \boldsymbol{\omega} - \mathbf{\Omega}$ (note that the magnitudes add, $|\boldsymbol{\omega}_m| = |\boldsymbol{\omega}| + |\mathbf{\Omega}|$). 

\begin{figure}
\begin{center}
\includegraphics[height=4.5cm]{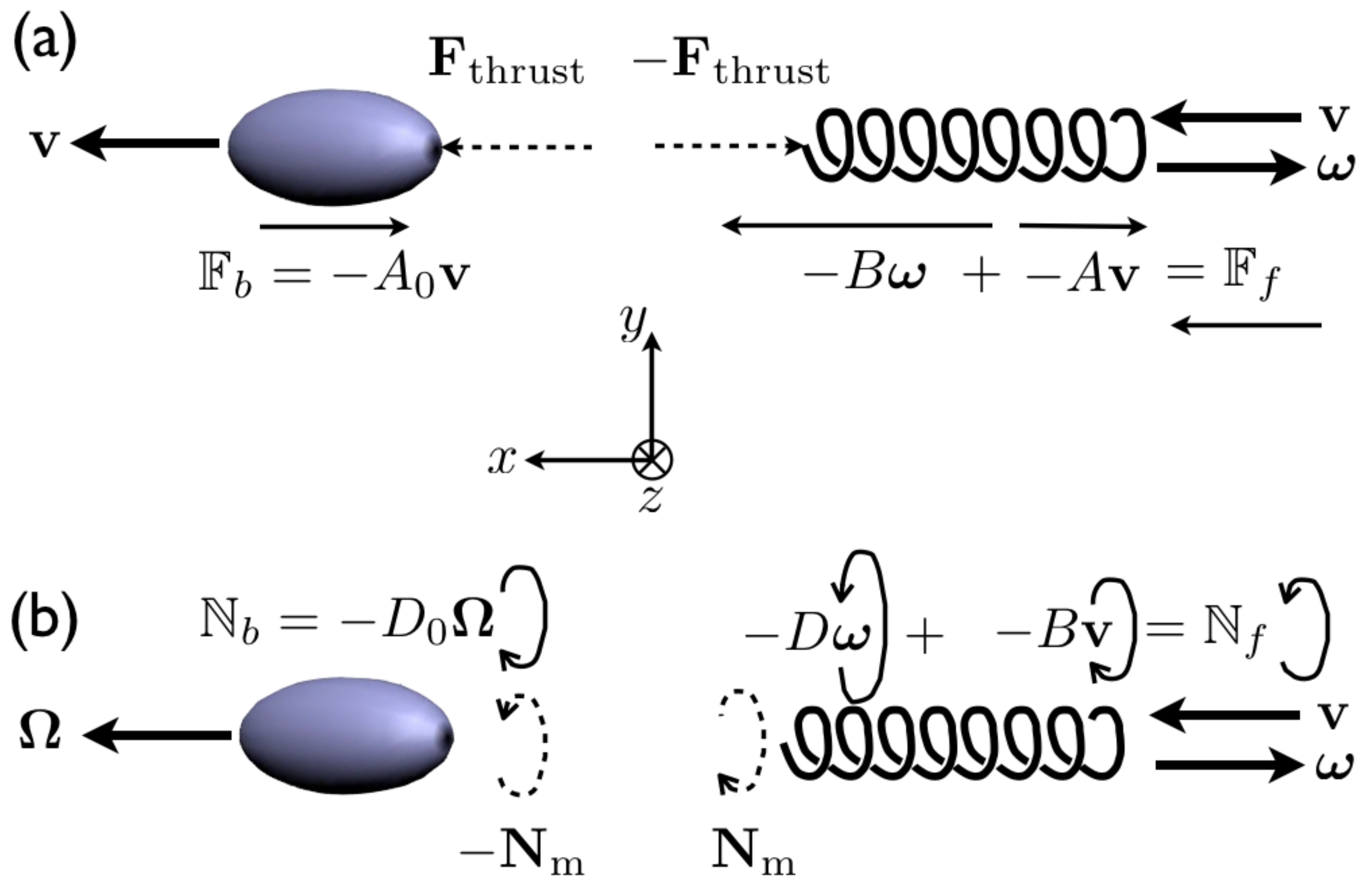}  
\hspace{1cm}\includegraphics[height=4.5cm]{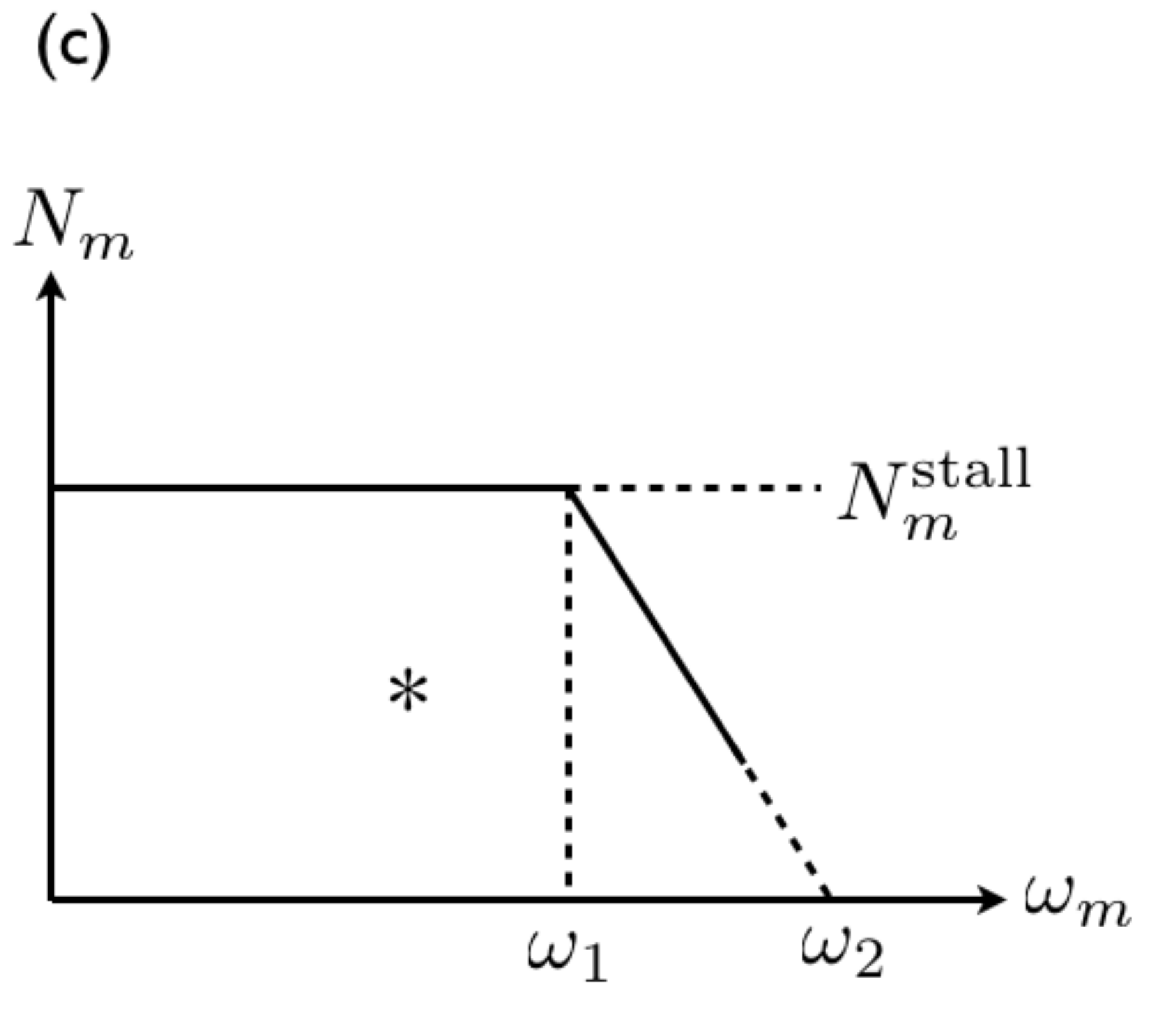}
\end{center}  
\caption{(a), (b) Forces ($F$) and torques ($N$) in a model bacterium swimming at velocity $\mathbf{v}$, Eqs.~(\ref{eq:bodyFN}) and (\ref{eq:helixFN}). The motor on the body applies a torque, $\mathbf{N}_{\rm m}$, to the flagellum, (b), rotating it at $\boldsymbol{\omega}$. Rotation-translation coupling means that the liquid exerts a force $-B\boldsymbol{\omega}$ on the flagellum, (a); the flagellum translating at $\mathbf{v}$ experiences a drag of $-A\mathbf{v}$. The net force exerted by the liquid on the flagellum, $\mathbb{F}_f$, is the thrust on the body, $\mathbf{F}_{\rm thrust}$, which balances the drag on the body, $\mathbb{F}_b$, to render it force free. The equal and opposite Newton III `partners' to $\mathbb{F}_f$ and $\mathbb{F}_b$ (not shown) makes up the force dipole exerted by the bacterium on the liquid. A similar analysis can be performed on the torques. (c) Schematic of the relation between motor angular speed, $\omega_m$, and motor torque, $N_m$; for {\it E. coli}, $N_m^{\rm stall} \approx 10^3$pN.nm, and, at room temperature, $\omega_1/2\pi \approx 200$Hz and $\omega_2/2\pi \approx 300$Hz \cite{BerryReview}. $\ast$ = approximate torque-speed values inferred from the measurements in \cite{WuEfficiency}.} \label{fig:coliforces}
\end{figure}

To the analyse this motion quantitatively, we need to set up forms of Eq.~(\ref{eq:linear}) with resistive matrices for the cell body and the flagellum, and then equate forces and torques on these two parts of the organism \cite{PurcellAJP,PurcellPNAS,WuEfficiency} \footnote{Note that by equating the forces and torques on an isolated body and an isolated flagellum, we are neglecting hydrodynamic interaction between these parts. Purcell \cite{PurcellPNAS} expressed the hope that such interaction may be weak; but whether this hope is well founded remains untested.}. The axis of our model organism, $x$ in Fig.~\ref{fig:coliforces}, is a principal axis for both the ellipsoidal cell body and the helical flagellum, so that along this axis, the whole organism is characterised by five resistive coefficients, $A_0$, $D_0$ for the cell body (for which $B_0 = 0$), and $A, B (=C), D$ for the flagellum. A great deal can be learnt about the motion by proceeding symbolically without using specific algebraic forms for these coefficients, which depend on the geometric parameters shown in Fig.~\ref{fig:coli}(a). The drag force and torque on the cell body and the flagellum are given by:
\begin{eqnarray}
\left( \begin{array}{cc}
\mathbb{F}_b \\
\mathbb{N}_b
\end{array}
\right) & = & 
-\left( \begin{array}{cc}
A_0 & 0 \\
0& D_0 \\
\end{array}
\right)
\left( \begin{array}{cc}
\mathbf{v} \\
\mathbf{\Omega}
\end{array}
\right)  \;\mbox{and} \label{eq:bodyFN}\\
\left( \begin{array}{cc}
\mathbb{F}_f \\
\mathbb{N}_f
\end{array}
\right) & = &
-\left( \begin{array}{cc}
A & B \\
B & D \\
\end{array}
\right)
\left( \begin{array}{cc}
\mathbf{v} \\
\boldsymbol{\omega}
\end{array}
\right),\,\,\mbox{with} \label{eq:helixFN}
\end{eqnarray}
\begin{equation}
\mathbf{v} = (v,0,0),\, \boldsymbol{\omega} = (-\omega,0,0), \,\mathbf{\Omega} = (\Omega,0,0) \label{eq:components}
\end{equation} 
relative to axes defined in Fig.~\ref{fig:coliforces}. Note that in Eqs.~(\ref{eq:bodyFN}) and {\ref{eq:helixFN}), $A_0, D_0, A, D>0$, and Fig.~\ref{fig:rottrans} shows that $B>0$ for a left-handed helical flagellum \footnote{We use the right-hand corkscrew sign convention for axial vectors throughout. Note that a right-handed helix under the same sign convention would give $B < 0$.}.

The whole organism needs to be force and torque free. For force balance, Fig.~\ref{fig:coliforces}(a), the drag on the cell body is straightforwardly $\mathbb{F}_b = -A_0 \mathbf{v}$. The drag force on the flagellum consists of two terms: $\mathbb{F}_f = -A\mathbf{v} - B\boldsymbol{\omega}$, coming from friction experienced by the helix as it translates with velocity $\mathbf{v}$, and the rotational-translation coupling arising from its angular velocity $\boldsymbol{\omega}$. Note that these two terms have opposite directions, owing to the fact that $\mathbf{v}$ and $\boldsymbol{\omega}$ are antiparallel; $-A\mathbf{v}$ is a `cost', and $-B\boldsymbol{\omega}$ is a `benefit'. Successful propulsion requires that $B|\boldsymbol{\omega}| = B\omega > A v$. A net $\mathbb{F}_f$ in the $+x$ direction provides the propulsive force, $\mathbf{F}_{\rm thrust}$, to overcome the drag on the body, $\mathbb{F}_b$, while the reaction of the body on the flagellum, $- \mathbf{F}_{\rm thrust}$ balances out $\mathbb{F}_f$ to give a force free flagellum. More simply, neglecting the Newton III pair of internal forces $\pm \mathbf{F}_{\rm thrust}$, a force-free organism requires $\mathbb{F}_b + \mathbb{F}_f = 0$, or $-(A_0 + A)\mathbf{v} -B\boldsymbol{\omega} = 0$, or in component form (recalling Eq.~\ref{eq:components})
\begin{equation}
(A_0 + A)v = B \omega. \label{eq:purcell1}
\end{equation}
The body and flagellum exert forces $-\mathbb{F}_b$ and $-\mathbb{F}_f$ on the liquid respectively, Fig.~\ref{fig:coliforces}(a). The magnitude of each force is $F_{\rm thrust}$, and the centres application are separated by $\sim \ell$, the flagellar length. The self propelled bacterium acts like a force dipole.

We now repeat a similar exercise to satisfy the torque-free condition, Fig.~\ref{fig:coliforces}(b). Again the torque exerted by the liquid on the flagellum has two terms, and there is an internal torque pair consisting of the torque exerted by the motor on the flagellum, and the reaction torque on the body (to which the motor is fixed), $\pm \mathbf{N}_m$. The torque-free condition requires $\mathbb{N}_b + \mathbb{N}_f = 0$, which, from Eq.~(\ref{eq:helixFN}) and putting in components, gives
\begin{equation}
D_0 \Omega = -Bv + D\omega. \label{eq:purcell2}
\end{equation}
Reference to Fig.~\ref{fig:coliforces}(b) shows that the body and flagellum exert equal and opposite torques on the liquid separated spatially by $\sim \ell$: the organism also acts as a `torque dipole'. 

Discussions of {\it E. coli} motility in terms of the model shown in Fig.~\ref{fig:coli}(a) typically proceed from  Eqs.~(\ref{eq:purcell1}) and (\ref{eq:purcell2}), and derive results such as the propulsion speed in terms of the motor frequency $\omega_m = \Omega + \omega$, etc.. But it is important to note that given a particular set of values for the five resistive coefficients $(A_0, D_0; A, B, D)$, these  {\it two} equations do not uniquely determine the {\it three} kinematic variables $(v, \Omega, \omega)$. To do so, we need extra information, which comes from experiment in the form of the measured relation between the motor torque, $N_{m}$, and the motor angular frequency, $\omega_m$, Fig.~\ref{fig:coliforces}(c) \cite{BerryReview}, which has much the same form as the torque-speed relation of many electric motors. Since $N_m = D_0\Omega$ (from torque-free body, see Fig.~\ref{fig:coliforces}(b)), the measured function $N_m(\omega_m)$ provides an independent relation between,  effectively,  $\Omega$ and $\omega$, which, together with Eqs.~(\ref{eq:purcell1}) and (\ref{eq:purcell2}), form a closed set for the unique determination of $(v,\Omega, \omega)$. 

Equations~(\ref{eq:purcell1}) and (\ref{eq:purcell2}) on their own lead to the result that
\begin{equation}
v =\left[ \frac{BD_0}{(A_0 + A)(D_0 + D) - B^2}\right] \omega_m.
\end{equation}
Thus, the propulsion speed is directly proportional to the motor speed. This can be traced back to the linearity of the governing equation of creeping flow, Eq.~(\ref{eq:stokes}). Moreover, the rather complicated constant of proportionality depends purely on the geometry of the body and flagella: the liquid viscosity cancels out. Finally, as expected, the propulsion relies on the finite rotation-translation coupling of the flagella: if $B = 0$, $v = 0$. 

To proceed further, we need concrete expressions for the five resistive coefficients. For an ellipsoid \footnote{Strictly, the results quoted are for ellipsoids with $b/a \gg 1$}, it can be shown (e.g. using singularity methods \cite{Chwang}) that 
\begin{eqnarray}
A_0 & = & \frac{4\pi \eta b}{\ln\left(\frac{2b}{a}\right) - \frac{1}{2} }, \label{eq:A0ellipse}\\
D_0 & = & \frac{16\pi \eta a^2 b}{3}.
\end{eqnarray}
Obtaining the resistive coefficients for a rigid helix built out of cylindrical elements is less straightforward, because the problem of determining the specific (i.e. per unit length) resistance coefficients $\xi_\parallel$ and $\xi_\perp$ (see Fig.~\ref{fig:rottrans}(a)) for a long cylinder in creeping flow has no solution (the so-called Stokes paradox). The mathematical malaise is immediately evident when we look at the form taken by the tangential and normal specific friction coefficients for the middle portion of a cylinder of radius $a$ and length $2\Delta$ when $\Delta/a \gg 1$:
\begin{equation}
\xi_\parallel  =  \frac{4\pi \eta}{2\ln \left( \frac{2\Delta}{a} \right) - 1}, \;
\xi_\perp = \frac{8\pi \eta}{2\ln \left( \frac{2\Delta}{a} \right) +1}. \nonumber
\end{equation}
Embarrassingly, these results depend on $\Delta$ \footnote{Note that in the limit of $\Delta/a\rightarrow \infty$, the ratio $\gamma_k = \xi_\parallel/\xi_\perp \rightarrow 0.5$ independent of $\Delta$.}, so that their use in any calculation depends on an apparently rather arbitrary choice of this parameter. Lighthill's interpretation of the approximations necessitated by the Stokes paradox suggests that in the case of a helical filament, in which the (linear) force density necessarily varies along the filament on the length scale of $\lambda$ \footnote{In fact, Lighthill was dealing with an undulating filament with wavelength $\lambda$; for our helical filament, the relevant length scale is $\approx 2\pi R$, Fig.~\ref{fig:coli}(a); but $2\pi R \approx \lambda$ in many cases.}, $\Delta$ should chosen so that the force density is effectively constant on this length scale, i.e. we should have $\Delta \ll \lambda$. A self-consistent argument in fact returns the value $\Delta \approx 0.09\lambda$. Thus, Lighthill suggests the forms
\begin{eqnarray}
\xi_\parallel & =  & \frac{4\pi \eta}{2\ln \left( \frac{c\lambda}{a} \right) - 1} \label{eq:xipara}  \\
\xi_\perp & = & \frac{8\pi \eta}{2\ln \left( \frac{c\lambda}{a} \right) +1}, \label{eq:xiperp}
 \end{eqnarray}
with $c \approx 0.18$. Crudely, one could interpret this as saying that we consider just under one fifth of one period of the helix as locally straight. 

Irrespective of the exact form used for $\xi_\parallel$ and $\xi_\perp$, the resistive coefficients for the helical filament defined in Fig.~\ref{fig:coli}(a) can be found in terms of these two coefficients \cite{WuEfficiency}:
\begin{eqnarray}
A & = & \xi_\perp  \ell \frac{1-\beta}{\sqrt{\beta}} \left( 1 + \gamma_k \frac{\beta}{1-\beta} \right), \label{eq:A}\\
B & = & \xi_\perp \ell \left( \frac{\lambda}{2\pi} \right) \frac{1-\beta}{\sqrt{\beta}} \left( 1 - \gamma_k  \right), \label{eq:B} \\
D & = & \xi_\perp \ell \left( \frac{\lambda}{2\pi} \right)^2  \left( 1 + \gamma_k \frac{\beta}{1-\beta} \right), \label{eq:D}
\end{eqnarray}
where $\gamma_k = \xi_\parallel / \xi_\perp$, and $\beta = \cos^2 \psi$ with $\psi = \tan^{-1} (2\pi R/\lambda)$ is the helix pitch angle.  
Note that the all important rotation-translation coupling coefficient, $B$, scales as $ 1 - \gamma_k $. If there is no anisotropy in the local friction coefficients, i.e. $\gamma_k = 1$, then $B = 0$. 

It appears that this model for self-propulsion in {\it E. coli} has only been rigorously tested experimentally once: Chattopadhyay et al. \cite{WuEfficiency} used an ingenious laser-tweezers set up to measure $A, B, D$ for the `effective flagellum' (i.e. the real flagellar bundle), calculated $A_0$ and $D_0$ for ellipsoids with dimensions to fit actual cells, and reported that their measurements agreed well with those calculated using Eqs.~(\ref{eq:xipara})-(\ref{eq:D}) and were consistent with Eqs.~(\ref{eq:purcell1}) and (\ref{eq:purcell2}). However, to arrive at this conclusion, Chattopadhyay et al. have to use a value of $c \approx 2.4$ in Eqs.~(\ref{eq:xipara}) and (\ref{eq:xiperp}). Recall the geometric interpretation of this admittedly somewhat arbitrary constant as the fraction of one period of the helix over which the filament can be considered locally straight, so that using a value of $c > 1$ does not make self-evident physical sense. Another cause for concern is that, as I have already remarked, given a set of specific values for $(A_0, D_0; A, B, D)$, Eqs.~(\ref{eq:purcell1}) and (\ref{eq:purcell2}) need to be supplemented with the experimentally-determined motor torque-speed curve, $N_m(\omega_m),$ to determine $(v, \Omega, \omega)$ uniquely. Chattopadyay et al. found $N_m = D_0\Omega \approx 500$pN.nm and $\omega_m = \Omega + \omega \approx  2\pi \times 135 \mbox{Hz}$. This pair of values (`$\ast$' in Fig.~\ref{fig:coliforces}(c)) does not fit with any of the measured {\it E. coli} torque-speed relations  \cite{BerryReview}, for which (see Fig.~\ref{fig:coliforces}(c) for definitions) $N_m^{\rm stall} \gtrsim 1000$pN.nm and, at $\approx 20^\circ$C, $\omega_1 \approx 2\pi \times 200$Hz and $\omega_2 \approx 2\pi \times 300$Hz.

Thus, the applicability of the `single effective flagellum' model to {\it E. coli} propulsion remains to be demonstrated. One hint  that all is not well if one neglects the multi-flagellated nature of {\it E. coli} is that the axial torque exerted on the cell body does not seem to scale linearly as the number of flagella \cite{BergTorque}. It may yet turn out that the simple model of Fig.~\ref{fig:coli}(a) can only be applied quantitatively to an organism like  {\it P. aeruginosa}, which indeed possesses only a single helical flagellum at its trailing pole. 

\subsubsection{Wild-type swimming}

Propelled by a flagellar bundle, a wild-type {\it E. coli} cell swims for about 1s in a more or less straight `run', and then one or more flagella would unbundle for about 0.1s because the driving motor reverses direction from CCW to clockwise (CW).  The bacterium tumbles, so that when the flagella re-bundle, it swims in a different direction \cite{BergBrown}. In the long time limit, the cell engages in a random walk. 

Bacteria are capable of sensing chemical species in their environment and either move towards a favourable species (an attractant) or away from an unfavourable species (a repellant). This is the phenomenon of chemotaxis. If, during a run, an {\it E. coli} cell detects that the concentration of an attractant is increasing at successive sampling time points, then its molecular machinery lowers the tumble probability, so that over time, the random walk is biased in the direction of increasing gradient of the attractant \footnote{Porous media with restricted mean free paths may therefore interfere with chemotaxis \cite{Croze}.} Chemotactic behaviour of this kind is widely believed to be one of the major driving forces for evolving motility in the first place. The mechanism for chemotaxis just described imposes a number of physical constraints on the cell's molecular machinery and its overall kinematics \cite{Bergchemo}. For our purposes, the most interesting constraint is that the cell must be able to swim in a straight line between temporal sampling points, which is about 1s. 

Having to swim in a straight line is a non-trivial requirement because a bacterium is an active {\it colloid}, i.e., it lives in an inherently noisy, or Brownian, environment. In particular, rotational Brownian motion randomises the orientation of the cell, so that a propulsion force directed along the axis of the cell body nevertheless gives rise to curved trajectories over long times, even without tumbles, as is the case with `smooth swimming' mutants. At first sight, this renders chemotaxis rather hopeless for {\it E. coli}. The rotational diffusivity of a prolate ellipsoid (semi-axes $a, a, b$, $b/a > 1$, Fig.~\ref{fig:coli}(a)) has the form
\begin{equation}
D_r = \left( \frac{k_B T}{8\pi \eta b^3} \right) \times f(b/a), \label{eq:Drot}
\end{equation}
where $f(b/a)$ is a numerical factor \footnote{First given by F. Perrin in the 1930s, available conveniently in, e.g., \cite{Yamaoka}.}. The bracketed term is recognised as the rotational diffusivity of a sphere of radius $b$, which, using $b\approx 1 \mu$m, allows us to estimate $D_r \approx 0.2$s$^{-1}$ for the cell body of {\it E. coli}. The mean squared angular deviation is 
\begin{equation}
\langle \theta ^2(t) \rangle = 2D_r t, \label{eq:Drot2}
\end{equation} 
so that we expect a directional deviation of order 0.6~rad, or just under $40^\circ$, after 1s of a putatively `straight' run. This is hardly straight, so that chemotaxis should not work! 

What we have forgotten is that the cell body is attached to a long flagellum. The combination is like an $L \approx 10 \mu$m rod (flagellum $\approx 8 \mu$m + cell body $\approx 2\mu$m). Equation~(\ref{eq:Drot}) reminds us that $D_r$ scales as the cube of the longest dimension of the object. Using $b = L/2 \approx 5 \mu$m  lowers $\sqrt{\langle \theta ^2(t) \rangle}$ by a factor of $\approx 5^{1.5}$ to  $\lesssim 4^\circ$ in 1s. Thus, the flagellum (or flagellar bundle) acts not only as a propellor, but also as a rudder. 

For a WT swimmer performing run and tumble, it is intuitively clear that the long-time motion of the cell is a random walk, so that the mean-squared displacement is
\begin{equation} 
\langle \Delta r^2 (t) \rangle = nD_{\rm eff} t,  \label{eq:Deff}
\end{equation}
where $n$ is the spatial dimension and $D_{\rm eff}$ is an effective diffusivity. More formally, consider a random walker taking a step $\mathbf{L}$ after $N$ steps. The displacement after $N+1$ steps is related to that after $N$ steps by $\mathbf{r}_{N+1} = \mathbf{r}_N + \mathbf{L}$. Squaring and averaging gives
\begin{equation}
\langle r_{N+1}^2 \rangle = \langle r_N^2 \rangle + 2 \langle \mathbf{r}_N \cdot \mathbf{L} \rangle + \langle L^2 \rangle. \nonumber
\end{equation}
In a random walk, $\langle \mathbf{r}_N \cdot \mathbf{L} \rangle = 0$. Thus, mathematical induction shows that $\langle r_N^2 \rangle = N\langle L^2 \rangle$. If the average time taken per step is $\langle \tau \rangle$, then this implies $\langle \Delta r^2 \rangle = \langle L^2 \rangle t/\langle \tau \rangle$, so that $D_{\rm eff} \simeq  \langle L^2 \rangle /\langle \tau \rangle$. For a run-and-tumble bacterium, we therefore expect 
\begin{equation}
D_{\rm eff} \simeq \bar{v}^2\langle \tau \rangle, \label{eq:Deff2}
\end{equation}
where $\bar{v}$ is the average swimming speed, and $\langle \tau \rangle$ is the average time between tumbles \footnote{Strictly, $D_{\rm eff} = D_0 + k\bar{v}^2\langle \tau \rangle$, where $k$ is some numerical constant, since Brownian (= thermal) diffusion cannot be `tuned off'. In practice $D_0 \ll D_{\rm eff}$ and is often neglected. \label{fn:Deffcaveat}}. This heuristic treatment is borne out by a formal calculation starting from the appropriate Langevin equation in which the cells are approximated as spheres \cite{Condat}. A different treatment  \cite{Loveley}  taking into account the finite duration of the tumbles, $\langle \tau_1\rangle$, and the finite average direction cosine between two successive runs, $\langle \cos \theta \rangle$, gives
\begin{equation}
D_{\rm eff} = \frac{\bar{v}^2 \langle \tau \rangle}{3\left(1-\langle \cos \theta \rangle \right)} \left( \frac{\langle \tau\rangle}{\langle \tau \rangle + \langle \tau_1 \rangle } \right). \label{eq:Loveley}
\end{equation}
Macroscopic measurements of the way a dilute colony of cells spreads out have indeed found moving fronts that propagate with diffusive dynamics, returning values of $D_{\rm eff} \gtrsim 10^2 \mu$m$^2$s$^{-1}$ \cite{BergTurnerDiff}, which is $\approx \bar{v}^2\langle \tau \rangle/3$ with $\bar{v} \approx 20 \mu$ms$^{-1}$ and $\langle \tau \rangle \approx 1$s. These values of $D_{\rm eff}$ are a few orders of magnitude higher than those of non-motile cells (for which $D_0 \sim 0.3 \mu$m$^2$s$^{-1}$). This has given rise to the idea of an active suspension of bacteria as a `hot' colloid, with a high `effective temperature'. We will see in the next section that the same description has been used of synthetic self-propelled particles, and will evaluate the merits of this `effective temperature' idea in Section~\ref{sec:generic}.

\begin{figure}
\begin{center}
\includegraphics[height=5cm]{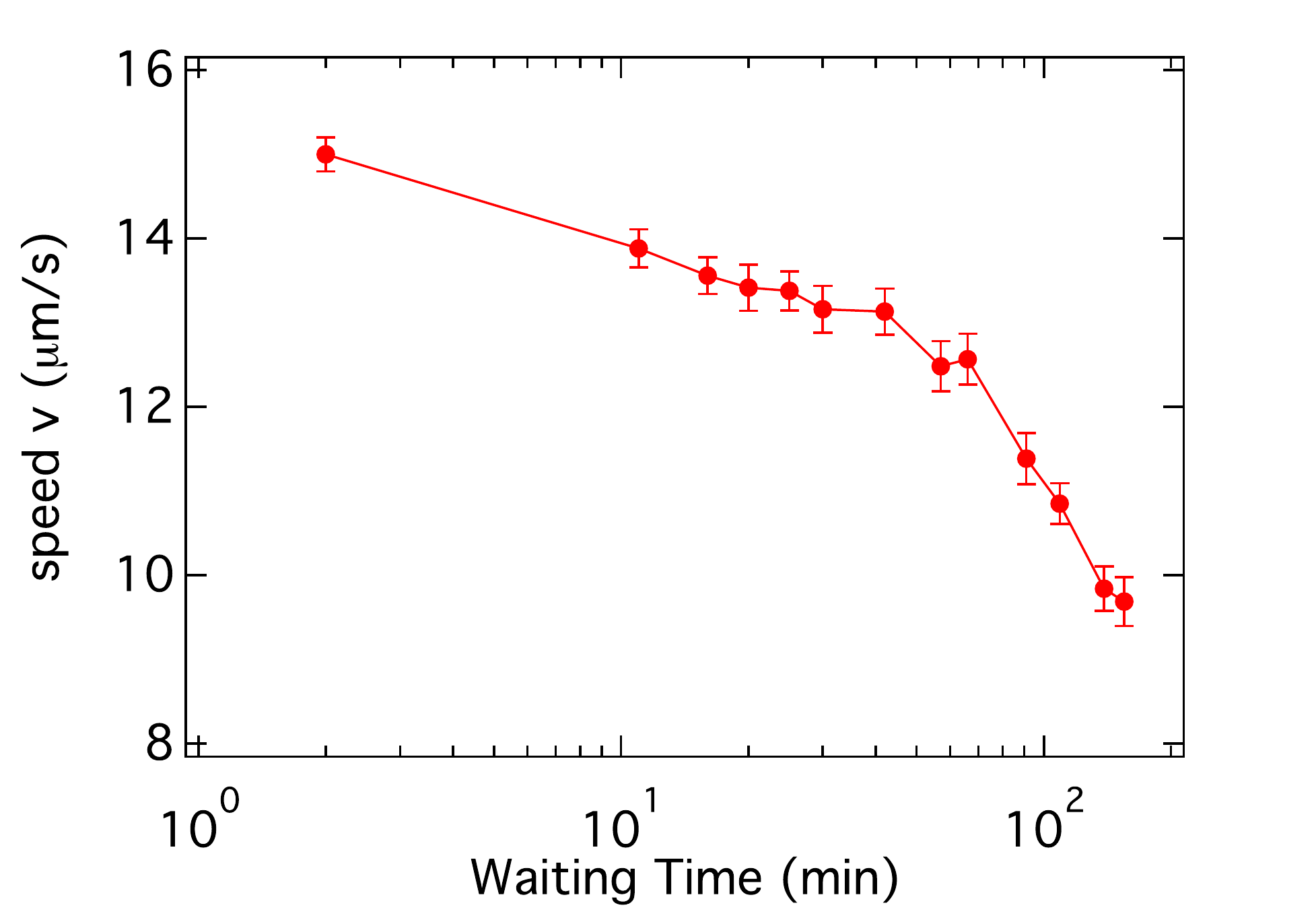}  
\end{center}  
\caption{The average swimming sped of {\it E. coli} AB1157 cells as a function of time elapsed since sealing into a capillary cell measured by differential dynamic microscopy (see Section~\ref{sec:ddm}). The bacteria  were grown in Lauria-Bertani broth to mid-exponential phase, harvested and re-suspended in motility buffer at $\approx 3 \times 10^8$ cells per ml. A $\approx 150 \mu$l drop was placed  in a $\approx 400 \mu$m deep capillary sealed at both ends for these measurements. (V. A. Martinez, unpublished data.)} \label{fig:speedtime}
\end{figure}

The swimming of WT {\it E. coli} is influenced by many environmental variables. This fact presents both an experimental challenge --- these variables have to be held reproducibly constant to obtain meaningful results, and an experimental opportunity --- once these effects are understood, they can be used to `tune' the motility of our natural active colloids. Most basically, a bacterium needs an energy source to remain motile. Typically, bacteria are dispersed in a `minimal motility buffer' for experiments. It is at first sight paradoxical that such a buffer does not typically contain any energy-rich molecules (hence `minimal'). But it turns out that the bacteria swim more rigorously in this situation, making use of internal resources. But this rigorous swimming is aerobic, and depends on the availability of oxygen in the cells' surrounding liquid medium, which decreases as a function of time in a sealed sample. Thus, the swimming speed may be expected to decrease with time, Fig.~\ref{fig:speedtime}. The precise time dependence of the motility seems sensitive to experimental details; thus another experiment at the same approximate cell density (though the strain used was not recorded) but different sample geometry found an abrupt speed transition as oxygen was exhausted \cite{Douarche}. If the sample cell is not sealed, but oxygen can diffuse in either through leaks or deliberately-left gaps, then chemotaxis (in this case known as oxytaxis) can take place, leading to pattern formation \cite{Douarche}. The response of {\it E. coli} motility to a host of other chemical and physical conditions was reported in a classic paper by Alder and Templeton in 1967 \cite{AdlerEnviron}, although there has been little systematic attempt since then to check and build on their work. An intriguing responses is that the tumble rate of the WT increases under illumination with blue light in chromophore-free motility buffer \cite{ParkinsonBlue}. All of these effect can be used to `tune' the behaviour of WT {\it E. coli}, making it a versatile tool for active colloid physics. 

From a physics point of view, surface and other confinement effects can also be regarded as `environmental tuning' of bacterial swimming. I will not review this somewhat more advanced topic in any detail in this introductory account. Suffice it to say that many of these effects, e,g. swimming in circles next to glass \cite{LaugaCircle} and air-water interfaces \cite{diLeonardoInterface} (but in opposite senses due to contrasting boundary conditions) and `driving on the right' in micro channels \cite{BergRight}, are hydrodynamic in origin, or at least strongly influenced by hydrodynamics. A convenient summary of these effects is available \cite{LaugaRev}. Other surface effects, such as the accumulation of motile bacteria at walls, may have non-hydrodynamic origins \cite{TangSurface}. A recent example of confinement effects in more complex geometry is the observation that mean-free-path restrictions in porous media (specifically, soft agar) can alter, and even turn off, chemotactic response \cite{Croze}. The biochemistry of the bacterial cell surface, of course, introduces all kinds of specific interactions with external surfaces, which in general must be taken into account in interpreting experiments \cite{BosSurface}.

\subsubsection{The genetic toolkit} 

The versatility of {\it E. coli} as a model active colloid is further enhanced by the availability of many motility mutants. We have already mentioned smooth swimming mutants in which tumbles are suppressed (e.g. HCB437 \cite{WolfeBerg}). There are also `tumbly' mutants in which runs are highly suppressed (e.g. RP2867 \cite{ParkinsonTumbly}). Mutants unable to synthesise certain iron-containing proteins show altered light-sensitivity to tumbling \cite{AdlerHeme}. More generally, a library of {\it E. coli} mutants with single gene knockouts can be obtained \cite{Keio}, e.g., giving access to a strain that does not synthesise flagella. While the microbiologist values these mutants for the insights they offer into the molecular genetics of bacteria, they are useful for our purposes because they allow us to `tune' swimming behaviour either statically or, more interestingly, as a function of time. 

The genetic tool kit for {\it E. coli} does not end with mutations to the WT genome. New `functionalities' can be added either by transducing plasmids containing particular genes, or by cloning new genes directly onto the bacterium's own genome \footnote{See \cite{DaleBook} for an introduction to the molecular biology of bacterial genetic manipulation.}. Using such techniques, it is possible, for example, to manufacture a strain of {\it E. coli} that, when its normal respiratory pathway is  poisoned (e.g. using azide), will only swim if illuminated because it contains a plasmid expressing the photosensitive protein proteorhodopsin \cite{Walter}.

\subsection{{\it E. coli} is not the only bug} \label{subsec:notonly} I have focussed {\it E. coli} not because it is by any means the simplest imaginable motile bacterium: e.g. the existence of a flagella bundle complicates matters, but because it is by far the most well studied and best understood motile micro-organism on the colloidal length scale. We have just seen one of the benefits of working with such a `model organism', namely, the availability of many bespoke mutants. But if physicists were to have a free choice on a `model' bug in the sense of something as simple as possible showing the essential physics we have been reviewing, we may wish to `order' a bacterium with a spherical cell body (`cocci') propelled by a single, rigid helical flagellum. It is interesting that there seems to be few motile cocci, though the reason is not known. Perhaps something like {\it P. aeruginosa} is the best we can do to approximate to the model bacterium shown in Fig.~\ref{fig:coli}(a). 

There is in fact a great variety of flagellation \footnote{The out-of-print atlas of electron micrographs by Leifson \cite{Leifson} is superb. Luckily, a free download is currently (2012) available at \url{http://archive.org/details/atlasofbacterial00leif}.}, and therefore motility, patterns in bacteria. For example, {\it Rhodobacter spheroids} bears a single flagellum, but it emerges {\it laterally}, i.e. perpendicular to the long axis of the cell body. The motor rotates only in one direction, and the bacterium reorients largely by intermittently stopping its propulsive motion and allowing rotational Brownian motion a free hand \cite{ArmitageRhodo}. Each cell of the magnetotactic bacterium {\it Magnetospirillum gryphiswaldense} contains up to 100 single-domain magnetite crystals so that a cell would orient in a magnetic field like a compass needle. The cell body is a right-handed spiral, and typically bears a featureless flagellum at each end. It apparently swims by rotating one or both flagella CCW so that the body rotates CW; propulsion here, as in other {\it Magnetospirillum} species, is due to the rotational-translational coupling of the cell body \cite{Erglis}. Spirochetes such as the {\it Leptospiracae} posses internal flagellum enclosed between a cell body and an outer sheath. Rotation of these flagella produces non-reciprocating distortions along the cell body to generate propulsion \cite{Charon}. Many spirochetes are virulent pathogens (Lyme disease, syphilis, etc.); this may be related to their ability to swim through highly viscous or viscoelastic media, such as the mucus covering the mammalian gastrointestinal tract \cite{Kaiser}. The physics of propulsion in viscoelastic media \cite{LaugaVisco} is a fascinating area that we cannot review here.

Finally, although it is definitely not a bacterium and not, strictly, colloidal either, the genus of biflagellated green algae {\it Chlamydomonas} may usefully be mentioned. Co-ordinated beating of the two flagella on each cell is the origin of self propulsion. In {\it C. reinhardtii}, which is a model organism for eukaryotic motility, the cell body is roughly spherical (mean diameter $\approx 10 \mu$m), and the flagella (length $\approx 10 \mu$m) beat at $\approx 50$Hz, propelling the cell with an average speed of $ \lesssim 10^2 \mu$ms$^{-1}$.	The basic propulsive physics of a beating flagellum is the same as that we have reviewed for a rigid helix, relying as it does on differential drag along and transverse to a cylindrical element \cite{LaugaRev,LighthillBook}. The organism uses a `two gear' mechanism to achieve the same effect as the `run and tumble' of {\it E. coli} \cite{GoldsteinGear}. One of the reasons for mentioning this organism is that the type of far-field flow it sets up (that of a `puller') is fundamentally different from that of most bacteria (which are `pushers'). We will return to this in Section~\ref{sec:generic} (see Fig.~\ref{fig:pusherpuller}). 

All other eukaryotic motile micro-organisms also have sizes outside the colloidal domain. Many generate motion using a small number of long, beating flagella, but others use a thin (compared to body size) layer of beating filaments (cilia) on their surfaces (what a theorist would call a `squirmer' -- a swimmer that moves by directly manipulating the velocity field on its surface). Discussion of cilia-driven locomotion is outside the scope of the present lectures (but see the recent review by Lauga and Powers \cite{LaugaRev}). 

\section{Synthetic swimmers} \label{sec:janus}

One of the most exciting advances in the last decade in colloid science is the ability to synthesise colloidal swimmers. Some are driven by external fields \cite{BibetteSwim,Tierno,Fischer}, often with bio-mimetic designs \cite{BibetteSwim,Fischer} that rely on similar physics to that we have already reviewed for micro-organismic propulsion. I will not discuss these systems any further, because the focus here is on {\it self-propelled} particles. One way to generate self propulsion is to use heterogeneous surface chemistry \cite{Howse,Sen}, most typically spherical Janus particles with two (equal or unequal) halves. These are the focus in the present section. Evidence to date suggests that Janus particles generate self propulsion via various kinds of auto- or self-phoresis, where `phoresis' refers to the motion of particles in gradients of all kinds (concentration, temperature, etc.). A particle with surface heterogeneities can generate a local gradient in which it then undergoes phoresis. The starting point for understanding such auto-phoresis is to understand phoresis in an externally-imposed gradient. 

\subsection{Phoretic motion} 

`Phoresis' (Greek {\it phorein} = `to carry') is the general expression for any sort of colloidal migration in gradients of all kinds: solute concentration, temperature, etc. \cite{AndersonRev2}. Classically, the gradient is externally imposed. A pedagogical account was given some time ago \cite{AndersonRev} (which I follow). A high-level discussion of the physics in terms of non-equilibrium thermodynamics is available \cite{Prost}. The key point, captured by the title of \cite{AndersonRev2}, is that the phenomenon is a case of {\it colloidal transport by interfacial forces}.  Below I will explain in detail diffusiophoresis in a gradient of neutral solutes, and review other cases more briefly. 

\subsubsection{Diffusiophoresis} \label{subsec:dph}

Consider an isolated colloidal particle in the presence of a neutral solute at bulk concentration $c_\infty$, Fig.~\ref{fig:surface}(a). We take the particle to be large enough that its surface can be considered locally flat. The interaction of the solute and the surface of the particle is described by a short-range potential (of mean force), $U(y)$, where the $y$ axis is the (outward) surface normal, giving rise to an excess surface energy per unit area of $\sigma(c)$. If the solute is attracted to the surface, $(U, \sigma) < 0$. The solute profile away from the surface is sketched for this case in Fig.~\ref{fig:surface}(a), which shows an excess surface concentration (number per area), characterised by
\begin{equation}
\Gamma = \int_0^\infty [c(y) - c_\infty] \;\mbox{d}y. \label{eq:gamma}
\end{equation}
This excess surface concentration is often normalised as an adsorption length,
\begin{equation}
{\cal K} = \frac{1}{c_\infty}\int_0^\infty [c(y) - c_\infty] \;\mbox{d}y, \label{eq:K}
\end{equation}
so that $|{\cal K}|$ is the thickness of the layer of bulk solution that contains as many solute molecules per unit area as the excess layer (hatched in Fig.~\ref{fig:surface}(a)); note that ${\cal K}$ has the opposite sign to the interaction potential $U(y)$, so that ${\cal K} < 0$ for repulsive interaction. 

\begin{figure}
\begin{center}
\includegraphics[height=4cm]{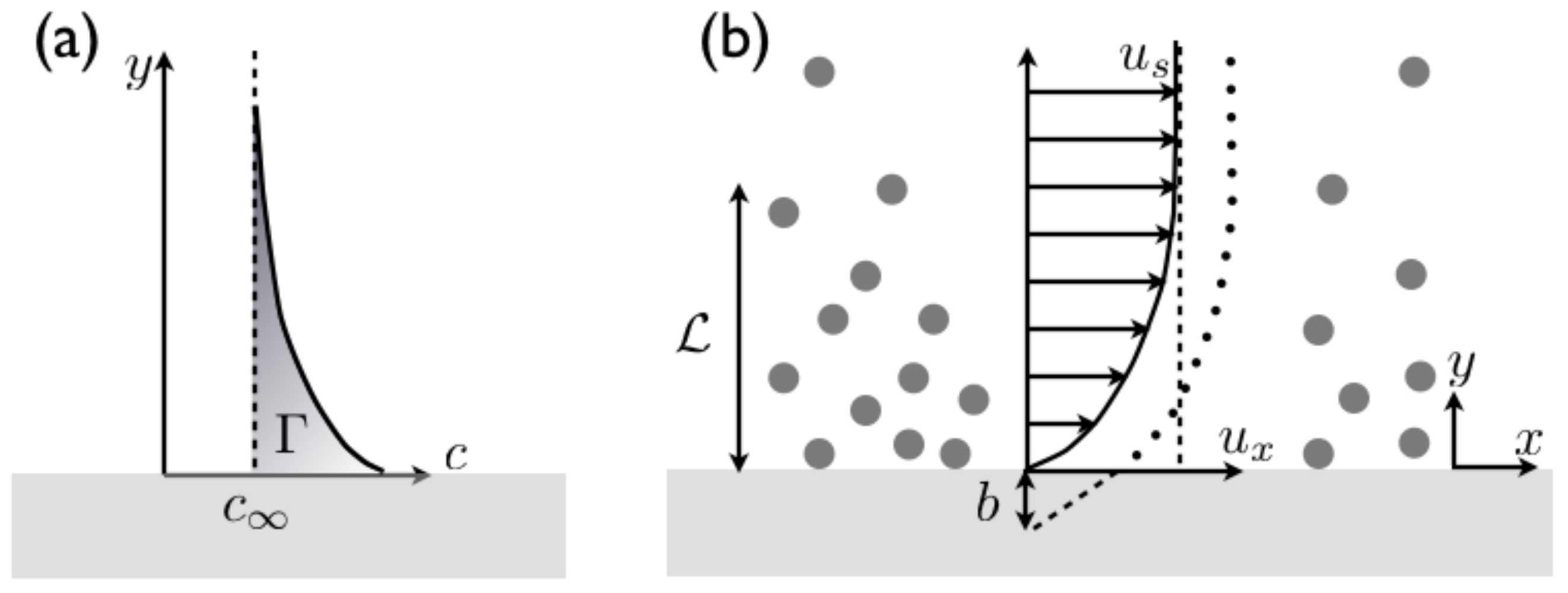}  
\end{center}  
\caption{(a) The excess surface concentration. In a plot of the solute concentration along the direction of the surface normal, $c(y)$, $\Gamma$ is the shaded area, Eq.~(\ref{eq:gamma}). (b) The origin of diffusiophoresis. There is a distribution of solutes (filled circles) away from the surface ($y$ direction) due to surface interactions. If there is a concentration gradient of solutes along $x$ (here, more concentrated on the left), a tangential pressure gradient results, which drives flow parallel to the surface. The flow profile, arrows, reaches an asymptotic value, $u_s$, at a distance governed by the range of the surface-solute interaction, $\simeq {\cal L}$, Eq.~(\ref{eq:L}). If there is slip on the surface with slip length $b$, then the flow profile is the dotted curve. Partly redrawn after \cite{Ajdari}.} \label{fig:surface}
\end{figure}

If now there is a solute concentration gradient, $\partial c_\infty(x)\partial x \neq 0$, one might suppose that a simple thermodynamic argument will predict the direction of particle migration and its speed. At position $x$, the particle (radius $a$) has surface free energy $G = 4\pi a^2 \sigma[c(x)]$ due to its interaction with the solute \footnote{We assume that $a$ is much smaller than the length scale over which $c_\infty$ varies. \protect\label{fn:length}}. Since the particle has different surface free energies at different positions $x$, it experiences a force $F = - \partial G/\partial x$ and moves with velocity $u = F/\xi$ where $\xi = 6\pi \eta a$:
\[
u = - \frac{1}{\xi} \frac{\partial G}{\partial x} = - \frac{2a}{3\eta} \left( \frac{\partial \sigma}{\partial c_\infty} \right) \left(\frac{\partial c_\infty}{\partial x} \right).
\]
The first bracket can be re-written in terms of the Gibbs equation
\begin{equation}
\frac{\partial \sigma}{\partial c_\infty} = - \frac{k_B T \Gamma}{c_\infty} = - k_B T {\cal K}.
\end{equation}
This equation is derived in textbooks  \cite{Adamson} \footnote{Briefly, for a surface of area $A$, standard manipulations give $S_s dT + Ad\sigma + N_s d\mu_s = 0$, where $N_s$, $S_s$ and $\mu_s$ are the number, entropy and chemical potential of surface solutes. Since $N_s = A \Gamma$, and $\mu_s$ is equal to the chemical potential in the bulk ($= k_B T \ln c_\infty$ in the dilute limit), with which the surface is in equilibrium, the Gibbs equation follows for constant $T$.}, but the {\it sign} is intuitively obvious: if the solute adsorbs (${\cal K} > 0$), then a higher bulk concentration gives rise to more adsorption and therefore a more negative $\sigma$, i.e. $\partial \sigma/\partial c_\infty < 0$ for $\Gamma > 0$. So finally, we predict
\begin{equation}
u = \frac{2k_B T}{3\eta} (a{\cal K}) \frac{\partial c_\infty}{\partial x}.  \;\;\; \mbox{(Wrong!)}\label{eq:wrongphoresis}
\end{equation}
While this argument is appealingly simple, it is wrong, although not completely. Equation~(\ref{eq:wrongphoresis}) has the right sign. The particle moves towards the higher concentration of solute if the latter is attracted to the particle. If the solute is repelled by the surface, the movement is in the opposite direction. But the algebraic form of Eq.~(\ref{eq:wrongphoresis}) is incorrect, essentially because the term in brackets that scales as (length)$^2$, is wrong. Diffusiophoresis relies on a rather  subtle coupling between surface forces and fluid dynamics that this thermodynamic argument simply fails to capture. I now explain how this arises.

The basic physics, Fig.~\ref{fig:surface}(b), is this. The surface applies force to the liquid via its interaction with the solutes within an interfacial layer of extent $\approx \lambda$. Thus, a concentration gradient of an ideal solute, $\nabla c_\infty(x)$, gives rise to a tangential pressure gradient $k_B T\nabla c_\infty(x)$. This drives flow in the decreasing pressure, or $-\nabla c_\infty$, direction in the interfacial layer. The balance of viscous stress, $\eta \nabla^2 u \simeq \eta u/\lambda^2$, and pressure gradient in the interfacial layer determines the flow profile, which rises from zero (no slip at the surface) to $\mathbf{u}_s \simeq - k_B T \lambda^2 \nabla c_\infty /\eta$ at a distance $\sim \lambda$ from the surface, Fig.~\ref{fig:surface}(b). Since $\lambda$ is a molecular length determined by the solvent-surface interaction, from the point of view of a micron-size particle (whose locally flat surface we are considering), this situation looks {\it as if} the fluid is slipping at its surface. Thus, $\mathbf{u}_s$ is known as the `slip velocity'. In the stationary frame of the liquid, the surface spontaneously translates with a diffusiophoretic velocity $ - \mathbf{u}_s$. The final result, Eq.~(\ref{eq:slip}), shows that our scaling arugment is essentially correct, but $\lambda^2$ is in fact the product of two slightly different molecular length scales: $\lambda^2 = {\cal KL}$. We have already encountered ${\cal K}$, Eq.~(\ref{eq:K}), which measures the strength of adsorption; ${\cal L}$ measures the range of the solute-surface interaction potential, Eq.~(\ref{eq:L}). We now turn to review the detailed calculation of $\mathbf{u}_s$. 

For an ideal solute, its distribution from the surface is given by $c(y) = c_\infty e^{-\beta U(y)}$. Suppose the concentration varies along $x$ over much longer length scales than those over which interfacial forces operate (cf. footnote~(\ref{fn:length})), so that equilibration of concentrations and pressures along $y$ is much faster than along $x$. Thus, the solute concentration profile is locally Boltzmannian ($\beta = 1/k_B T$):
\begin{equation}
c(x,y) \approx c_\infty(x) e^{-\beta U(y)}. \label{eq:localc}
\end{equation} 
A solute at $y$ experiences a force $-\partial U(y)/\partial y$ through its interaction with the surface, which is transmitted to the solvent; force balance in the $y$ direction requires 
\begin{equation}
-\frac{\partial p}{\partial y} + c \left(-\frac{\partial U}{\partial y}\right)= 0. 
\end{equation}
These two equations solve to \footnote{Or integrate directly the Gibbs-Duhem relation, $dP = cd\mu$ with $\beta \mu = \ln c$ to get $p(y) - p_\infty = \int c (\partial \mu/\partial c) dc = k_B T \int_{\infty}^y dc$; Eq.~(\ref{eq:pressure}) follows from Eq.~(\ref{eq:localc}). D. Frenkel pointed this out to me.}
\begin{equation}
\beta [p(x,y) - p_\infty ]= c_\infty(x) \left[ e^{-\beta U(y)} - 1 \right]. \label{eq:pressure}
\end{equation}
The {\it tangential} variation in pressure due to the variation of $c_\infty$ along $x$ drives solvent flow. The viscous stress of this flow balances the tangential pressure gradient, i.e. 
\begin{equation}
-\frac{\partial p}{\partial x} + \frac{\partial}{\partial y}\left(\eta \frac{\partial u_x}{\partial y}\right)= 0, \label{eq:anderson}
\end{equation}
to be solved subject to the following boundary conditions:
\begin{eqnarray}
u_x(y = 0) & = & 0 \;\;\; \mbox{(no slip on the surface)}\label{eq:bc1} \\
\left(\frac{\partial u_x}{\partial y}\right)_{y \rightarrow \infty} & = & 0  \;\;\;\mbox{(no velocity gradient in the bulk).}\label{eq:bc2}
\end{eqnarray}
Qualitatively, the profile of this flow can immediately be sketched, Fig.~\ref{fig:surface}(b). It rises from zero at the surface until it reaches an asymptotic value of $u(x) = u_s$ at some distance $\approx {\cal L}$, which we expect to be of the order of the range of $U(y)$ (see Eq.~(\ref{eq:L})). To obtain $u_s$, we proceed as follows. Integrating Eq.~(\ref{eq:anderson}) twice, we find
\begin{eqnarray}
\frac{\partial u_x}{\partial y} &  = & \frac{k_B T}{\eta} \frac{\partial c_\infty}{\partial x} \int_\infty^Y \left[ e^{-\beta U(y^\prime)} - 1\right] dy^\prime \label{eq:surface1} \\
u_x(y)  & = & - \frac{k_B T}{\eta} \frac{\partial c_\infty}{\partial x}  \int_0^y dY \int_Y^\infty \left[ e^{-\beta U(y^\prime)} - 1\right] dy^\prime. \label{eq:surface2}
\end{eqnarray}
The order of the limits in Eq.~(\ref{eq:surface1}), reflecting Eq.~(\ref{eq:bc2}), is swapped in Eq.~(\ref{eq:surface2}), which introduces a minus sign. Since we expect $u(x, y)$ to have reached its asymptotic value $u_s$ at $y \gtrsim {\cal L}$, Fig.~\ref{fig:surface}(b), we introduce very little error by taking $u_s = u_x(y \rightarrow \infty)$, i.e.
\begin{equation}
u_s   =  - \frac{k_B T}{\eta} \frac{\partial c_\infty}{\partial x}  \int_0^\infty dY \int_Y^\infty \left[ e^{-\beta U(y^\prime)} - 1\right] dy^\prime. \label{eq:surface3}
\end{equation}
We evaluate the integrals by first defining $g(Y) = \int_Y^\infty \left[ e^{-\beta U(y^\prime)} - 1\right] dy^\prime$, and noting that $g^\prime(Y) = dg(Y)/dY = -[e^{-\beta U(Y)} - 1]$ \footnote{Recall that generally $\frac{d}{da} \int_a^b f(x)dx = -f(a)$.}. Then the $Y$ integral in Eq.~(\ref{eq:surface3}) can be performed by parts. Since $ [Yg(Y)]_0^\infty= 0$, we obtain
\begin{equation}
\int_0^\infty g(Y) dY = -  \int_0^\infty Y g^\prime(Y)dY = \int_0^\infty Y[e^{-\beta U(Y)} - 1]dY.
\end{equation}
Finally, then, we find that
\begin{eqnarray}
u_s & = & - \frac{k_B T}{\eta} ({\cal KL}) \frac{\partial c_\infty}{\partial x},   \;\;\;\mbox{with}\label{eq:slip}\\
{\cal L} &=& \frac{\int_0^\infty y[e^{-\beta U(y)} - 1]dy}{\int_0^\infty [e^{-\beta U(y)} - 1]dy}  =   \frac{\int_0^\infty y[e^{-\beta U(y)} - 1]dy}{\cal K} . \label{eq:L}
\end{eqnarray}
The last identity follows from substituting Eq.~(\ref{eq:localc}) into  Eq.~(\ref{eq:K}).

Equation~(\ref{eq:slip}) predicts that, in the stationary frame of the surface, the coupling between surface forces and fluid dynamics leads to a flow of solvent from high to low solute concentration if the solute is attracted to the surface. In the stationary frame of the liquid, the surface therefore translates up the solute concentration gradient. This agrees with our thermodynamic derivation, Eq.~(\ref{eq:wrongphoresis}), as far as the direction of migration is concerned \footnote{The sign difference between Eqs.~(\ref{eq:wrongphoresis}) and (\ref{eq:slip}) is simply due to switching between the stationary frame of the liquid and of the solid surface or particle.}. But the different physics is brought out by the form of the (length)$^2$ term: $aK$ and ${\cal LK}$ respectively. Both formulations agree that a strong interaction between solute and surface is important: a large adsorption length ${\cal K}$ gives rise to strong diffusiophoresis. But  Eqs.~(\ref{eq:slip}) and (\ref{eq:L}) show that this is not enough. If the potential $U(y)$ is of infinitesimal range, then ${\cal L} \rightarrow 0$ and therefore $u_s \rightarrow 0$. \footnote{In the limit of a weak potential, ${\cal L} \sim \int_0^\infty yU(y) dy$, i.e. it scales as the first moment of the potential. Clearly, ${\cal L} = 0$ for a potential of infinitesimal range}. For significant diffusiophoresis, we require a finite ${\cal L}$, which in turn requires an interaction potential of finite range, giving a {\it diffuse interfacial layer} of at least a few solute molecules thick.

Although the interfacial layer must be finite for diffusiophoresis, its thickness is still very small compared to any colloidal particle, ${\cal L} \ll a$. From the point of view of the particle, therefore, that the fluid velocity has the finite value $u_s$ at a short distance $\approx {\cal L}$ from its surface can be interpreted effectively as {\it slip}. From this point of view, phoresis in general, and diffusiophoresis in particular, is caused by the interaction between surface forces and flow in such a way that gives a finite (effective) slip velocity at the surface. 

Equation~(\ref{eq:slip}), which gives the slip velocity relative to an infinite, planar surface, can be applied to calculate the velocity of a particle in a concentration gradient of neutral solutes. The simplest case is that of a particle of arbitrary shape whose characteristic dimension, $L$, is much larger than the length scales set by solute-surface interactions, viz., $L \gg {\cal K}, {\cal L}$. In this case, a full calculation \cite{AndersonRev} shows that Eq.~(\ref{eq:slip}) (with a change of sign from switching frames of reference) in fact gives the diffusiophoretic velocity (now written for a gradient in a general direction):
\begin{equation}
\mathbf{v}_{\rm dph} =  \frac{k_B T}{\eta} ({\cal KL}) \nabla c_\infty. \label{eq:vdph}
\end{equation}
We now apply Eq.~(\ref{eq:vdph})  to the case of solutes that interact with the surface only through excluded volume \cite{AndersonRev2}, i.e. $U(y) = 0$ for $y \geq \delta$ and $U(y) = \infty$ for $y < \delta$, where $\delta$ is the radius of a solute particle. We find, using Eqs.~(\ref{eq:K}) and (\ref{eq:L}) that ${\cal K L} = -\delta^2/2$, so that
\begin{equation}
\mathbf{v}_{\rm dph}^{\rm hard} =  -\frac{k_B T}{\eta} \delta^2 \nabla c_\infty, \label{eq:vdphHard}
\end{equation}
i.e. particles migrate down a gradient of hard solutes.

One might be tempted to derive Eq.~(\ref{eq:vdphHard}) using the argument that the osmotic pressure difference ($\Delta \Pi$) between the two halves of a particle (radius $a$) drives diffusiophoresis: the force on the particle is $\simeq \Delta \Pi \times$(surface~area) $\simeq - [(k_B T \nabla c) a] \pi a^2$, so that  
\begin{equation}
\mathbf{v}_{\rm dph}^{\rm hard} \simeq - \frac{[(k_B T \nabla c) a] \pi a^2}{\pi\eta a} \simeq -  \frac{k_B T}{\eta} a^2 \nabla c.\;\;\;\mbox{(Wrong!)} \label{eq:wrong2}
\end{equation} 
This fallacious result has the right sign, but wrongly identifies the relevant length scale to be the particle size, $a$, whereas in fact, it is the length scale of the solute, $\delta$, that matters; indeed, for hard solutes, $K = {\cal L} = \delta$, Eq.~(\ref{eq:vdphHard}).

In our discussion so far, we have assumed that the liquid obeys a strictly stick boundary condition on the particle surface, Eq.~(\ref{eq:bc1}). Slip on the surface \footnote{This must be distinguished from the {\it apparent} slip, $u_s$, that drives phoresis,  Fig.~\ref{fig:surface}.} enhances $u_s$, Fig.~\ref{fig:surface}(b). Formally, if we use a `Navier' boundary condition, i.e. the velocity at the surface is given by $u_x(y = 0) = b(\partial u_x/\partial y)_{y = 0}$, instead of Eq.~(\ref{eq:bc1}), we now have \cite{Ajdari} 
\begin{equation}
\mathbf{v}_{\rm dph} =  \frac{k_B T}{\eta} ({\cal KL}) \left( 1 + \frac{b}{\cal L} \right) \nabla c_\infty. \label{eq:Ajdari}
\end{equation}
While ${\cal L}$ is a small-molecular length (order $10^{-10}$m), a hydrophobic surface in water can display a `slip length' $b$ of the order $10^{-8}$m \cite{Cottin}; $(1 + b/{\cal L})$ can therefore be considerable.

For a charged surface interacting with cations and anions in the solvent \cite{AndersonRev}, the electrostatic interaction between the surface and ions (of both signs) in the solvent allows the former to exert a pressure on the liquid via the latter in a diffuse `electric double layer', whose thickness is set by the Debye screening length, $\kappa^{-1}$, where
\begin{equation}
\kappa^2 =   \frac{1}{\epsilon_0 \epsilon k_B T}\sum_j n_j q_j^2 .  \label{eq:debye}
\end{equation}
Here, $\epsilon$ is the dielectric constant of the solvent, and $n_j$ and $q_j$ are the number density and charge (in electronic units) of the ions respectively.  A gradient in ion concentrations generates flow in the double layer in much the same way as for neutral solutes. The resulting $u_s$ is always towards lower electrolyte concentration \footnote{Loosely, the attraction between the surface and ions of the opposite sign `wins' over the repulsion with ions of the same sign: overall, the electrolyte adsorbs; cf. Eq.~(\ref{eq:vdph}).}. In general, anions and cations have different diffusivities, $D_+ \neq D_-$, so that the bulk ionic concentration gradient leads to an electric current. To prevent bulk charge separation, an electric field spontaneously arises in the bulk to generate a counter current. This electric field also generates tangential flow, the direction of which depends on the sign of 
\begin{equation}
\tilde\beta = (D_+ - D_-)/(D_+ + D_-) \label{eq:beta}
\end{equation}
and the sign of the surface (zeta) potential, $\zeta$. The total slip velocity now has so-called `chemophoretic' and `electrophoretic' components \cite{AndersonRev2}:
\begin{equation}
u_s = - \frac{k_B T}{\eta} \kappa^{-2} [\nu_c(\zeta) + \tilde\beta\zeta \nu_e] \frac{\partial \ln C_\infty}{\partial x}, \label{eq:slip2}
\end{equation}
where both numerical coefficients $\nu_c, \nu_e$ are positive and $\nu_e$ is a constant. 

To conclude our discussion of diffusiophoresis, it is worth emphasising that a finite concentration gradient alone, $|\nabla c_\infty| \neq 0$, is insufficient. There must be non-zero solute-surface interaction, $K{\cal L} \neq 0$ or finite $\kappa^{-2}$, for the phenomenon to occur.  

\subsubsection{Electrophoresis} Electrophoresis is the migration of charged colloids in an electric field. We have already referred to this phenomenon implicitly in our discussion of the $\tilde\beta$-dependence of the $\nu_e$ term in Eq.~(\ref{eq:slip2}). There is a sense in which electrophoresis illustrates more directly than any other phoretic phenomenon the interfacial nature of gradient-driven migration. A charged particle plus its diffuse electric double layer is a neutral object. It is unclear at first sight why such a neutral body should move in an electric field. The answer lies in the fact that the body as a whole is not rigid due to the presence of a diffuse interface. The simplest treatment that brings out the essential physics is as follows. 

The fixed charge on the surface (normal along $y$) is balanced by a net space charge in the diffuse layer (thickness $\approx \kappa^{-1}$, charge density $\rho_e$) due to different concentrations of cation and anions. An electric field (or potential gradient) $\mathbf{E} = (E, 0,0)$ acts on the space charge to exert a force per unit volume of $\rho_e E$ on the liquid. Balancing viscous and electrostatic force densities gives
\begin{equation}
\eta \frac{\partial^2 u_x}{\partial y^2} + \rho_e E = 0. \label{eq:electro}
\end{equation}
This equation needs to be supplemented by Poisson's equation, which relates the local electrical potential, $\Psi$, and the charge distribution in the double layer
\begin{equation}
\frac{\partial^2 \Psi}{\partial y^2} = - \frac{\rho_e}{\epsilon_0 \epsilon }. \label{eq:poisson}
\end{equation} 
Equations~(\ref{eq:electro}) and (\ref{eq:poisson}) are solved subject to the hydrodynamic boundary conditions Eqs.~(\ref{eq:bc1}) and (\ref{eq:bc2}) and the electrostatic boundary condition that $\Psi$ is equal to the surface, or zeta, potential at $y = 0$:
\begin{equation}
\Psi (y = 0)  =  \zeta \label{eq:Ebc}.
\end{equation}
The result is
\begin{equation}
u_x = \frac{\epsilon_0 \epsilon}{\eta} [\Psi(y) - \zeta] E,
\end{equation}
so that the `slip velocity' at $y \rightarrow \infty$ (in practice, $y \gg \kappa^{-1}$ would do) is given by
\begin{equation}
u_s = \frac{\epsilon_0 \epsilon \zeta}{\eta} E.
\end{equation}
In this simplest treatment, if the net space charge in the double layer is positive, corresponding to a negatively-charged particle, the solvent slips in the direction of $\mathbf{E}$, so that the particle moves in the direction of $-\mathbf{E}$, as a bare negative particle would. 

Since the electric field applies equal and opposite forces on the particle and the diffuse double layer, the net force on the neutral composite object undergoing electrophoretic migration is zero. Thus, the composite object exerts no force on the liquid outside a layer of thickness $\kappa^{-1}$ away from the particle surface. The lowest order far-field flow around a particle in electrophoresis is therefore not that due to a force monopole (or stokeslet, Eq.~(\ref{eq:seen})), but must be a higher-order multipole. This contrasts with a particle sedimenting under gravity, but is qualitatively comparable with motile micro-organisms (cf. Fig.~\ref{fig:noforce}). Quantitatively, however, the far field flow due to phoresis scales as $v \sim r^{-3}$ \cite{AndersonFlow}, which is a faster decay that that of swimming {\it E. coli} or {\it Chlamydomonas} (which, as I have already pointed out, look like force dipoles in the far field, so that $v \sim r^{-2}$).

\subsubsection{Thermophoresis}

The migration of particles in temperature ($T$) gradients (the Soret effect) is well attested experimentally. Theoretically, the thermophoretic drift velocity of a particle can be written as \cite{DerjaguinThermo} 
\begin{equation}
\mathbf{u}_{\rm thph} = - \frac{2 \nabla T}{\eta T} \int_0^\lambda y \Delta h(y) dy, \label{eq:thph}
\end{equation}
where $\Delta h$ is the excess enthalpy per unit volume of the solvent (including any solutes) due to interaction with the surface in the interfacial region of thickness $\lambda$ \footnote{Once again, this result is valid if the particle radius satisfies $a \gg \lambda$.}. This formulation has obvious similarities with that we have used for diffusiophoresis -- compare the integral in Eq.~(\ref{eq:thph}) with the numerator of Eq.~(\ref{eq:L}) \footnote{Note that because the excess specific enthalpy rapidly approaches zero outside the interfacial region, we may replace the upper limit in the integral in Eq.~(\ref{eq:thph}) by $\infty$.}. In particular, note that a finite temperature gradient, $\nabla T \neq 0$, is insufficient. Once again, finite interaction between the solvent/solutes and the particle surfaces necessary for a finite $\Delta h$. 

The relationship between these microscopic interactions and $\Delta h$ is subtle. This subtlety is reflected in the experimental observation that thermophoresis is notoriously sensitive to conditions. In particular, even small modifications to surface interactions seem to result in particles changing the direction in which they migrate (up or down $\nabla T$). The detailed microscopic mechanism of thermophoresis is therefore still a matter of on-going research \cite{Braun,PiazzaThermo,PiazzaThermo2}. Note, however, that this sensitivity also offers the possibility of an exquisite degree of control in the laboratory.

\begin{figure}
\begin{center}
\includegraphics[width=6.5cm]{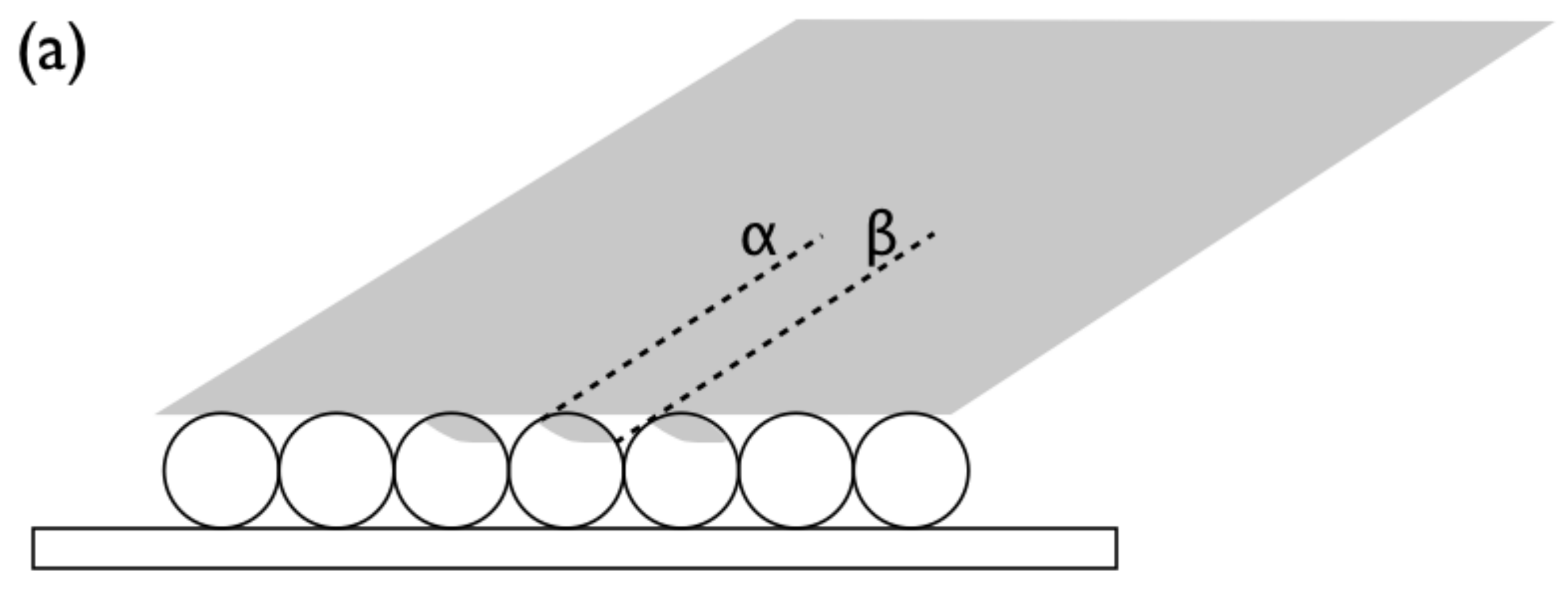}  
\includegraphics[width=6.5cm]{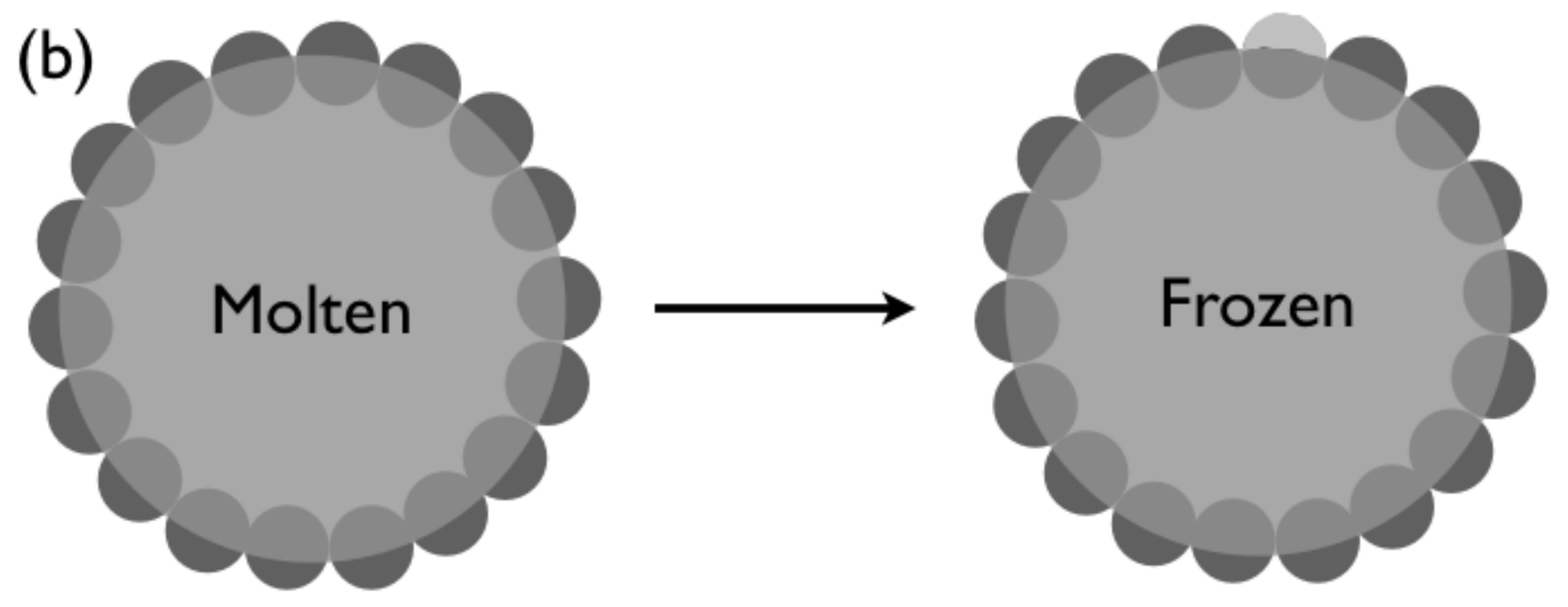}  
\end{center}  
\caption{Schematic of Janus particle synthesis. (a) Evaporating or sputtering metal onto a monolayer of particles coated onto, e.g., a glass slide. The combined effects of self shadowing (dashed line $\alpha$) and shadowing by a neighbour (dashed line $\beta$) control the shape of the coated patch on each particle (shown on central three particles). The yield is low because Janus particles are prepared one monolayer at a time. (b) One method for bulk synthesis of Janus particles \cite{GranickJanusBulk}. Particles are used to stabilise an emulsion of molten wax droplets in water; then the temperature is lowered to freeze the wax. The resulting `colloidosomes' are easily manipulated, and the exposed surfaces of the particles can be functionalised (symbolised by top particle with lighter upper half). The emulsion has a macroscopic amount of interfaces, hence the bulk yield. } \label{fig:janusprep}
\end{figure}

\subsection{Janus particles and auto-phoresis}

The synthesis of Janus particles, first proposed by P.-G. de Gennes in his Nobel lecture \cite{deGennesNobel}, is now almost routine \cite{Pawar,HuJanusRev}. In a widely-used technique, Fig.~\ref{fig:janusprep}(a), a metal is evaporated or sputtered onto a monolayer of particles on a substrate. The degree of coverage can be tuned by the angle of incidence of the deposited species, with normal incidence giving more or less hemispherical covering. A serious drawback is yield -- each batch prepares a single monolayer of particles. Various bulk techniques have been demonstrated. One scheme \cite{GranickJanusBulk}, Fig.~\ref{fig:janusprep}(b), involves using particles to stabilise an emulsion of molten wax drops in water, and then solidifying the wax to give droplets with frozen-in particles at the interface, whose exposed surfaces can then be functionalised. The macroscopic amount of interfaces in the original emulsion turns this into a technique for synthesising gramme quantities of Janus particles. 

\begin{figure}[t]
\begin{center}
\includegraphics[height=4cm]{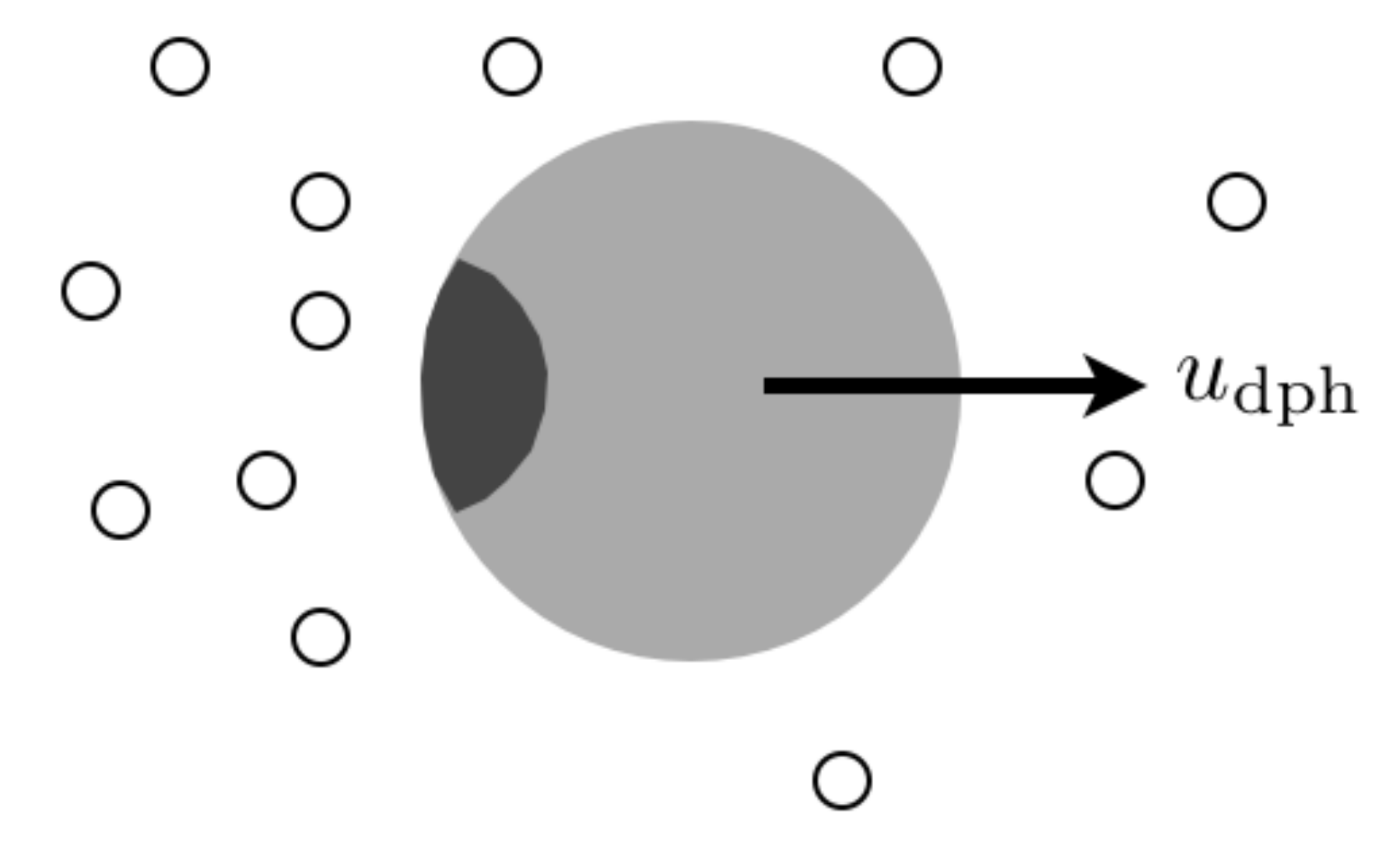}   
\end{center}  
\caption{Schematic of autodiffusiophoresis. A catalytic patch (dark) on a particle catalyses a reaction in the solvent, giving rise to a concentration gradient of products (open circles). The particle moves in the direction shown if the interaction of the product with the particle surface is repulsive, Eq.~(\ref{eq:vdphHard}). The particle and reaction products are not drawn to scale: the latter needs to be much smaller than the former for Eq.~(\ref{eq:vdphHard}) to be applicable. Redrawn after \protect\cite{GolestanianJanus}.} \label{fig:self}
\end{figure}

Janus particles are the focus of much current attention. For some, they offer novel routes to colloidal self assembly \cite{GranickJanusRev}. Our interest is in their use to generate self propulsion. The heterogeneous surface chemistry produces a local gradient of some kind, in which the particle undergoes phoretic migration. In self-diffusiophoresis, a surface chemical reaction on a patchy particle creates a spatial gradient of products, which drives diffusiophoresis \cite{GolestanianJanus}, Fig.~\ref{fig:self}. To obtain $u_{\rm dph}$ using Eq.~(\ref{eq:vdph}), we need an expression for $\nabla c$, which requires solving the diffusion equation for a particle with surface sources. This calculation is in general complex, and a number of regimes are possible depending on reaction mechanisms and rate constant(s) \cite{EbbensSize}. In the case of a simple one-step reaction giving rise to a uni-molecular product that causes the local concentration gradient, a scaling argument proceeds as follows. Product molecules diffuse away from the catalytic patch with a flux $\simeq |D_P \nabla c|$. In a steady state, the number of product molecules diffusing away per unit time, $dN_p/dt \simeq -|D_P \nabla c| \times a^2$, must be matched by the rate of production of fresh ones, $dN_p/dt \simeq \tau_f^{-1}$, the `firing rate' of the catalytic surface. Thus,
\begin{equation}
|D_P \nabla c|a^2 \simeq \frac{1}{\tau_f} \Rightarrow |\nabla c| \simeq \frac{1}{D_Pa^2\tau_f}.
\end{equation}
I now express $\tau_f$ in terms of more directly determinable quantities for the reaction 
\begin{equation}
\mbox{H}_2\mbox{O}_2 \stackrel{\mbox{\tiny Pt}}{\longrightarrow} \mbox{H}_2\mbox{O} + 0.5\, \mbox{O}_2 \label{eq:reaction}
\end{equation} 
at a platinum (Pt) surface in water \footnote{We are making the approximation that H$_2$O is very similar to H$_2$O$_2$, so that the only result of the reaction relevant to diffusiophoresis is the generation of O$_2$ molecules.}. In a one-step approximation to the reaction kinetics, the rate of O$_2$ production is governed by \footnote{Throughout, $[\ldots]$ denotes concentration.}:
\begin{equation}
\frac{d[\mbox{O}_2]}{dt} = k_1[\mbox{H$_2$O$_2$}]. \label{eq:kinetics}
\end{equation} 
If [H$_2$O$_2$] is given in volume fraction~\% and $d[\mbox{O}_2]/dt$ is measured in number of molecules per unit area per unit time, then $k_1$ is the (in principle measurable) areal rate of production of O$_2$ at a Pt surface in neat H$_2$O$_2$, so that $k_1[\mbox{H$_2$O$_2$}]A \simeq \tau_f^{-1}$, where $A \simeq a^2$ is the area of the catalytic patch on a particle of radius $a$. Substituting into Eq.~(\ref{eq:vdph}) gives
\begin{equation}
\left|u_{\rm dph}^{\rm self}\right|^{\rm (I)} \simeq \frac{k_B T}{\eta} |K{\cal L}| \frac{k_1}{D_P} [\mbox{fuel}], \label{eq:JanusSpeed}
\end{equation}
where we have generalised from H$_2$O$_2$ to a generic `fuel' whose behaviour can be approximated by some equivalent to (\ref{eq:reaction}) and (\ref{eq:kinetics}). (The superscript `(I)' indicates that this is one of three regimes of behaviour; the other two will be introduced a little later on.) A full calculation \cite{GolestanianJanus,Jones} gives a numerical prefactor of $1/4$ for a half-covered particle.

The first experimental attempt to realise this scheme for self propulsion made use of reaction~(\ref{eq:reaction}) on polystyrene spheres (diameter $1.62 \mu$m) half coated with Pt dispersed in water-H$_2$O$_2$ mixtures \cite{Jones}. Particles next to a glass surface were tracked to measure their mean-squared displacement (MSD) \cite{Jones,Bocquet}. At [H$_2$O$_2$] = 0, the MSD was purely diffusive, i.e. $\langle \Delta r^2(t) \rangle = 4D_0 t$, with $D_0 = k_B T/6\pi \eta a$. For [H$_2$O$_2]\neq 0$, the short-time MSD was ballistic, i.e. $\langle \Delta r^2 \rangle = v^2t^2$, evidencing self propulsion. At longer times, the MSD made a transition to diffusive behaviour, i.e. Eq.~(\ref{eq:Deff}) with $n = 2$. The physics here is the same as that which limits the ability of an {\it E. coli} cell to swim in a straight line. In the long time limit, rotational diffusion, Eq.~(\ref{eq:Drot2}), randomises the `aim' of of a self-propelled colloid. We therefore expect that the active part of the effective diffusivity to be given by $D_{\rm eff} - D_0 \simeq v^2 \tau_r$, where $\tau_r = D_r^{-1}$ is a characteristic time for rotational diffusion \footnote{Compare also Eq.~(\ref{eq:Deff2}) describing the effective diffusion of run-and-tumble {\it E. coli} cells.}. Solution of the relevant Langevin equation gives the numerical factor: 
\begin{equation}
D_{\rm eff} = D_0 + \frac{v^2 \tau_r}{6}. \label{eq:Deff3}
\end{equation} 

There is no obvious chemical reason to treat the interaction between O$_2$ and either Pt or polystyrene as any other than repulsive, in which case the situation depicted in Fig.~\ref{fig:self} should apply, i.e. we expect these particles to `swim' with the polystyrene end pointing forward. In the one case where direction of motion has been explicitly reported \cite{EbbensDirection}, this was indeed found to be the case, supporting the self-diffusiophoretic mechanism. Moreover, at low enough [H$_2$O$_2$] ($\lesssim 2 \%$ in the first experiments \cite{Jones}), it was found that the measured propulsion speed, $v$, was proportional to [H$_2$O$_2$], as predicted by Eq.~(\ref{eq:JanusSpeed}). At higher concentrations, the dependence of $v$ on [H$_2$O$_2$] becomes sublinear. This has been explained by postulating that reaction~(\ref{eq:reaction}) takes place in two steps: 
\begin{equation}
\mbox{Pt}\;+\mbox{H}_2\mbox{O}_2 \stackrel{k_1}{\longrightarrow} \mbox{Pt}\!:\!\mbox{H}_2\mbox{O}_2 \stackrel{k_2}{\longrightarrow}
\mbox{H}_2\mbox{O} + 0.5\, \mbox{O}_2 + \;\mbox{Pt}.\label{eq:reaction2}
\end{equation} 
The probability (per unit area per unit time) that a reactive site on Pt becomes bound to H$_2$O$_2$ scales as $k_1$[H$_2$O$_2$] \footnote{Throughout, [H$_2$O$_2$] is measured in volume fraction \%.}, and the probability that the bound state, `Pt:H$_2$O$_2$', decomposes scales as $k_2$ \cite{EbbensSize,Jones}. The approximation leading to Eq.~(\ref{eq:JanusSpeed}) assumes that the second reaction is much faster than the first, so that $k_1$[H$_2$O$_2$] controls the overall rate of oxygen production. At higher [H$_2$O$_2$], the first reaction is saturated, so that it is the rate of decomposition that is rate limiting, so that $ \tau_f^{-1} \simeq k_2 a^2$, giving \cite{EbbensSize}
\begin{equation}
\left|u_{\rm dph}^{\rm self}\right|^{\rm (II)} \simeq \frac{k_B T}{\eta} |K{\cal L}| \frac{k_2}{D_P}. \label{eq:JanusSpeed2}
\end{equation}
The transition from regime I, Eq.~(\ref{eq:JanusSpeed}), to regime II, Eq.~(\ref{eq:JanusSpeed2}), explains the sublinear dependence of the observed propulsion speed on [H$_2$O$_2$].

\begin{figure}
\begin{center}
\includegraphics[height=6cm]{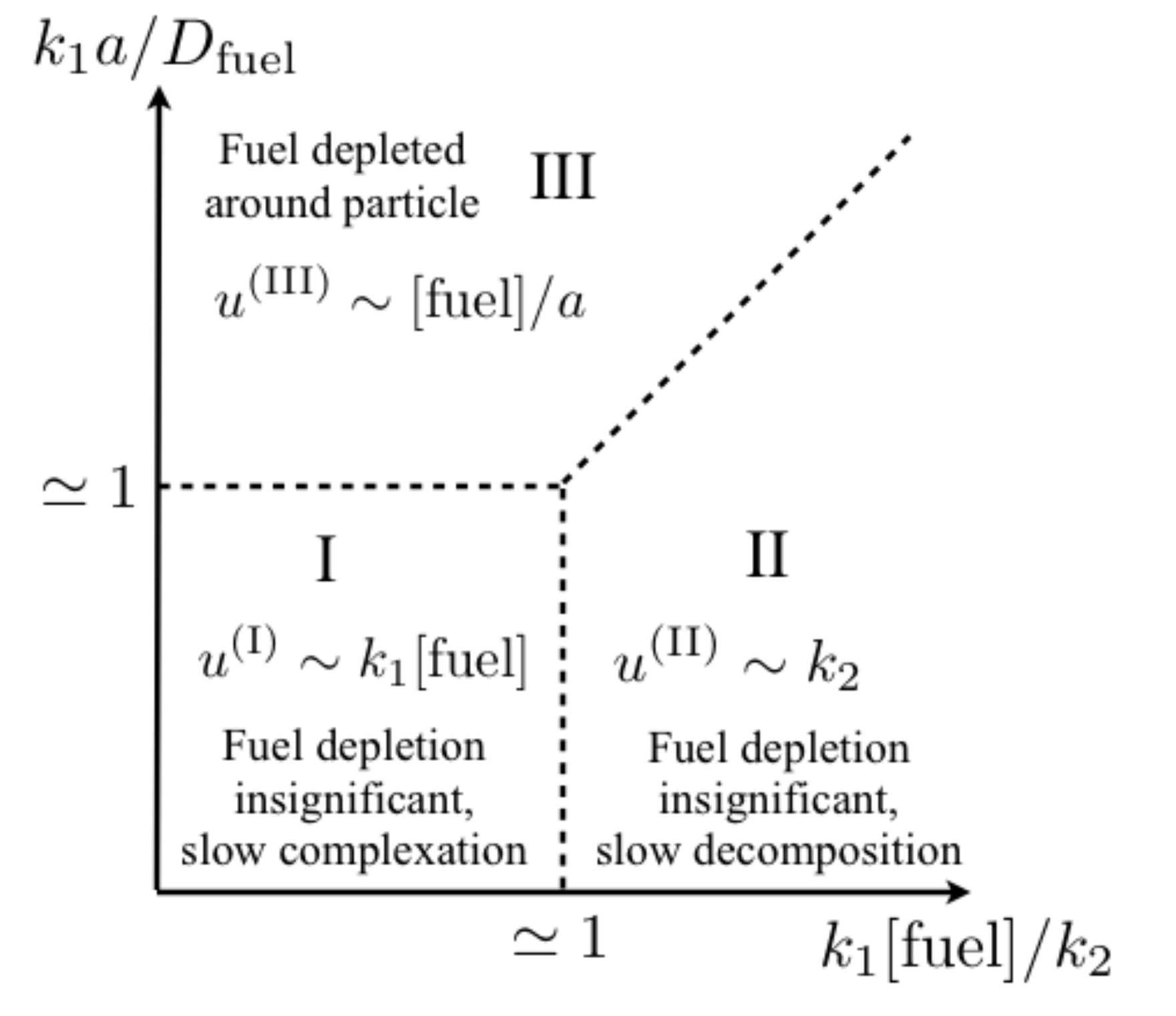}   
\end{center}  
\caption{Schematic of different propulsion regimes predicted for a self-diffusiophoretic Janus particle dispersed in a solution `fuel' that is decomposed to give a product according to some scheme analogous to reaction~(\ref{eq:reaction2}), where fuel~=~H$_2$O$_2$ and product~=~O$_2$. The rates $k_1$ and $k_2$ are defined in reaction scheme (\ref{eq:reaction2}). Drawn after \protect\cite{EbbensSize}.} \label{fig:EbbensSize}
\end{figure}

In the most recent experiments \cite{EbbensSize}, a dependence of the propulsion speed on particle size has been observed for particles in the range $0.2 \mu\mbox{m} \leq a \leq 5 \mu\mbox{m}$, which is not predicted by either Eq.~(\ref{eq:JanusSpeed}) or Eq.~(\ref{eq:JanusSpeed2}), with the data being consistent with $v \sim a^{-1}$. This can be understood by solving the full reaction-diffusion problem round a Janus particle with a catalytic patch \cite{EbbensSize}. The physics is as follows. The reaction at the catalytic patch depletes a region around each particle of reactants (here H$_2$O$_2$); this depletion zone is of extent $\simeq a$, so that the diffusive flux transporting reactants towards the catalytic patch is $\simeq D_R [$H$_2$O$_2]/a$, where $D_R$ is the reactant diffusion coefficient. Size-dependent self-phoretic velocity occurs when this flux is the rate limiting step, and not $k_1$ or $k_2$, i.e. when the particle is large and/or the reaction rates are fast. Quantitatively, this requires $D_R [\mbox{H}_2\mbox{O}_2]/a \ll k_1 [\mbox{H}_2\mbox{O}_2]$ and $D_R [\mbox{H}_2\mbox{O}_2]/a \ll k_2$. In this regime for a generalised fuel,
\begin{equation}
\left|u_{\rm dph}^{\rm self}\right|^{\rm (III)} \simeq \frac{k_B T}{\eta} |K{\cal L}| \frac{[\mbox{fuel}]}{a}. \label{eq:JanusSpeed3}
\end{equation} 

We can now revisit Eqs.~(\ref{eq:JanusSpeed}) and (\ref{eq:JanusSpeed2}) in terms of the physics introduced in the last paragraph and estimate their regimes of validity. Both of them require  a fast flux of reactants onto the reactive site, i.e. $D_R [\mbox{H}_2\mbox{O}_2]/a \gg k_1 [\mbox{H}_2\mbox{O}_2]$, or $k_1 a/D_R \ll 1$. Regimes I and II are then distinguished by whether the first (complexation) or second (decomposition) reaction in Eq.~(\ref{eq:reaction2}) is rate limiting: $k_1[\mbox{H}_2\mbox{O}_2]/k_2 \ll 1$ (regime I) or $k_1[\mbox{H}_2\mbox{O}_2]/k_2 \gg1$ (regime II). The three regimes are shown schematically in Fig.~\ref{fig:EbbensSize} in the parameter space of dimensionless particle size, $k_1 a/D_R$, and reaction rates, $k_1[\mbox{fuel}]/k_2$.

To date, there is some experimental support for the framework summarised in Fig.~\ref{fig:EbbensSize} \cite{EbbensSize,Jones,EbbensDirection}. It is likely that this will be the dominant mechanism under some experimental conditions. However, it is as yet unclear whether this is in fact the dominant mechanism for all cases of Pt-coated polystyrene particles dispersed in H$_2$O$_2$ solutions so far reported. There are two main reasons for caution: the nature of the particles, and the proximity of surfaces. The polystyrene particles used in different experiments \cite{EbbensSize,Jones,Bocquet,EbbensDirection} are not all the same. This matters because polystyrene colloids are stabilised by charge, and the Pt-catalysed decomposition of H$_2$O$_2$ may involve ionic intermediates  (see Eqs.~(\ref{eq:ionic1}) and (\ref{eq:ionic2}) in Section~\ref{subsec:otheractive} and \cite{McKeePt}). It is therefore possible that surface charges, which often differ between preparations, may play a role in determining propulsion  (see Section~\ref{subsec:dph}). Significantly, the direction of propulsion has been checked explicitly in only one reported case, and there is no study of the effect of solvent ionicity. It would therefore be interesting to check the speed and direction of motion of different Janus polystyrene particles with known surface properties (especially the $\zeta$ potential) in different salt concentrations. Secondly, and partly following on from the first point, phoresis is known to be subject to significant wall effects \cite{AndersonRev2}, especially when electrostatics is involved. All experiments on Pt-coated Janus polystyrene colloids to date have been performed next to surfaces. Convincing demonstration of any mechanism would require the study of bulk motion and/or elucidation of wall effects

Irrespective of propulsion mechanism, Pt-coated polystyrene beads with H$_2$O$_2$ as fuel is a useful model system in which the propulsion speed is readily controllable by [H$_2$O$_2$]. A number of complications exist, however. First, the product here is oxygen, which has limited solubility in water. It ultimately comes out as gas, which gathers as bubbles in any sealed sample chamber, hindering observations and experiments. Secondly, metal-coated Janus particles in general, and Pt-coated particles in particular \footnote{Pt is  the third densest naturally-occurring element.}, are heavier at one end. The gravitational potential energy difference between a particle of radius $a$ with a half-coating of metal (density $\rho_m$) of thickness $\delta \ll a$ pointing coating up and coating down in a solvent of density $\rho_s$ is 
\begin{equation}
\Delta U = 2\pi a^3\delta (\rho_m - \rho_s)g. \label{eq:gravity}
\end{equation} 
For $\delta = 10$nm, $\rho_m = 21.6$g/cm$^3$ (Pt), $\rho_s = 1$g/cm$^3$ (water), $a = 1\mu$m and $g = 9.81$m/s$^2$, $\Delta U \approx 3k_B T$, which is substantial. 

The gravitational effect can be alleviated by using other metals, provided the relevant catalytic properties are still present \cite{McKeePt}. On the other hand, a self-propelled Janus particle that is `gravitactic' should be interesting in its own right, see Section~\ref{subsec:control}. The most obvious way to circumvent the `gas problem' is to use alternative chemistries that do not involve gaseous products. For example, silver-coated Janus silica particles are self-propelled when irradiated with UV light \cite{SenSilver} because of the reaction
\begin{equation}
3\mbox{H}_2\mbox{O}_2 + 3 \mbox{Ag} \stackrel{\mbox{\tiny UV}}{\longrightarrow} 3\mbox{Ag}^++ 2\mbox{OOH}^- + 2 \mbox{H}_2\mbox{O}. \label{eq:SenSilver}
\end{equation}
Since Ag$^+$ diffuses 5 times faster than OOH$^-$, $\tilde\beta = \frac{2}{3}$,  Eq.~(\ref{eq:beta}), and a strong contribution from the $\nu_e$ term in Eq.~(\ref{eq:slip2}) can be expected, as well as dependence on solvent ionicity. 

A different way to circumvent the `gas problem' is to suspend Janus particles in a just-subcritical binary liquid mixture \cite{BechingerJanus}, specifically, a mixture of lutidine and water (LuW). Like many other hydrogen-bonded liquid mixtures, Lu-W is miscible at low temperatures, but demixes above a lower critical solution temperature (LCST, here $T_c = 33^\circ$C), Fig.~\ref{fig:Bechinger}(a). Buttinoni et al. synthesised gold-coated polystyrene Janus particles and suspended these in a LuW mixture at critical composition just below the LCST, $(C_c, T_0)$ in Fig.~\ref{fig:Bechinger}(a). Fluorescence imaging shows that illumination with a laser heats the gold side of each particle to above $T_c$, Fig.~\ref{fig:Bechinger}(b), bringing about demixing. Depending on whether the gold cap is functionalised to be hydrophobic or hydrophilic, the Lu-rich phase preferentially gathers on the gold side or the polystyrene side, creating a local concentration gradient that can be imaged directly using suitable dyes. Self-diffusiophoresis results without the evolution of gas. An added advantage is that no fuel is consumed -- the necessary energy for activity comes from the absorption of photons from the light source. 

\begin{figure}
\begin{center}
\includegraphics[height=4.5cm]{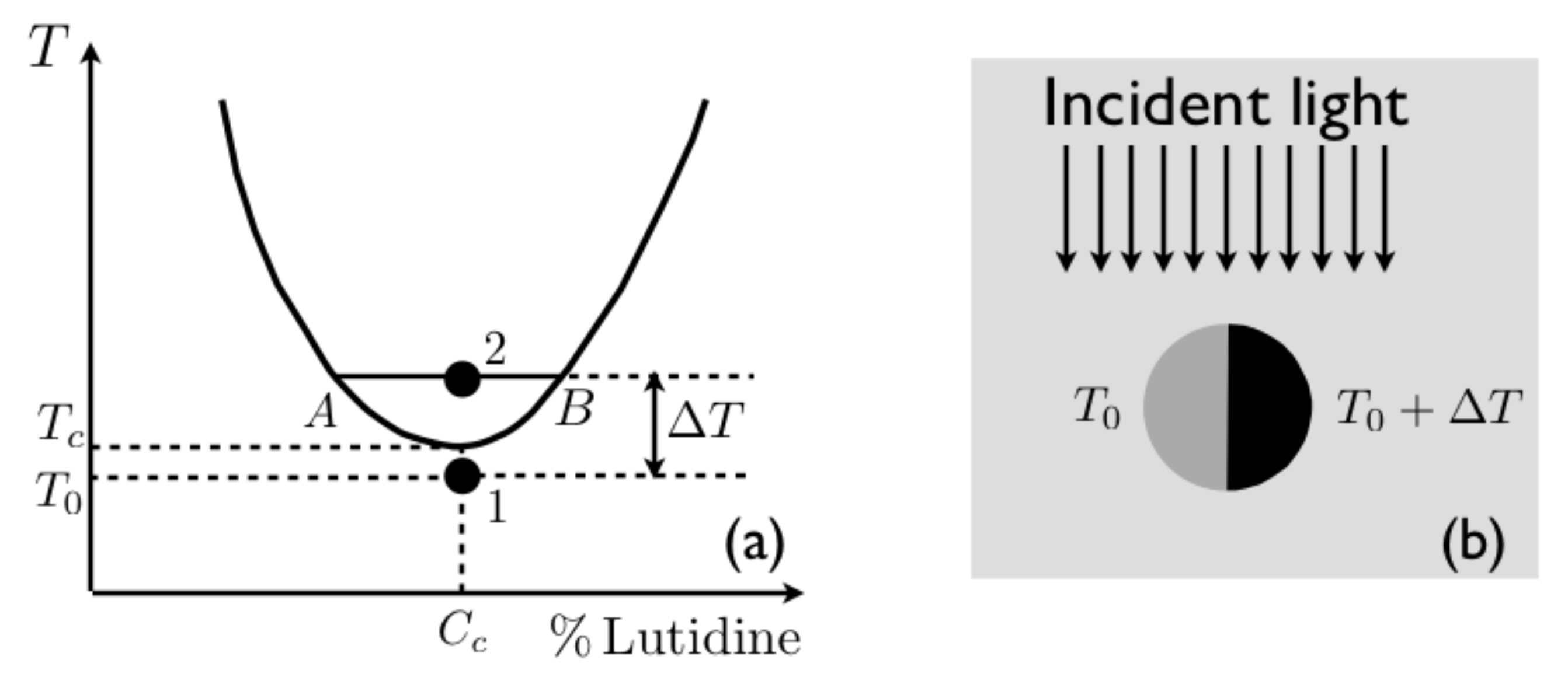}   
\end{center}  
\caption{Janus particle in binary liquid mixture \protect\cite{BechingerJanus}. (a) A schematic of the phase diagram of the lutidine-water system. Janus particles are dispersed at state point 1 just below the critical point $(C_c, T_c)$. (b) A polystyrene particle half coated in gold (black). Absorption of incident light causes heating on the gold side, raising the local temperature by $\Delta T$, taking the liquid there to state point 2, see part (a), so that it demixes into a water-rich phase $A$ and a lutidine-rich phase $B$. Phase $A$ ($B$) will gather preferentially at the gold cap if it is functionalised to be hydrophilic (hydrophobic), creating a concentration gradient for self-diffusiophoresis. } \label{fig:Bechinger}
\end{figure}

While there is a temperature gradient in the propulsion mechanism schematised in Fig.~\ref{fig:Bechinger}, it is not thermophoresis but diffusiophoresis that propels the particle forward -- the temperature differential places the binary liquid mixture on either side of a particle in different state points on the phase diagram. Indeed, when Buttinoni et al. repeated their experiments in pure water, no propulsion occurred \cite{BechingerJanus}. However, self-thermophoresis can be used to generate propulsion in Janus particles under the right conditions \cite{JiangThph}.

\subsection{Other systems} \label{subsec:otheractive} While there are many different self-propelled systems of spherical Janus particles, other synthetic active colloids exist. Here I briefly introduce three. 

First, the demonstration of self propulsion in bimetallic platinum-gold (PtAu) nano-rods dispersed in H$_2$O$_2$ in fact pre-dated the observation of motility in Janus spheres \cite{PaxtonRod}. There has been significant uncertainty over the propulsive mechanism, the initial suggestion being that it was due to flow driven by surface tension gradient (Marangoni effect).  Mounting experimental evidence, however, suggests that propulsion is in fact due to catalytic self-electrophoresis \cite{SenElectro}. With Pt less electronegative than Au (electronegativities of 2.3 and 2.5 respectively on the Pauling scale), the following may occur:
\begin{eqnarray}
\mbox{Anode (Pt)}: &\;&\;\; \mbox{H}_2\mbox{O}_2  \rightarrow  2\mbox{H}^+ + 2\mbox{e}^- + \mbox{O}_2 \label{eq:ionic1}\\
\mbox{Cathode (Au)}:&\; & \;\; 2\mbox{H}^+  +  2\mbox{e}^- + \mbox{H}_2\mbox{O}_2   \rightarrow 2\mbox{H}_2\mbox{O} .\label{eq:ionic2}
\end{eqnarray}
The electron flow from Pt to Au within the rod must be balanced by a proton flow within the electric double layer in the same direction. In the rest frame of the rod, this results in an electro-osmotic flow of solvent form Pt to Au, so that in the rest frame of the liquid, the rod is propelled from in the Au to Pt direction, as observed. 

The second system I will mention does seem to `swim' by the Marangoni effect, but relies on a novel coupling with the flow field generated by self-propulsion; it is also interesting because each particle carries its own fuel \cite{HerminghausActive}. These swimmers are in fact too large to be in the colloidal domain. The `particles' are $\simeq 100 \mu$m diameter water droplets (equivalent to nanoliters) in squalane (an oil). Dissolved in the oil is the surfactant mono-olein, which is present in excess ($> 30 \times$ the critical micellar concentration) for the purpose of stabilising the oil droplets. Bromine dissolved in the water serves as the fuel. It is able to {\it trans}-brominate and therefore saturate the double bond in a mono-olein molecule \footnote{The reaction needs to occur continually because of the loss of brominated molecules to the bulk, where its concentration is very low (initially zero).}. The product is a less effective surfactant than the parent mono-olein. Non-uniformities in the distribution of brominated and non-brominated molecules on the surface of a droplet will result in surface tension gradients that can drive flow. The resulting flow interacts with the surface distribution via hydrodynamic coupling. Pleasingly, under the right conditions, this coupling can sustain the non-uniform distribution of surfactants necessary to generate the flow field in the first place. The fact that propulsion in this system relies on tangential surface deformations (surface tension gradients) makes it bio-mimetic on a coarse-grained level to many eukaryotic micro-organisms like {\it Paramecium} or {\it Volvox}; collectively these are known as `squirmers'. 

A `parasitic' mode of biological propulsion inspired the third system of artificial swimmers that I will mention here. When pathogenic bacteria of the genus {\it Listeria} enter a host eukaryotic cell, it moves by the polymerisation of the host's actin (fuelled by the host's ATP supply) \cite{StevensRev}. This `actin rocket' mode of propulsion has been mimicked in synthetic colloids. Polystyrene beads with ActA protein (a catalyst for actin polymerisation) tethered to the surface are self-propelled by a trailing tail of polymerisation in the presence of ATP, giving  speeds in the range $\simeq 0.01$-$0.1 \mu$ms$^{-1}$ \cite{CameronActin}. 

It is interesting to point out that by either carrying their own `fuel' \cite{HerminghausActive} or having the ability (in principle at least) to use a biologically abundant source of `fuel' (actin + ATP) \cite{CameronActin}, the previous two systems have demonstrated how to achieve independence from a source of non-natural fuel in the external medium. Such independence is vital if medical `nano-robots' are to move from the realm of science fiction into real-life technologies. 

Many variations on the themes already introduced in the last two sections are available in the literature, such as the use of biological enzymes rather than metallic catalysts \cite{Vicario}, together with propulsion modes that I cannot review here. Since synthetic active colloids is still a relatively young field, one must also expect completely novel propulsion modes to continue to appear, the absolutely essential requirements being only the consumption of fuel (`active') coupled to some form of asymmetry (`propulsion'), perhaps not necessarily structural \cite{GolestanianChucker,AllenChucker}, and certainly not necessarily externally imposed \cite{HerminghausActive}. 

\subsection{Controlling synthetic swimmers} \label{subsec:control}

Left to itself, a particle self-propelled at speed $u$ will undergo a random walk with $D_{\rm eff} \simeq u^2D_r^{-1}$, because its `aim' is randomised by rotational diffusion (with diffusivity $D_r$). It is both scientifically interesting and technologically fruitful to consider how one might render the motion of such active colloids responsive to its environment in general, and therefore use the environment to control and perhaps direct such motion. In a concentrated suspension of active particles, other particles, of course, form part of the `environment' of any one particular swimmer. We defer many-body effects to Section~\ref{sec:generic} and focus here on controlling single-particle motion. 

Different levels of environmental control can be envisaged. The most basic is `on-off'. This can be rather easily achieved whenever motility is switched on by an external light field \cite{SenSilver,BechingerJanus,JiangThph}. Since light can be flexibly patterned in space and time, this functionality opens up the possibility of light-choreographed self assembly of active colloids. The dependence of swimming speed on fuel concentration (e.g. \cite{Jones}) introduces a finer level of control than simple `on-off'. The possible dependence of motility on particle size, Fig.~\ref{fig:EbbensSize}, and on surface hydrophobicity via solvent slip, Eq.~(\ref{eq:Ajdari}), give further possibilities for designing `smart' active colloids. Thus, for instance, one might imagine synthesising active colloids using microgel particles, whose size is temperature sensitive \cite{Schild}; and the technology for creating surfaces with switchable hydrophobicity exists \cite{LangerSwitch}. Such possibilities appear not to have been exploited to date in the active colloids field. 

Other kinds of environmental control mimic various `tactic' motion of prokaryotic \cite{Schlegel} and eukaryotic  \cite{BrayBook} cells. In chemotaxis, cells travel up or down chemical gradients, $\nabla c$; phototactic cells, often photosynthetic, track gradients in light intensity. Gravitaxis and magnetotaxis are not about gradients; instead, cells with these abilities move along gravitational or magnetic field lines respectively. In both gravitaxis and magnetotaxis, the cell experiences a torque whenever some cellular axis (typically aligned closely to the average swimming direction) is misaligned with the field lines. 

Mimicking gravitaxis or magnetotaxis is conceptually the simplest. We have already seen that Pt-coated polystyrene Janus particles are in principle gravitatic, because the Pt coating makes the particle heavier on one side, Eq.~(\ref{eq:gravity}). Such particles can therefore be used to model aspects of the the complex swimming behaviour of gravitatic organisms such as {\it Chlamydomonas} \footnote{Strictly, the coupling between viscous and gravitational torques produces what is known as {\it gyrotaxis} in this organism.}. This feature has not been exploited to date. Designing magnetotactic self-propelled colloids is conceptually straightforward -- a magnetic moment has to be incorporated, e.g. in nickel-striped Pt-Au nano-rods \cite{SenNickel}. 

Chemotaxis has been observed in Pt-Au rods, which move up a concentration gradient of H$_2$O$_2$ \cite{SenChemo}. Simulations suggest that the mechanism hinges on the faster propulsion speed in higher [H$_2$O$_2$]. If this is correct, then chemotaxis should operate quite generally, e.g., in regimes I and III in Fig.~\ref{fig:EbbensSize}. Particles self-propelled by reaction~(\ref{eq:SenSilver}) are phototactic up a UV light gradient \cite{SenSilver}. The mechanism is indirect. Reaction~(\ref{eq:SenSilver}) generates ions. In stronger UV more ions are generated, so that in a gradient of UV illumination, there is also a concentration gradient of ions, giving rise to diffusiophoresis according to Eq.~(\ref{eq:slip2}). 

Other interesting ways to control active colloids have been reviewed by Hong et al. \cite{Sen}. I close this brief survey by mentioning that surface and other confinement effects can also be mentioned under `environmental control'. Since phoresis is an intrinsically surface phenomenon, all methods of self propulsion relying on any form of self-phoresis should be very sensitive to the proximity of interfaces \cite{AndersonRev2}. Indeed, the majority of experiments to date have been carried out next to glass surfaces (likely negatively charged unless treated in some specific way). The effect of the presence of such surfaces has neither been systematically investigated nor exploited for `control' purposes to date. 

\section{Characterising active colloids} \label{sec:ddm}

Whether one chooses to do experiments with bacteria or Janus particles, an active suspension needs to be characterised. The `standard' measurements are, of course, still needed, e.g. light scattering or microscopy to find the size and shape. In many cases, this step is already more challenging than characterising passive colloids. Thus, for example, many self-propelled particles are non-spherical: this applies to the majority of motile bacteria, and to an increasing number of synthetic active colloids. In characterising electrical properties, the existence of a periplasm (or `cell wall') \cite{Schlegel} porous to ions complicates the analysis of electrophoresis \cite{BosSurface}. More generally, since most, if not all, self propelled particles to date rely on structure asymmetry, the potential of mean force between two such particles will necessarily be anisotropic; characterising such interactions in active colloids is a challenge that so far has largely been neglected. Since phoresis depends sensitively on interfacial properties, measurements of surface characteristics to a much higher level of details and precision may be needed to ensure reproducibility from experiment to experiment. However, none of these requirements is intrinsic to characterising active colloids, which requires minimally the measurement of the speed distribution, $P(v)$, and the fraction of motile particles, $\alpha$, in an active suspension. 

\subsection{Tracking} Conceptually, the simplest way to characterise an active suspension is to image the 3D trajectory of $N$ particles, $\mathbf{s}_i(t), 1 \leq i \leq N$. Such 3D tracking can yield very detailed information, e.g. the statistics of the run-and-tumble swimming behaviour of {\it E. coli} \cite{BergBrown}. The velocity of particle $i$ is obtained from
\begin{equation}
\mathbf{v}_i(t) = \frac{d \mathbf{s}_i(t)}{dt} = \lim_{\delta t \rightarrow 0} \frac{\mathbf{s}_i(t + \delta t) - \mathbf{s}_i(t)}{\delta t} \approx \frac{\mathbf{s}_i(t + \delta t) - \mathbf{s}_i(t)}{\delta t} ,
\end{equation}
so that a velocity distribution can be built up, subject to uncertainties introduced by finite time and space resolutions (the smallest $\delta t$ and $\delta \mathbf{s}$ resolvable)

In practice, 3D tracking is seldom used, because it requires specialist equipment and/or data processing. For example, one could use servo-control of the objective to keep a single particle in focus \cite{BergBrown}. In principle, fast confocal microscopy, which has been used to image particles moving under flow \cite{RutReview}, and holographic microscopy \cite{LeeHolographic} can also be used to track active colloids, but have not in fact been so used to date. A more straightforward method relies on the fact that defocussing transforms the image of a particle into a ring whose size is proportional to the amount of defocussing \cite{WuFocus}. This has been used, e.g., to study the motion of bacterial cells as they approach walls \cite{TangSurface}. 

Tracking in 2D is much more straightforward and is therefore widely practised. It can be used to study motion that is confined to a plane, such as bacteria in a thin film \cite{LibchaberEnhanced} or next to a wall. In the case of synthetic particles, all tracking studies to date appear to have been in 2D (\cite{Jones} is one example). Two-dimensional tracking can also be used to obtain projected information of bulk motion in 3D. Taking the $xy$ plane to be the imaging plane and $z$ to be the optic axis of a microscope, it is clear that 2D tracking of bulk motion yields the distribution $P(v_{xy})$, e.g. of swimming {\it Chlamydomonas} \cite{GoldsteinEnhanced}. 

Tracking, of course, depends on particles that are visible in a microscope. In many cases, some form of contrast enhancement is necessary, e.g. by phase modification (phase contrast and Normarski differential interference contrast (DIC), reviewed in \cite{ElliotRev} in the context of colloid microscopy) or by blocking unscattered light (dark field). Such contrast enhancement methods work for both bacteria and synthetic particles. The particles can also be rendered fluorescent. In the case of bacteria, this can be done by incorporating plasmids with genes for the synthesis of green fluorescent protein (GFP) or one of its other-colour variants \footnote{Note that GFP fluorescence depends on oxygen, albeit only at a low level \cite{TsienRev}.}. In the case of synthetic particles, using metal-coated Janus fluorescent beads offers a means for discerning the direction of motion (only the un-coated portion of each particle glows), although DIC can also be used for this latter purpose without fluorescence \cite{GranickJanusRot}, even for particles in close proximity.

I will not discuss the details of how to process movies to obtain mean-squared displacements (and therefore speeds), which involves particles identification and tracking. The issues are similar to those encountered in tracking passive colloids, especially colloids in flow, for which a number of accounts are available \cite{RutReview,CrockerGrier,JenkinsConfocal} and relevant software is available on-line \footnote{By E. R. Weeks and J. C. Crocker at \url{http://www.physics.emory.edu/~weeks/idl/}.}. Instead, I turn to consider an issue that is peculiar to active colloids. In a population of active colloids, there is usually a fraction of particles that do not `swim'. In a bacterial suspension, these may be cells having damaged flagella from the washing procedure. In a synthetic population, this could be due to particles with defective coating. Using tracking to separate these two sub-populations, which we may call `diffusers' and `swimmers', turns out to be non-trivial. 

In principle, one might imagine proceeding as follows. After time $t$, the MSD of a diffuser with diffusivity $D$ will be $2nDt$ in $n$ dimensions and that of a swimmer with speed $v$ will be $v^2t^2$. These cross at $t_0 = 2nD/v^2$. If we perform tracking over some period $t_1 > t_0$, then all tracks with MSD $\langle r^2 \rangle > 2nDt_0$ (or, equivalently, $> v^2t_0^2$) belong to swimmers, and the other tracks belong to diffusers. In practice, this is not feasible because $D$ and $v$ are neither known {\it a priori} nor sharply defined quantities, but are unknown distributions to be determined. Under these circumstances, the measured motile fraction, $\alpha$, will be dependent on $t_1$, the choice of which is necessarily a rather subjective exercise. 

A more algorithmic method has been successfully implemented in 2D \cite{Gaston}. Each trajectory is split into short segments of duration $\Delta t$ over which an average swimmer moves $\approx 1$~pixel, Fig.~\ref{fig:tracking}(a). First, the mean angle $\langle |\theta| \rangle$ between successive segments is calculated; $\langle |\theta| \rangle=\pi/2$ for a random walk and $\langle |\theta| \rangle = 0$ for a straight swimmer. Then, using the trajectory's start-to-end distance $L$, duration ${\cal T}$, and the mean segment length $\Delta r_{2D}(\Delta t)$, we calculate the parameter $N_c=\frac{L/{\cal T}}{\Delta r_{2D}/ \Delta t}$, which is the ratio of linear speed (time to move between two points in a straight line) to the mean curvilinear speed (speed along the actual trajectory). It is clear that $N_c = 0$ for a random walk in the limit ${\cal T} \rightarrow \infty$ and $N_c = 1$ for a straight swimmer. Tracking of mixed swimmers and non-swimming bacteria near a wall \cite{Gaston} returned two well-separated clusters in the $(N_c, \langle |\theta| \rangle)$ plane, from which motile and non-motile populations could be separated and the respective $P(v)$ and $D$ extracted via fitting of the MSD calculated from each population of trajectories. Using this method on projected 2D trajectories of bacteria swimming in the bulk proves less successful, Fig.~\ref{fig:tracking}(b): there is a more or less continuous distribution in $(N_c, \langle |\theta| \rangle)$ space that is difficult to divide into separate populations of swimmers and diffusers, and therefore to calculate $P(v)$ and $\alpha$, without some degree of arbitrariness.

\begin{figure}
\begin{center}
\includegraphics[height=3.5cm]{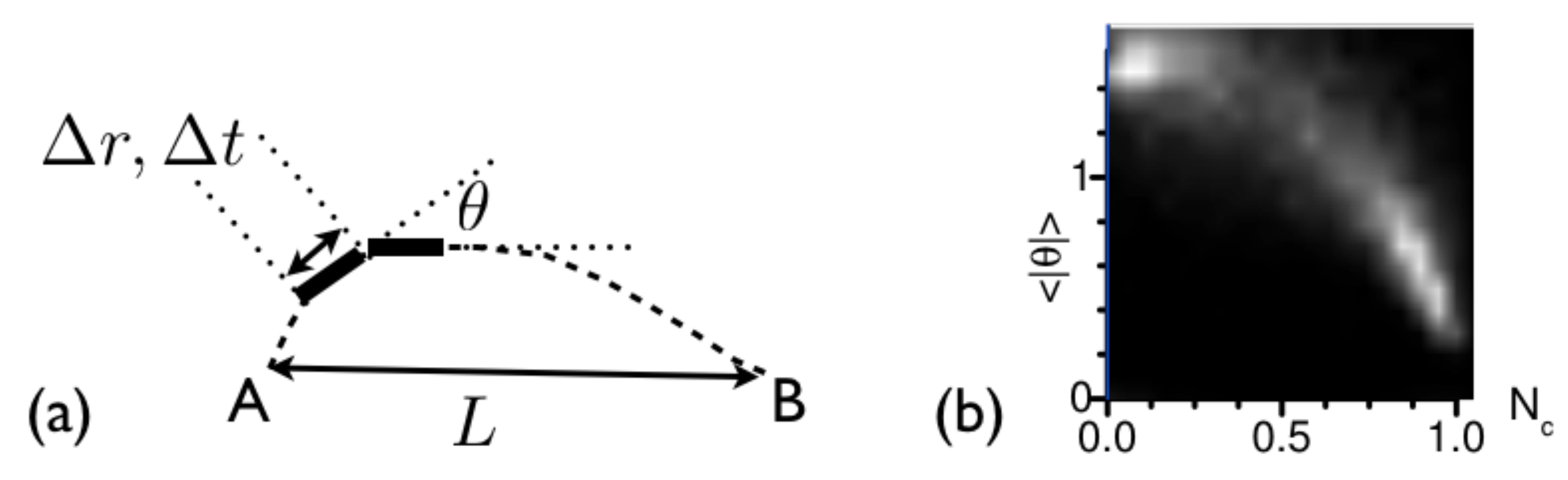}   
\end{center}  
\caption{Separating swimmers from diffusers using tracking \cite{Gaston}. (a) Each trajectory, A to B along the dashed curve, is subdivided into segments (bold) of (fixed) length $\Delta r$. Two statistics are calculated: the average angle between segments, $\langle \theta \rangle$, and $N_c = (L/{\cal T})/(\Delta r/\Delta t)$, where ${\cal T}$ is the time taken to travel along the (dashed) trajectory from A to B, $L$ is the start-to-end distance of this trajectory, and $\Delta t$ is the time taken to traverse segment $\Delta r$. Applying this algorithm to tracks of {\it E. coli} moving next to a surface produced two well-separated populations in $(N_c, \langle |\theta| \rangle)$ space \cite{Gaston}. (b) Applying the same algorithm to projected 2D tracks of {\it E. coli} swimming in 3D, however, gives a more or less continuous distribution in $(N_c, \langle |\theta| \rangle)$ space that cannot easily be separated into distinct populations of swimmers and diffusers \cite{MartinezDDM}. } \label{fig:tracking}
\end{figure}

Thus, while tracking, especially in 3D, can produce very detailed information on the motility of single particles, there are problems in implementing it as a routine method for measuring perhaps the most useful two statistics characterising a population of self-propelled particles: the speed distribution, $P(v)$, and the fraction of swimmers, $\alpha$.

\subsection{Dynamic light scattering}

A potentially high throughput method for characterising active colloids with good averaging is dynamic light scattering (DLS), which measures the intermediate scattering function (ISF), $f(\mathbf{q},\tau)$, where $\mathbf{q}$ is the scattering vector \cite{BernePecora}. For an ergodic system (i.e. one that explores the whole of configuration space on the time scale of an experiment), the ISF decays from $f=1$ at $\tau = 0$ to $f=0$ over some characteristic time $\tau_0$, which characterises the average dynamics of the system at length scale $\simeq 2\pi/q$ (where $q = |\mathbf{q}|$). For $N$ identical, non-interacting particles, 
\begin{equation}
f(\mathbf{q},\tau) = \left\langle e^{i\mathbf{q}\cdot\Delta \mathbf{r}(\tau)} \right\rangle, \label{eq:ISF}
\end{equation}
where $\Delta \mathbf{r}(\tau)$ is the time-dependent displacement of a particle, and $\langle \ldots \rangle$ denotes ensemble averaging, which, for a non-interacting system, can be replaced by an average over time. For isotropic, non-interacting spherical particles (radius $a$) diffusing in a liquid of viscosity $\eta$ at absolute temperature $T$,
\begin{equation}
f(q,\tau) = e^{-Dq^2\tau}, \label{eq:ISFdiff}
\end{equation}
where $D = k_B T/6\pi\eta a$ is the single-particle diffusivity. The isotropic nature of the motion renders the ISF dependent only on the magnitude of $\mathbf{q}$; the $q^2$ dependence is characteristic of diffusion. For a non-interacting collection of particles travelling at speed $v$ in straight lines with uniformly distributed directions, Eq.~(\ref{eq:ISF}) evaluates to
\begin{equation}
f(q,\tau) = \frac{\sin(qv\tau)}{qv\tau} \equiv \mbox{sinc} (qv\tau). \label{eq:ISFswim}
\end{equation}
Once more, the isotropic nature of the motion means that the ISF depends on the magnitude of $\mathbf{q}$; but now the scaling is linear, reflecting the ballistic nature of the motion. 

For a mixed population of isotropic diffusers (fraction $1-\alpha$)  and swimmers (fraction~$\alpha$), where the Brownian fluctuations of both are described by the same (thermal) diffusivity $D$, and the swimmers have a speed distribution $P(v)$, the ISF is a proportionate linear combination of Eqs.~(\ref{eq:ISFdiff}) and (\ref{eq:ISFswim}) averaged over speeds for the latter \cite{Stock}:
\begin{equation}
f(q,t) = (1-\alpha) e^{-Dq^2 \tau} + \alpha e^{-Dq^2\tau} \int_0^{\infty} P(v) \;\mbox{sinc}(qv\tau) dv. \label{eq:ISFmixed}
\end{equation}
To understand this form of the ISF, we use a Schulz distribution 
\begin{equation}\label{eq:schulz}
P(v) = \frac{v^Z}{Z!} \left(\frac{Z+1}{\bar{v}}\right)^{Z+1} \text{exp}\left[-\frac{v}{\bar v}\left(Z+1\right)\right],
\end{equation}
where $Z$ is related to the variance $\sigma ^2$ of $P(v)$ via $\sigma =\bar{v}(Z+1)^{-1/2}$, so that the integral in Eq.~(\ref{eq:ISFmixed}) can be analytically performed \cite{PuseySchulz}
\begin{equation}\label{eq:analytical_integral}
\int_0^\infty \!\!\!P(v) \;\mbox{sinc}{(q v \tau)} dv = \left(\frac{Z+1}{Zq\bar{v}\tau}\right) \frac{\text{sin}(Z \text{tan} ^{-1}\Lambda)}{(1+\Lambda ^2)^{Z/2}},
\end{equation}
where $\Lambda=(q \bar{v} \tau)/(Z+1)$. Using this result, we plot in Fig.~\ref{fig:ISFeg} an example ISF calculated at $q=1\;\mu{\rm m}^{-1}$ using typical {\it E. coli} motility parameters in Eqs.~(\ref{eq:ISFmixed})-(\ref{eq:analytical_integral}). The ISF shows a characteristic two-stage decay. The integral in Eq.~(\ref{eq:ISFmixed}) due to the straight-line motion of swimmers dominates the first, faster, process, while the purely diffusive first term due to the Brownian motion of non-swimmers dominates the second, slower, process.

\begin{figure}
\begin{center}
\includegraphics[height=5cm]{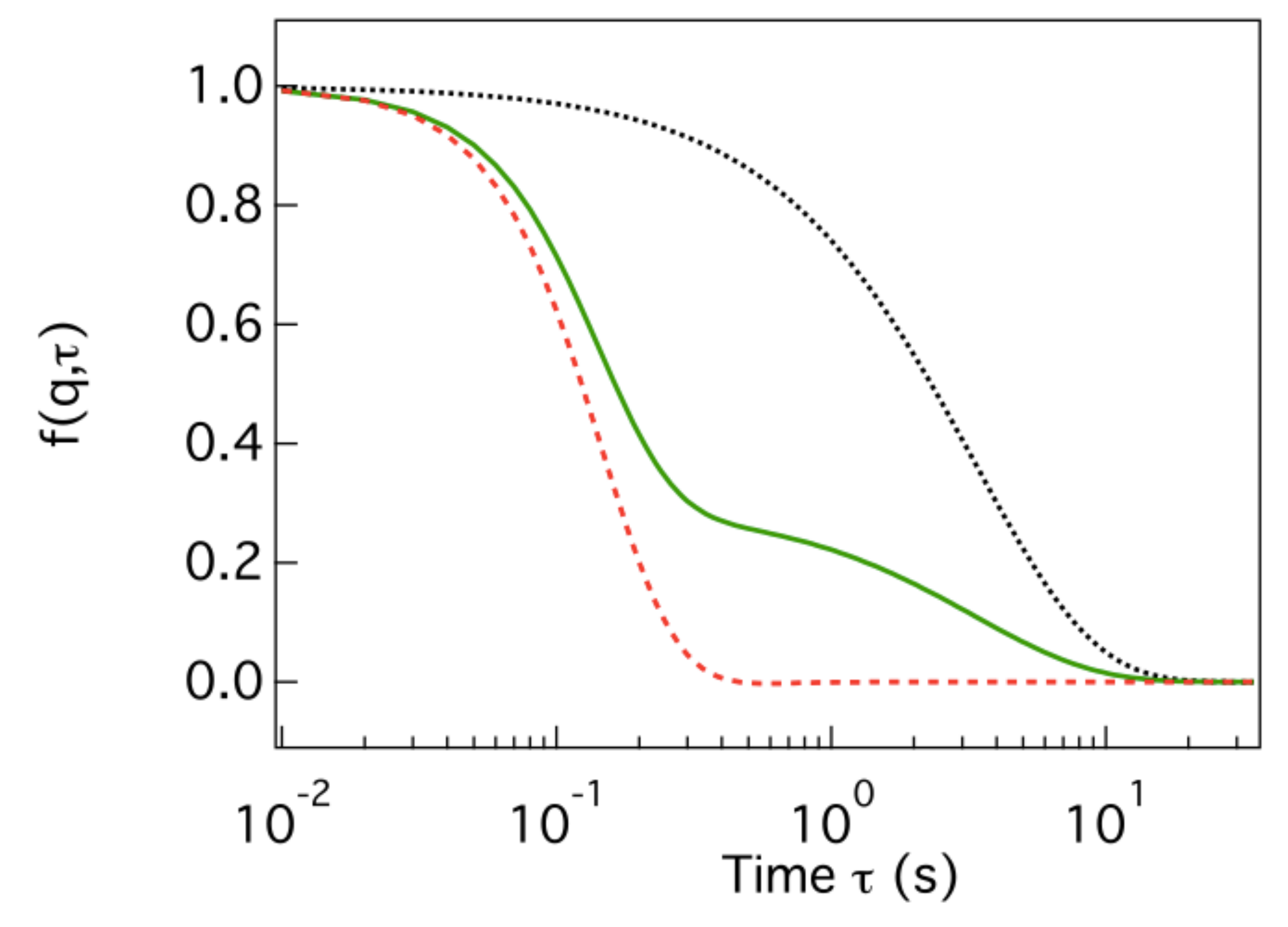}   
\end{center}  
\caption{Equation~(\ref{eq:ISFmixed}) plotted at $q = 1 \mu\mbox{m}^{-1}$ using the Schulz distribution and parameters typical of a population of swimming {\it E. coli}, $\bar{v} = 15\;\mu{\rm m s}^{-1}$, $D = 0.3\;\mu{\rm m}^2{\rm s}^{-1}$, and $\alpha = 0.7$ (continuous line), together with the swimming (dashed) and diffusive (dotted) components.} \label{fig:ISFeg}
\end{figure}

From visual inspection of this $f(q,\tau)$, the relative amplitudes of the fast and slow processes can be estimated to be $\approx 7:3$, which gives an estimated $\alpha \approx 0.7$. The length scale probed at this $q$ is $\ell \simeq 2\pi/q \approx 6\;\mu$m. Either by extrapolating the mixed ISF or by reference to the ISF for pure swimmers, it can be estimated that the fast process decays completely in $\tau_{\rm swim} \approx 0.5$~s. An order of magnitude estimate of the swimming speed is then $v \simeq \ell/\tau_{\rm swim} \approx 12\;\mu{\rm ms}^{-1}$. The slower, diffusive, process decays completely in $\tau_{\rm diff} \approx 20$~s, so that an estimate of the diffusion coefficient of the non-swimmers can be obtained from $6D\tau_{\rm diff} \simeq \ell^2$, giving  $D \approx 0.35\;\mu{\rm m}^2{\rm s}^{-1}$. These are credible estimates of the parameters used to generate this ISF: $\bar{v} = 15\;\mu{\rm m s}^{-1}$, $D = 0.3\;\mu{\rm m}^2{\rm s}^{-1}$, and $\alpha = 0.7$.

Figure~\ref{fig:ISFeg} suggests that DLS can be used to determine $D$, $\alpha$ and $P(v)$. Indeed, there had been attempts to characterise a population of motile {\it E. coli} by measuring its ISF using DLS. The method was, however, found to be not viable for an interesting reason. The magnitude of the scattering vector, $q$, is related to the scattering angle, $\theta$, via
\begin{equation}
q = \frac{4\pi}{\lambda} \sin \left(\frac{\theta}{2}\right),
\end{equation}
where $\lambda$ is the wavelength of the light in the medium of the suspension. In conventional DLS, $\theta \gtrsim 20^\circ$, so that the length scale probed is $\ell \lesssim 1 \mu$m, which is the size of the cell body. At this length scale, motions other than diffusion and swimming, especially a cell body `wobble', contributes towards decorrelating $f(q,\tau)$. These motions are poorly understood, and even the inclusion of the simplest model for the `wobble' \cite{Boon}  generates an ISF that contains too many fitting parameters for unambiguous data fitting. Characterising synthetic active colloids using DLS has not yet been attempted; it is unknown whether extraneous motions on the $\simeq 1 \mu$m scale would prove equally problematic. 

\subsection{Differential dynamic microscopy (DDM)}

This method can be used to obtain the ISF of a population of motile {\it E. coli} at low enough $q$ so that diffusion and swimming dominate its decay \cite{WilsonDDM}. DDM was first implemented \cite{CerbinoDDM1} to obtain the ISF of a dilute suspension of diffusing passive spherical polystyrene particles in water, and Eq.~(\ref{eq:ISFdiff}) was fitted to give $D$ and hence the particle radius. Two samples, with $2a = 73$nm and 420nm, were studied under bright-field imaging conditions, and the particles were barely visible because of a lack of contrast (the 420nm case) or invisible because the size was below optical resolution (the 73nm case). A narrow condenser aperture confers a degree of spatial coherence, and allows the capture of the near-field speckle pattern and the analysis of its fluctuations to obtain the ISF \cite{CerbinoDDM2} (see also Piazza's lectures in this volume). The physics is therefore  similar to that underpinning DLS (hence the title of \cite{CerbinoDDM2}: `Scattering information obtained by optical microscopy \ldots'). 

In the first demonstration of DDM for characterising bacterial motility \cite{WilsonDDM}, different imaging conditions were used. Although the data analysis algorithm (see below) was identical to the earlier work on colloids \cite{CerbinoDDM1,CerbinoDDM2}, the physics was probably different. Here, a $\times 10$ phase contrast objective was used to resolve each bacterium as a phase-dark object (refractive index $\simeq 1.4$ for {\it E. coli}) covering just over 1 pixel in the image. 

To perform DDM, a time sequence of images are taken, $I(\mathbf{r},t)$, where $\mathbf{r}$ refers to a position in the image plane. The basic object for analysis in all versions of DDM to date, $g(\mathbf{q},\tau)$, is the power spectrum of the Fourier transform of difference images, $D(\mathbf{r},\tau)$:
\begin{eqnarray}
D(\mathbf{r},\tau) & = & I(\mathbf{r},t+ \tau) - I(\mathbf{r},t), \label{eq:deltaI}\\
F_D(\mathbf{q},\tau) & = &  \int D(\mathbf{r},\tau) e^{i\mathbf{q}\cdot \mathbf{r}} d\mathbf{r}, \label{eq:FTdeltaI}\\
g(\mathbf{q},\tau) & = & \left\langle |F_D(\mathbf{q},\tau)|^2\right\rangle. \label{eq:powerspec}
\end{eqnarray}
The physics of DDM used in \cite{WilsonDDM} is as follows \cite{Reufer}. Figure~\ref{fig:mathias}(a) shows a one-dimensional schematic of the image of an object \footnote{For concreteness we may consider this the phase dark image of a bacterium.} located at $\mathbf{r}^0 = (x^0,0)$: $I(\mathbf{r},t) = I_0 - A(\mathbf{r},t)$. If there are no interference effects, the intensity pattern for $N$ identical objects at initial positions $\mathbf{r}_1^0,\mathbf{r}_2^0,..,\mathbf{r}_{\rm N}^0$ is $I(\mathbf{r},t=0)=I_0-\sum_{\rm j=1}^N  A(\mathbf{r}-\mathbf{r}_{\rm j}^0)$. During time $\tau$ the $j$-th object moves from $\mathbf{r}_{\rm j}^0$  to  $\mathbf{r}_{\rm j}(\tau)$. From Eqs.~(\ref{eq:deltaI}) and (\ref{eq:FTdeltaI}) we obtain
\begin{eqnarray}
D(\mathbf{r},\tau) &=&\sum_{\rm j}^N \left[ A(\mathbf{r}-\mathbf{r}_{\rm j}^0)- A(\mathbf{r}- \mathbf{r}_{\rm j}(\tau)) \right]. \label{eq:Dr}\\
F_D(\mathbf{q},\tau)&=& \tilde{A}(\mathbf{q})  \sum_{\rm j}^N \left[ {\rm e}^{-{\rm i}\,\mathbf{q} \cdot \mathbf{r}_{\rm j}^0}\,-{\rm e}^{-{\rm i}\,\mathbf{q}\cdot \mathbf{r}_{\rm j}(\tau)} \right], \label{FDq1}
\end{eqnarray}
where $\tilde{A}(\mathbf{q})$ is the Fourier transform of $A(\mathbf{r})$ \footnote{Thus, $\tilde{A}(\mathbf{q}) =$ (form factor of the object) $\times$ (contrast function of the microscope).}, Fig.~\ref{fig:mathias}(b).

\begin{figure}
\begin{center}
\includegraphics[width=12cm]{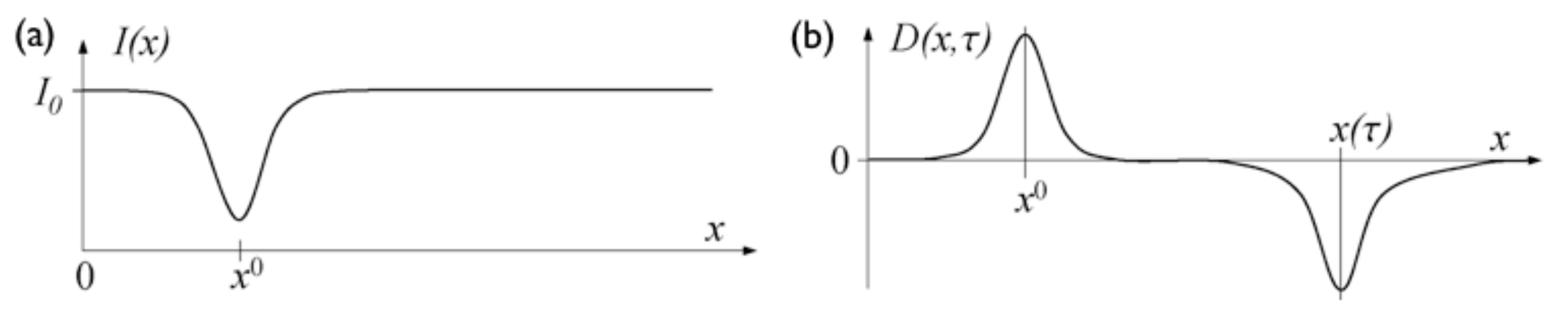}  
\end{center}  
\caption{Schematic of (a) the image of an object and (b) the difference image, Eq.~(\ref{eq:Dr}).} \label{fig:mathias}
\end{figure}

The ISF is the dynamic structure factor normalised by the static structure factor:
\begin{eqnarray}
f(\mathbf{q},\tau) &  = &\frac{S(\mathbf{q,\tau})}{S(\mathbf{q})} \label{eq:fqt}\\
S(\mathbf{q,\tau}) & = & \frac{1}{N} \left\langle\sum_{\rm j,k}^N  {\rm e}^{-{\rm i}\,\mathbf{q}\cdot \left[\mathbf{r}_{\rm j}(\tau)-\mathbf{r}_{\rm k}^0\right]}\right\rangle \label{eq:Sqt} \\
S(\mathbf{q}) &=& \frac{1}{N} \left\langle\sum_{\rm j,k}^N  {\rm e}^{-{\rm i}\,\mathbf{q}\cdot \left[\mathbf{r}_{\rm j}(\tau)-\mathbf{r}_{\rm k}(\tau)\right]}\right\rangle= \frac{1}{N} \left\langle\sum_{\rm j,k}^N {\rm e}^{-{\rm i}\,\mathbf{q}\cdot \left[\mathbf{r}_{\rm j}^0-\mathbf{r}_{\rm k}^0\right]}\right\rangle, \label{eq:Sq}
\end{eqnarray}
where $\langle \ldots \rangle$ denotes ensemble averaging. The power spectrum, Eq.~(\ref{eq:powerspec}) \footnote{ Note that in this equation, $\langle \ldots \rangle$ denotes averaging over sets of $\{\mathbf{r}_1^0,..,\mathbf{r}_N^0; \mathbf{r}_1(\tau),..,\mathbf{r}_N(\tau)\}$ with different initial times. However, in an ergodic system this is equal to an ensemble average.} is given by
\begin{eqnarray}
&\;&g(\mathbf{q},\tau) = \left|\tilde{A}(\mathbf{q})\right|^2 \times  \\
& \; & \left\langle\sum_{\rm j,k}^N  \left[{\rm e}^{-{\rm i}\,\mathbf{q}\cdot (\mathbf{r}_{\rm j}^0-\mathbf{r}_{\rm k}^0)} - {\rm e}^{-{\rm i}\,\mathbf{q}\cdot (\mathbf{r}_{\rm j}^0-\mathbf{r}_{\rm k}(\tau))} -{\rm e}^{-{\rm i}\,\mathbf{q}\cdot (\mathbf{r}_{\rm j}(\tau)-\mathbf{r}_{\rm k}^0)} + {\rm e}^{-{\rm i}\,\mathbf{q}\cdot (\mathbf{r}_{\rm j}(\tau)-\mathbf{r}_{\rm k}(\tau))}\right]\right\rangle \nonumber 
\end{eqnarray}
Comparison with Eqs.~(\ref{eq:fqt})-(\ref{eq:Sq}) gives
\begin{equation}
\label{g1}
g(\mathbf{q},\tau)= \left|  \tilde{A}(\mathbf{q}) \right|^2\, \left[ S(\mathbf{q})-S^*(\mathbf{q,\tau})-S(\mathbf{q,\tau})+S(\mathbf{q})  \right],
\end{equation}
where $^\ast$ denotes complex conjugation. Dividing through by $S(\mathbf{q})$, we obtain
\begin{equation}
\begin{aligned}\label{eq:DDM1}
g(\mathbf{q},\tau)=2 N \left|\tilde{A}(\mathbf{q})\right|^2  S(\mathbf{q}) \left[ 1-{\rm Re}\big( f(\mathbf{q,\tau}) \big) \right]+ B(q),
\end{aligned}
\end{equation}
where Re$(z)$ denotes taking the real part of the complex quantity $z$, and we have added $ B(q)$ to take into account of noise (presumed isotropic) that stems from the camera. This result is completely general. Simplifications can be made if the system is isotropic, in which case Re$(f(\mathbf{q},\tau)) = f(q,\tau)$, and dilute, so that $S(q) = 1$, whereupon \footnote{Throughout, for analytical convenience, we have equated positions in the image and sample, i.e. a magnification of unity. Non-unit magnification does not affect the results.}
\begin{eqnarray}
g(q,\tau) & = & A(q)\left[1 - f(q,\tau)\right] + B(q), \;\mbox{with} \label{eq:DDM2}\\
A(q) & = & 2N \left|\tilde{A}(\mathbf{q})\right|^2.
\end{eqnarray}
This result can also be derived by a more coarse-grained argument starting from the proportionality between the image intensity and the number density of particles \cite{WilsonDDM}. 

Equation~(\ref{eq:DDM2}) tells us that if we calculate $g(\mathbf{q},\tau)$ from an image sequence,  Eqs.~(\ref{eq:deltaI})-(\ref{eq:powerspec}), then we have access to the ISF. If a low magnification objective is used, then the image covers a large field of view, so that the lower limit of its Fourier transform reaches to $q \ll 1 \mu\mbox{m}^{-1}$. Quantitatively the pixel size of the detector and the magnification determine the inverse pixel size $k$ (in pixels/$\mu$m), which, together with the image size $L$ (in pixels) determines the $q$ range; in particular, $q_{\rm min} = 2\pi k/L$. Thus, for example, in \cite{WilsonDDM}, a $\times 10$ phase-contrast objective and an image size of $L = 500$ gave $k = 0.712$~pixels/$\mu$m, and $0.01 \mu\mbox{m}^{-1} \lesssim q \lesssim 2.2 \mu\mbox{m}^{-1}$. The lower reaches of this range are well away from the inverse cell body dimension \footnote{Note that we assume the long flagella are not resolved and do not contribute to our analysis.}. Thus, the ISF obtained from fitting $g(q,\tau)$ via Eq.~(\ref{eq:DDM2}) can be fitted using Eq.~(\ref{eq:ISFmixed}) to give $D$, $\alpha$ and $P(v)$. 

\begin{figure}
\begin{center}
\includegraphics[height=4cm]{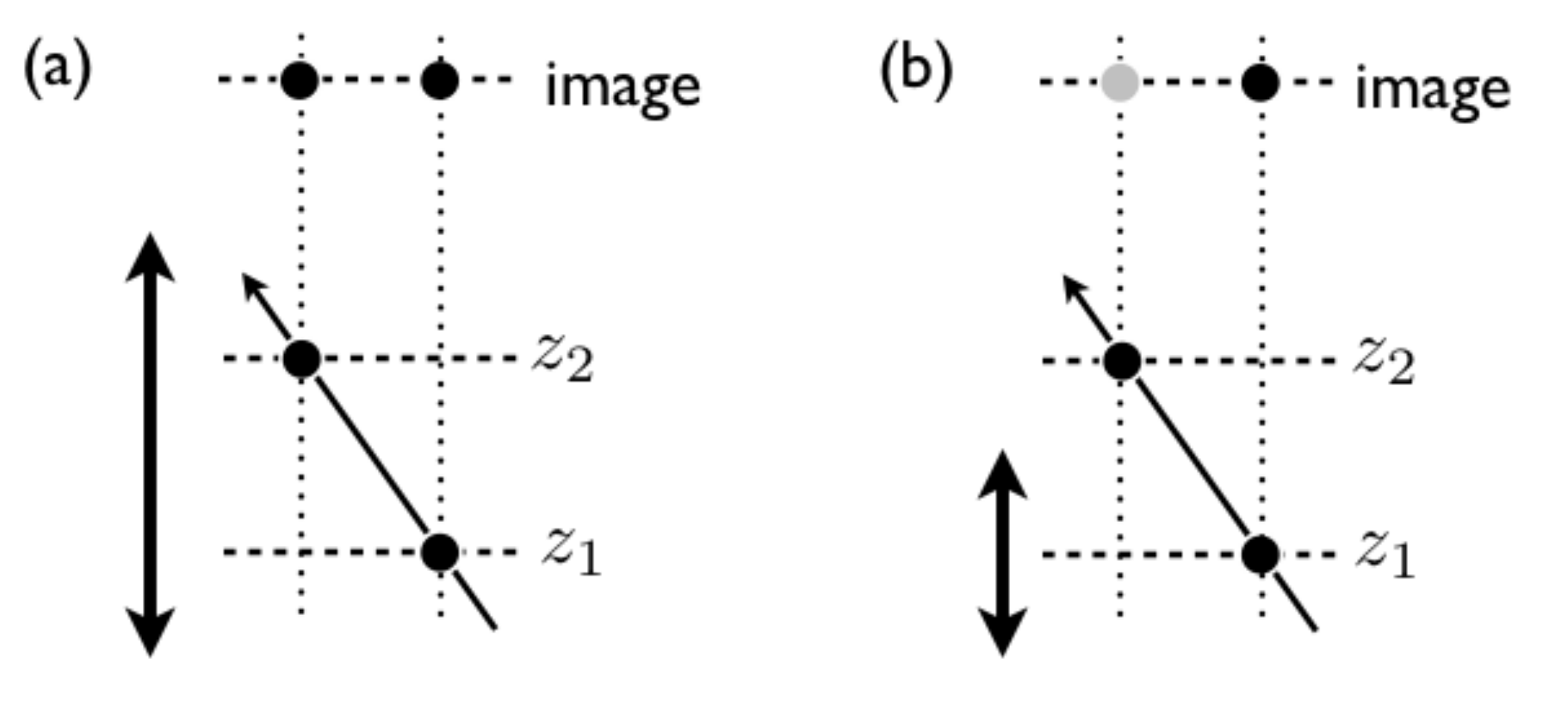}   
\end{center}  
\caption{Depth of field (bold double arrows) in DDM. A particle moves along the arrow from depth $z_1$ to depth $z_2$ in the sample. (a) For a large depth of field, its image at the two positions are of equal intensity, which is assumed in our derivation of Eq.~(\ref{eq:DDM2}), cf. Fig.~\ref{fig:mathias}. (b) For a small depth of field, the image will have little intensity when the particle is at $z_2$. } \label{fig:depth}
\end{figure}

The use of a low-magnification objective turns out to be important for another reason: it gives a relatively large depth of field (in \cite{WilsonDDM}, the $\times 10$ objective gave a $\simeq 40 \mu$m depth of field). This is important because the above analysis depends on the assumption that the intensity pattern in the image plane due to a particle is {\it not} dependent on its position along the optic axis ($z$), Fig.~\ref{fig:depth}, at least not during the time scale, ${\cal T}$, it takes $f(q,\tau)$ to decay (${\cal T} \simeq 10$s in Fig.~\ref{fig:ISFeg}). This requires a large enough depth of field (where `large' must be considered relative to the swimming speed being probed) so that particles do not disappear during ${\cal T}$ due to the $z$-component of their motion. 

Experimental details for using DDM to characterise populations of motile micro-organisms, such as choosing camera frame rate and data fitting, have been given recently \cite{MartinezDDM}. The ISF obtained for motile {\it E. coli} indeed show the two-stepped decay visible in the theoretical plot in Fig.~\ref{fig:ISFeg}. The method has been validated by comparison with tracking, and by analysing simulation data using the DDM algorithm \cite{WilsonDDM,MartinezDDM}. DDM has not yet been used to study synthetic active colloids. 

It should be remarked that real-space tracking and reciprocal-space methods (whether DLS or DDM) are complementary. Indeed, since reciprocal-space methods measure the ISF, Eq.~(\ref{eq:ISF}), a parametrised model for the particle motion, $\Delta\mathbf{r}(\tau)$, is needed if the measured ISF is to be fitted to yield motility information. Such a model, of which Eq.~(\ref{eq:ISFmixed}) for {\it E. coli} is an example, can only be obtained from tracking. A new system of active colloids must therefore first be tracked; but once a model is available from tracking, DDM (or, potentially, DLS) can be used to give high-throughput characterisation with good averaging. While a parameterised model used to fit the ISF necessarily reduces the amount of information obtainable compared to tracking, the utility of a high-throughput method such as DDM is illustrated by data such as those plotted in Fig.~\ref{fig:speedtime} (where each data point, which averages over $\simeq 10^4$ organisms, required only $\simeq 2$~minutes of observation). 

\section{The generic physics of motile particle suspensions}  \label{sec:generic}

I have now described in some detail two types of self-propelled colloids: natural ones, in the form of motile bacteria (specifically, {\it E. coli}), and synthetic ones in the form of particles with heterogeneous surface chemistry suspended in `fuel'. I have also explained how to characterise their speed distributions. The purpose of doing all of that, of course, is to enable well-defined experiments to study new physics in suspensions of such intrinsically non-equilibrium agents. Excellent reviews already exist in this area \cite{Sriram,CatesReview2012,Romanczuk}. Here I will give a overview with the primary purpose of suggesting to the reader a conceptual `map'. To this end, I will first describe a number of experimental observations that can be interpreted within a framework generalised from equilibrium statistical physics, using the concepts of an effective temperature or an effective interaction potential. Then I go on to explain why this approach must ultimately fail, because active suspensions do not have to satisfy the principle of detailed balance, and give example phenomena illustrating this point. Finally I review how the non-stokeslet hydrodynamics of individual swimmers will give rise to novel rheology and collective swimming behaviour. 

\subsection{Effective temperature and potential} A key experiment that  elucidated the physical nature of the colloidal state was Perrin's measurement of sedimentation equilibrium, Eq.~(\ref{eq:barometric}), in a dilute suspension of particles in the earth's gravitational field. Indeed, the gravitational height, Eq.~(\ref{eq:gravheight}), determines the size of the largest particles (radius $a_{\rm max}$) that can be counted as colloidal: $a_{\rm max} \lesssim z_0$. When suspensions become more concentrated, interparticle interactions become important. Ultimately, these interactions lead to phase transitions: hard spheres crystallise, and colloids with attractive interactions can phase separate to give the equivalent of gas (vapour), liquid and crystalline solid phases. The study of these colloidal phase transitions has formed a key part of colloid physics in the last three decades. In this section, I introduce active analogues of sedimentation equilibrium and phase transitions; current data suggest that the ideas of `effective temperature' and `effective potential' can help make sense of the phenomenology.  

\subsubsection{Active sedimentation equilibrium} \label{subsubsec:sed}
The sedimentation equilibrium in a passive colloidal suspension aries from flux balance: sedimentation at velocity $-v_s$ generates a flux of $J_s = -n(z)v_s$, while the resulting concentration gradient generates a diffusive flux of $J_D = -D_0\partial n(z)/\partial z$. In thermal equilibrium, the principle of detailed balance holds, so that there can be no net flux. Thus, $J_s + J_D = 0$, which solves to give Eq.~(\ref{eq:barometric}), with 
\begin{equation}
z_0 = \frac{D_0}{v_s}. \label{eq:gravheight2}
\end{equation}
Substituting $v_s = 2\Delta\rho g a^2/9\eta$ (where $\Delta \rho$ is the density difference between the particle and the solvent) and $D_0 = k_B T/6\pi\eta a$, we recover Eq.~(\ref{eq:gravheight}). 

Self-propelled particles, whether smooth swimming bacteria \cite{Condat,Loveley,BergTurnerDiff} or active Janus colloids \cite{Jones}, undergo random walks due to a combination of self-propulsion and rotational Brownian motion characterised by an effective diffusivity, $D_{\rm eff}$, Eq.~(\ref{eq:Deff}), that is many times their passive (or thermal) difusivity $D_0$. One may therefore expect that active colloids should show a barometric sedimentation equilibrium with $D_{\rm eff}$ taking the role of $D_0$ in Eq.~(\ref{eq:gravheight2}). Experimentally, this was indeed found to be the case in a dilute (i.e. non-interacting) suspension of self-propelled Janus particles \cite{Bocquet}, Fig.~\ref{fig:effective}(a). 

\begin{figure}
\begin{center}
\includegraphics[height=4.5cm]{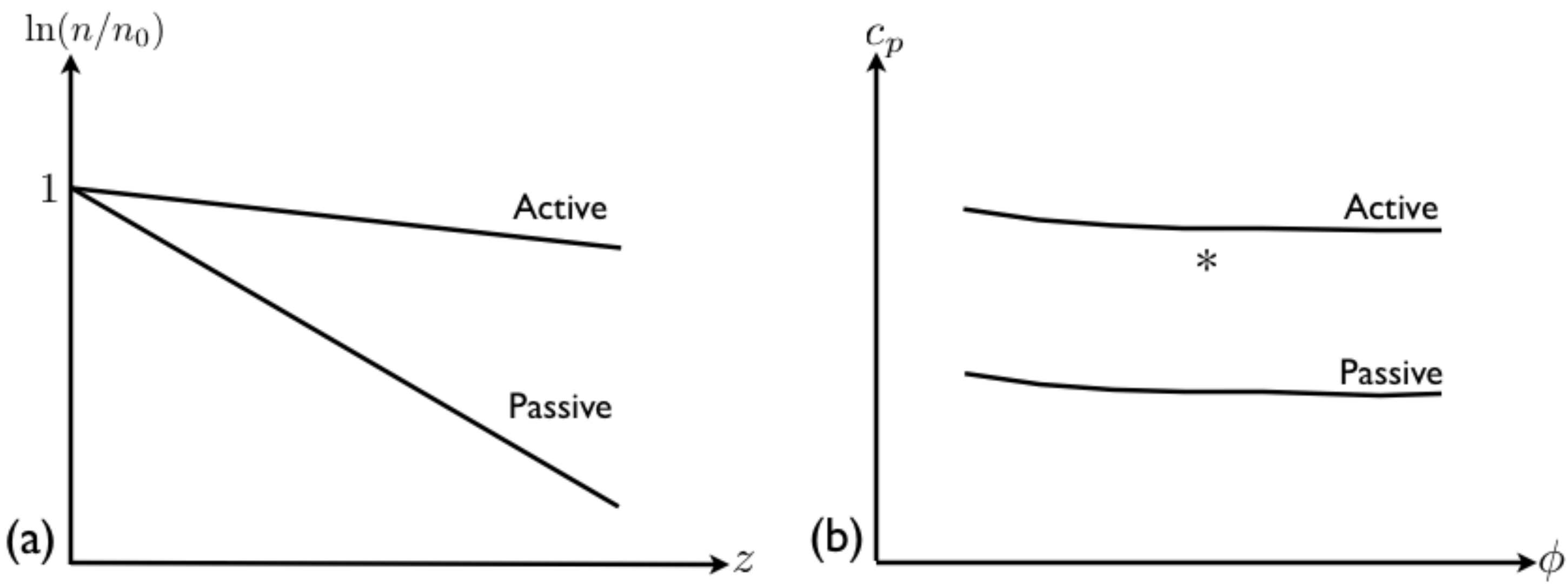}   
\end{center}  
\caption{Effective temperature and potential. (a) Schematic summary of the experimental observations of sedimentation equilibrium in suspensions of active Janus particles \cite{Bocquet}. Equation~(\ref{eq:barometric}) is satisfied in each case, but the active particles display an enhanced sedimentation height~$z_0$. (b)~Schematic summary of the phase boundaries in systems of passive and active {\it E. coli} (volume fraction $\phi$) driven to phase separation by the depletion attraction induced by non-adsorbing polymers (concentration $c_p$) \cite{Jana}. For each phase boundary, samples with compositions below the boundary remain single-phase, while those above the boundary would phase separate into coexisting colloidal gas (dilute) and liquid (concentrated) phases. Self-assembled `rotors' are formed just below the active phase boundary, e.g. at composition $\ast$.} \label{fig:effective}
\end{figure}

This observation suggests the definition of an `effective temperature', $T_{\rm eff}$, via a generalised Stokes-Einstein relation 
\begin{equation}
D_{\rm eff} = \frac{k_B T_{\rm eff}}{6\pi\eta a},
\end{equation}
so that $T_{\rm eff}/T = D_{\rm eff}/D_0$. In this language, a suspension of self-propelled particles that has reached steady state in a gravitational field mimics a passive colloid, only that the active suspension behaves as if it were a lot hotter as far as sedimentation equilibrium is concerned. The latter qualification is important: we shall see shortly that an `effective temperature' is only useful {\it in the context for which it is defined}. A relationship between the effective temperature for sedimentation equilibrium and the propulsion speed can now be found using Eq.~(\ref{eq:Deff3}):
\begin{equation}
\frac{D_{\rm eff}}{D_0} = \frac{T_{\rm eff}}{T} = 1 + \frac{v^2 \tau_r}{6D_0} = 1 + \frac{2}{9} \mbox{Pe}^2, \label{eq:TPe}
\end{equation}
where we have used the relationship between the rotational relaxation time and rotational diffusivity, $\tau_r = D_r^{-1} = 4a^2/3D_0$, and defined a P\'eclet number
\begin{equation}
\mbox{Pe} = \frac{a^2/D_0}{a/v} = \frac{va}{D_0},
\end{equation}
which compares the time taken for diffusion and self propulsion to cover distance $a$, so that $\mbox{Pe} \gg 1$ means that propulsion dominates. Equation~(\ref{eq:TPe}) has also been verified experimentally using Pt-coated Janus polystyrene particles dispersed in H$_2$O$_2$ \cite{Bocquet}, where the fastest particles reach $\mbox{Pe} \approx 5$, so that $T_{\rm eff} \approx 2000$K. 

Presumably, sedimentation equilibrium in suspensions of WT run-and-tumble {\it E. coli} can be treated within the same framework using Eq.~(\ref{eq:Deff2}) or (\ref{eq:Loveley}) (but see footnote~\ref{fn:Deffcaveat}), although this has not yet been verified experimentally. One of the practical difficulties in a putative experiment of this kind is the tendency for the bacterium to accumulate at surfaces over the time scales needed to establish a steady state.  

\subsubsection{Active aggregation and phase separation} \label{subsubsec:aggregate}
The success of the effective temperature approach in describing the sedimentation equilibrium of non-interacting active colloids prompts the question: would a similar approach work in describing the equivalent of phase transitions in an {\it interacting} system of active colloids? Experimental evidence from a suspension of {\it E. coli} bacteria driven to phase separation by the addition of non-adsorbing synthetic polymers \cite{Jana} suggests that the answer may be `yes'. In this system, the polymer induces an attraction between the particles via the `depletion' mechanism \cite{HenkBook,PoonReview}. The centre of a polymer molecule does not approach the surface of a particle closer than a distance given roughly by the former's radius of gyration, $r_g$, because doing so would lead to significant loss of configuration entropy for the polymer coil. The consequent total exclusion of polymers from the region between the surfaces of two nearby particles creates a net osmotic pressure pushing the particles together. The range and depth of this effective attraction between two particles are controlled by the size ($\approx 2r_g$) and the concentration \footnote{Strictly, the activity; but in a dilute suspension, activity $\approx$ concentration.} of the polymer, $c_p$, respectively. 

In a suspension of passive particles that are polydisperse and/or somewhat non-spherical, it is known that the addition of enough non-adsorbing polymer (or, equivalently, at a high enough depletion potential $U_{\rm dep}$) leads to phase separation into coexisting colloidal gas and liquid phases, with disordered arrangements of particles at different concentrations separated by a sharp interface \cite{HenkBook,PoonReview}. Adding non-adsorbing polymers to a suspension of non-motile {\it E. coli} bacteria gives rise to exactly this phenomenon \cite{JanaSoft}. The position of the phase boundary is consistent with the depletion mechanism. 

For motile bacteria, it was found that {\it more} polymer, i.e. a deeper $U_{\rm dep}$, was needed to cause phase separation, Fig.~\ref{fig:effective}(b). This observation can be explained by taking into account the fact that bacteria bound together by the depletion attraction can swim against this attraction. The {\it effective} depletion potential between two active particles, $U_{\rm dep}^{\rm eff}(r)$, is therefore shallower than the $U_{\rm dep}(r)$ for corresponding passive particles. Quantitatively, the effective potential in the active case can be obtained by shifting the passive depletion potential until force balance is satisfied, i.e.:
\begin{equation}
-\left.\frac{\partial U_{\rm dep}^{\rm eff}(r) }{\partial r}\right|_{\rm contact} = -\left.\frac{\partial U_{\rm dep}(r) }{\partial r}\right|_{\rm contact} + F_a(v),
\end{equation}
where $F_a(v)$ is the propulsive (or active) force pulling two bound particles apart when they are swimming at speed $v$. Using $v$ in the range of measured swimming speeds gives a credible account of the experiments, although a full theory needs to average over $P(v)$.  

Note that in this picture, activity reduces the interparticle attraction, but it is the thermodynamic temperature, $T$, that controls the fluctuations around the minimum of this effective potential, $U_{\rm dep}^{\rm eff}(r)$. However, it is known that the position of the gas-liquid phase boundary is controlled by the second virial coefficient \cite{Noro}
\begin{equation}
B_2 = 2\pi \int r^2 \left[ 1 - e^{-U(r)/k_B T}  \right]\;dr.
\end{equation}
In other words, the two phase boundaries in Fig.~\ref{fig:effective} should collapse into one (universal) curve if the vertical axis is $B_2$. Thus, it would be just as valid to explain the observations summarised schematically in this figure by leaving the depletion potential unchanged, and finding an effective temperature $T_{\rm eff} > T$ that leaves $B_2$ invariant. Under this perspective, the data reported in \cite{Jana} can be accounted for by choosing $T_{\rm eff} \approx 2T$. Note that the effective temperature for the diffusive (and therefore gravitational sedimentation) behaviour of the bacteria used in this work would be $T_{\rm eff} \approx 10T$, illustrating the crucial point that `effective' thermodynamic quantities are only valid and useful for the specific contexts in which they are defined. 

\begin{figure}
\begin{center}
\includegraphics[height=4.5cm]{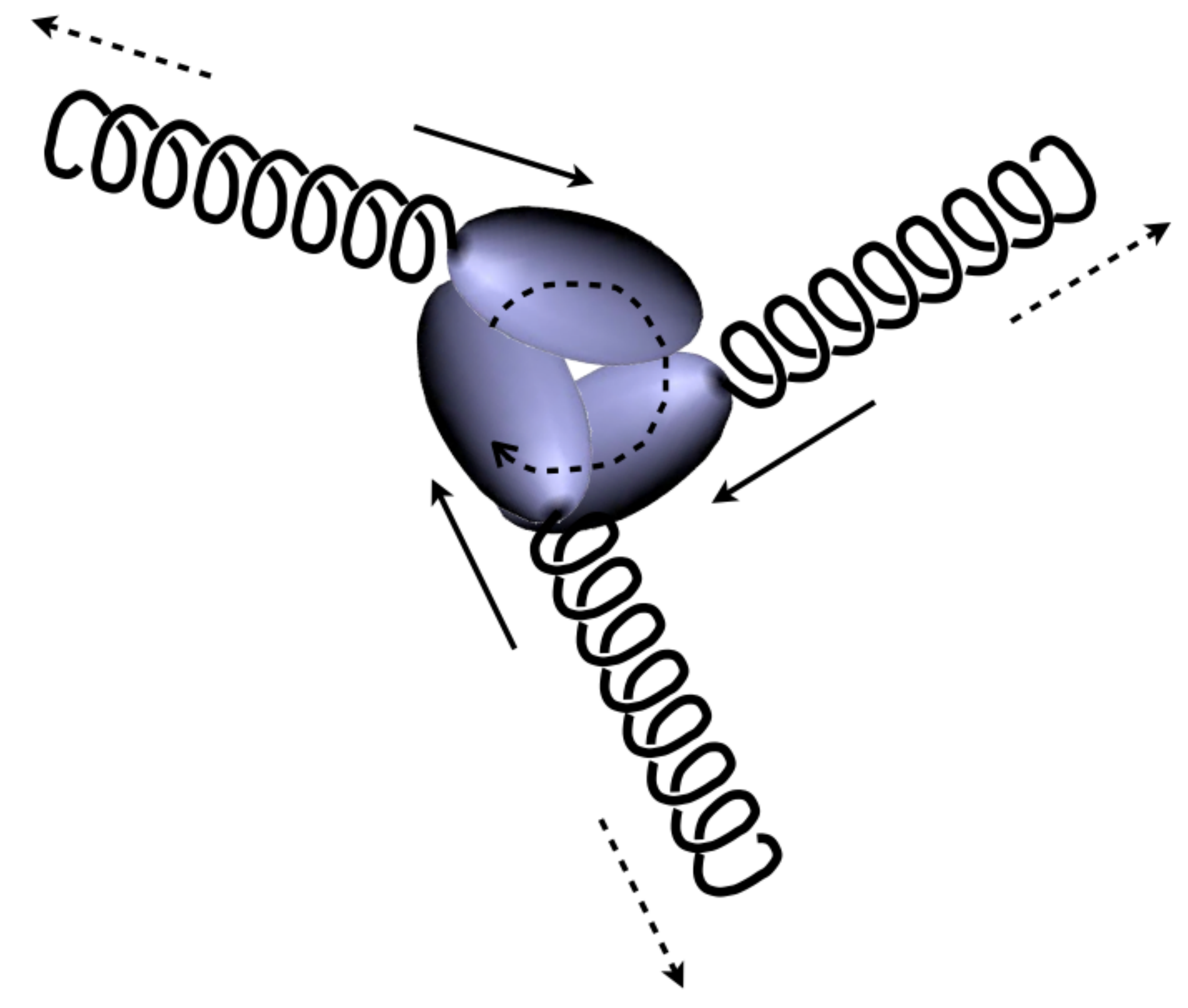}   
\end{center}  
\caption{Self-assembled rotors in bacteria with short-range attraction. The propulsion force of each bacterium (full arrow) exerts a net torque about the cluster centre. The sum of these torques in general is non-zero. If the structure of the cluster is fixed, then the sense of the rotation in the body frame of the cluster is constant. The model leading to Eq.~(\ref{eq:cluster}) neglects the flow that is generated by the force exerted by each flagellum on the liquid (dotted arrow), which exerts a drag on the cluster that to some extent cancels the push exerted by the propulsion force (full arrows). } \label{fig:rotate}
\end{figure}

An effective potential/temperature framework does not account for all observations in a mixture of motile bacteria and non-adsoring polymer. In both passive and active systems, pre-transition clusters of bacteria occur at polymer concentrations just below the phase boundary, Fig.~\ref{fig:effective}(b). In the active system, however, these clusters themselves were found to be  active: they translate and rotate. In particular, the rotation of any one cluster was uni-directional, even though the axis underwent thermal fluctuation. This can be explained by the non-zero net torque exerted on each cluster by its active constituent bacteria, Fig.~\ref{fig:rotate}. 
Quantitatively, suppose only the cells on the surface of a cluster (radius $R$) contribute to these torques \footnote{This could be due to steric restrictions to the flagellar movement of interior cells, and/or because in reality, there are non-motile cells present, which preferentially reside in the interior.}, and that the propulsion directions of these cells are randomly distributed. Each cell contributes a propulsion force $F_a$, and a torque $F_a R$. The number of surface cells $N_s \sim R^2$, but the sum of $N_s$ random vectors scales as $\sqrt{N_s}$, so that the total torque is $\Gamma \simeq N_s F_a R \sim R^2$. The drag torque and angular speed of  a cluster, $\Omega_c$, are related by $\Gamma = \xi \Omega_c$, where $\xi \sim R^3$. Putting all of this together, this model predicts that 
\begin{equation}
\Omega_c \sim R^{-1}, \label{eq:cluster}
\end{equation} 
which is indeed close to what is observed experimentally. The caption of Fig.~\ref{fig:rotate} explains why this derivation is not entirely  correct on a quantitative level, but the physical picture, which is rather general, remains valid. Thus, the assembly of such rotors should be a generic feature of the clustering active colloids. Thus, for example, the aggregation of synthetic active Janus particles \cite{BocquetCluster}, where the clustering is due to a totally different mechanism from depletion (perhaps chemotaxis), also gives rise to rotating clusters (Lyd\'eric Bocquet, private communication).

\subsection{No detailed balance} \label{subsec:detailed}

In my derivation of the sedimentation height, Eq.~(\ref{eq:gravheight2}), I appealed to the principle of detailed balance, which states that in an equilibrium thermodynamic system, transitions between any two states (`old', $o$, and `new', $n$) obey:
\begin{equation}
{\cal N}(o) \pi(o \rightarrow n) = {\cal N}(n)\pi(n \rightarrow o), \label{eq:detailed}
\end{equation}
where ${\cal N}$ denotes the density of states and $\pi$ denotes the transition probability between states. Two consequences follow from  Eq.~(\ref{eq:detailed}): the Boltzmann distribution, and the requirement that there is no net flux. Detailed balance no longer holds in non-equilibrium systems, so that an approach based on generalising equilibrium results and `effective' parameters must break down. I will now survey two examples of such break down in active colloids, introducing in the process a theoretical framework under which such phenomena can be treated in a unified way. Within the same framework, it is possible to show that there are circumstances under which one may expect that a mapping to an `effective' equilibrium system with detailed balance should work. My presentation follows \cite{CatesReview2012}, where more details and extensive references can be found.

As a first example, I introduce a model for sedimenting swimmers. A 1D model contains the essential physics \cite{TailleurEPL}. Independent particles travel left and right along the $x$-axis with velocities $v_L$ and $v_R$. The time-dependent probability densities of these two populations are $R(x,t)$ and  $L(x,t)$, and they `tumble' (i.e. make an attempt to change direction) with rates $\alpha_L$ and $\alpha_R$. The equations of motion for $R$ and $L$ are
\begin{eqnarray}
\dot{R} & = & -(v_R R)^\prime - \frac{\alpha_R}{2}R + \frac{\alpha_L}{2}L \label{eq:Tailleur1}\\ 
\dot{L} & = & (v_L L)^\prime + \frac{\alpha_R}{2}R - \frac{\alpha_L}{2}L, \label{eq:Tailleur2}
\end{eqnarray}
where $\dot{(\ldots)}$ and $(\ldots)^\prime$ denote $t$- and $x$-derivatives respectively \footnote{The factors of $\frac{1}{2}$ arise because only half of the `tumbles' lead to a change in direction.}. Sedimentation is modelled with $v_R = v + v_s$ and $v_L = v - v_s$, where $v$ is the propulsion velocity of each particle and $v_s$ is the sedimentation speed. Equations~(\ref{eq:Tailleur1}) and (\ref{eq:Tailleur2}) can be solved exactly. Without sedimentation, $v_s = 0$, and for a `symmetric tumbler', $\alpha_L = \alpha_R = \alpha$, the long-time motion is diffusive, with $D_{\rm eff} = v^2/\alpha$ (cf. Eqs.~(\ref{eq:Deff}) and (\ref{eq:Deff3})). With $v_s \neq 0$, the steady-state probability distribution of all particles, $P(x) = R(x) + L(x)$, is given by
\begin{eqnarray}
P(x) & = & P(0) e^{-x/\lambda}, \;\mbox{with} \label{eq:Tailleur3}\\
\lambda & = & \frac{v^2 - v_s^2}{\alpha v_s}.\label{eq:Tailleur4}  \label{eq:lambda}
\end{eqnarray}
Note that even though an exponential distribution still obtains, Eq.~(\ref{eq:Tailleur3}), the gravitational height $\lambda \neq D_{\rm eff}/v_s$, i.e. an approach based on generalising the equilibrium thermodynamic framework breaks down. The latter approach is seen to be an approximation in the limit $v \gg v_s$. As $v \rightarrow v_s^{(+)}$, the system approaches `gravitational collapse': the particles simply do not swim fast enough to overcome gravity, and should all sediment out, $\lambda \rightarrow 0$ \footnote{The inclusion of thermal (Brownian) motion in a full model will `soften' this singularly and give a finite gravitational height of the normal colloidal kind.}. Working in 3D complicates the analysis but does not change the qualitative conclusion. The presence of a speed distribution similarly does not alter the prediction of `gravitational collapse' -- in any real active colloid, there will undoubtedly be a maximum propulsion speed; this contrasts with the Maxwell-Boltzmann distribution of speeds in, say, a gas, where (in classical mechanics) arbitrarily high speeds are possible. Finally, to map the model to real systems, the interpretation of $\alpha$ in {\it E. coli} is straightforward, while for synthetic self-propelled colloids, one may expect $\alpha \sim D_r$ (see Eq.~(\ref{eq:Deff3})). 

The `gravitational collapse' predicted by Eq.~(\ref{eq:Tailleur4}) is unlikely to be seen in suspensions of bacteria or self-propelled polystyrene Janus particles in the earth's gravitational field, since in both cases $v/v_s \gg 1$ -- hence the success of a quasi-equilibrium description for the sedimentation equilibrium of Janus swimmers \cite{Bocquet}. However, this limit can probably be studied in suspensions of bimetallic nano-rods \cite{PaxtonRod}. We may estimate the sedimentation speed of a 370nm diameter, $1 \mu$m long gold rod along its long axis by approximating it as an ellipsoid and using Eq.~(\ref{eq:A0ellipse}), which gives $v_s \approx 4 \mu\mbox{ms}^{-1}$. The range of propulsion speeds achieved when these rods are half coated with Pt is $4 \lesssim v \lesssim 8 \mu\mbox{ms}^{-1}$, depending on H$_2$O$_2$ concentration \cite{PaxtonRod}, so that $1 \lesssim v/v_s \lesssim 2$.

\begin{figure}
\begin{center}
\includegraphics[height=3cm]{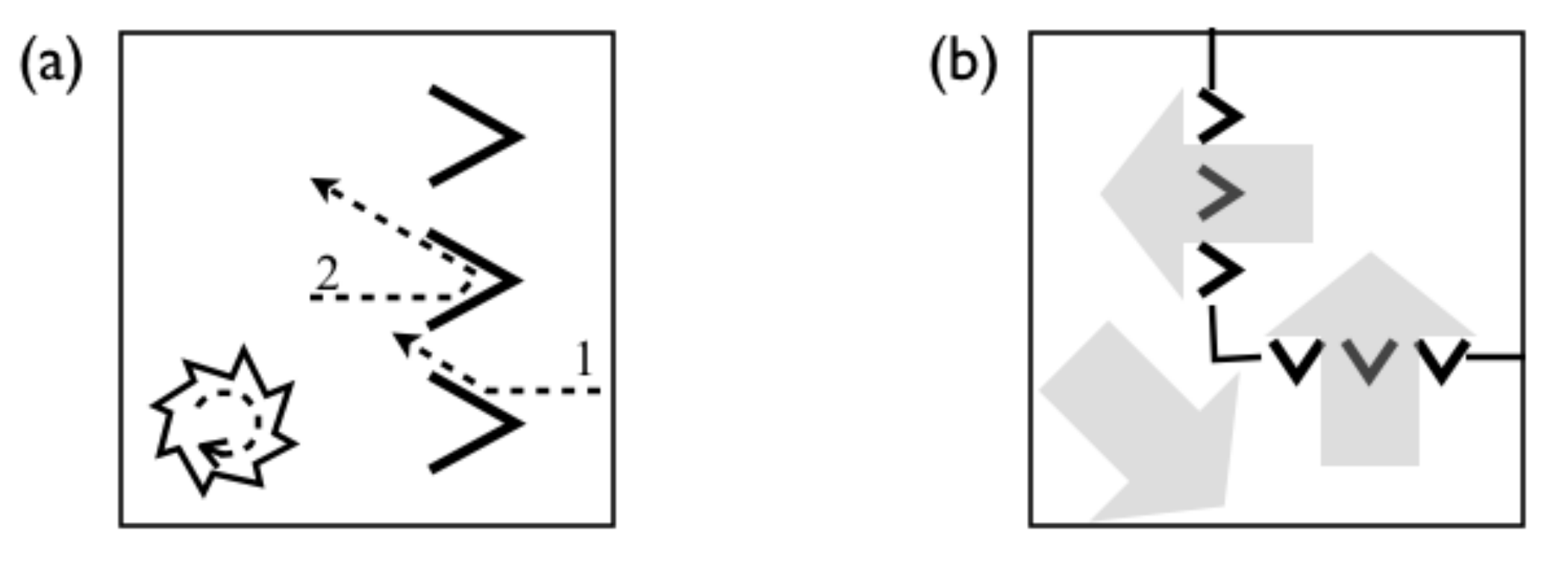}   
\end{center}  
\caption{Rectification. A WT {\it E. coli} bacterium swimming to a wall has a tendency to `hug' it until its next tumble. (a) This `wall hugging' tendency leads to `rectification' at an array of funnel gates: cell 1 probably goes through, but cell 2 is likely turned back. The array of funnel gates can therefore be modelled as a spatial region favouring left tumbles. Inset: Wall hugging means that such a gear wheel will spin as shown in a bath of swimming bacteria. (b) This array of funnel gates should give rise to a steady state flux shown by the arrows. Redrawn after \cite{CatesReview2012}. }\label{fig:rectify}
\end{figure}

The second example of the breakdown of a quasi-equilibrium framework is `rectification', where the non-applicability of the no-flux constraint becomes manifest. Partly for hydrodynamic reasons, a swimming WT {\it E. coli} bacterium encountering a wall tends to swim along it until its next tumble. This `wall hugging' tendency has been used to `rectify' bacterial motion experimentally using micro-fabricated `funnel gates' \cite{Galajda}; the essential physics is self explanatory from Fig.~\ref{fig:rectify}(a), leading to a spontaneous concentration of bacteria in the right-hand chamber in the case shown starting from a uniform density of cells. The same physics also means that a gear wheel, Fig.~\ref{fig:rectify}(a) inset, in a bath of swimming {\it E. coli} will spin clockwise, as observed in experiments \cite{Gear} \footnote{Note the the rotation generated here has quite a different origin to the self-assembled rotors observed as self-assembled pre-transition clusters \cite{Jana} already mentioned in Section~\ref{subsubsec:aggregate}.}.

Such rectification can be modelled within the same framework already introduced to model sedimentation, Eqs.~(\ref{eq:Tailleur1}) and (\ref{eq:Tailleur2}) \cite{TailleurEPL}. Observe first that in the situation shown in Fig.~\ref{fig:rectify}(a), the array of funnels can be treated as a region of space where left tumbles are favoured. This  suggests the use of a space-dependent tumble rate. Specifically, outside a small region on the axis, $\alpha_L = \alpha_R$, while inside a small region of width $2\epsilon$, $\alpha_L = \alpha_R + 2\alpha_c$. This model indeed predicts rectification, with the steady-state ratio of particle densities to the left and right of the $2\epsilon$-wide region given by $\exp[2 \epsilon \alpha_c/v]$. 

It can be shown quite generally \cite{ProstRectify} that rectification is only possible if spatial asymmetry is accompanied by time-asymmetric trajectories, the latter clearly implying the violation of detailed balance. Here, the funnel gates provide the spatial asymmetry, while the wall-hugging leads to trajectories that are time-asymmetric. This violation of detailed balance occurs on a mesoscopic (= singe cell) level in the experiment shown schematically in Fig.~\ref{fig:rectify}(a). It has been suggested \cite{CatesReview2012} that macroscopic manifestation of detailed-balance violation can be achieve using the arrangement shown in Fig.~\ref{fig:rectify}(b), where a macroscopic steady-state flux should occur even for a low concentration of cells. This or similar experiments have not been attempted to date, but will provide a clear visual demonstrations of the non-equilibrium nature of active suspensions. 

I have now reviewed two examples in which a mapping to an `effective' equilibrium system with detailed balance fails. In the sedimentation of active colloids, an exponential steady-state density profile still holds, Eq.~(\ref{eq:Tailleur3}), but the gravitational height is no longer given by the ratio of the system's effective diffusivity to the sedimentation speed, Eq.(\ref{eq:lambda}), unless the latter is much smaller than the propulsion speed. In a funnel gate system, the wall-hugging tendency of bacteria gives rise to time-asymmetric trajectories and therefore rectification, Fig.~\ref{fig:rectify}(a), and hence the possibility of macroscopic fluxes, Fig.~\ref{fig:rectify}(b). The question naturally arises: can anything general be said about the circumstances under which the mapping of an active colloid to an `effective' system obeying detailed balance may work? The framework of the model formalised in Eqs.~(\ref{eq:Tailleur1}) and (\ref{eq:Tailleur2}) can be generalised to answer this question.  This is best introduced by way of reviewing a third prediction from this model: a new mechanism for aggregation in active colloids. 

The physical basis of this new mechanism is rather intuitive. First, it can be shown that the symmetric version of Eqs.~(\ref{eq:Tailleur1}) and (\ref{eq:Tailleur2}), with symmetric but space-dependent velocities and tumble rates $v_L = v_R = v(x)$ and $\alpha_L = \alpha_R = \alpha(x)$, predicts the following steady-state density distribution of swimmers $P(x) = L(x) + R(x)$:
\begin{equation}
P(x) = P(0)\frac{v(0)}{v(x)}. \label{eq:steady}
\end{equation}
This is an unusual result from the point of view of equilibrium statistical physics, since the steady-state distribution is velocity dependent: unconstrained by detailed balance, run-and-tumble swimmers can accumulate where they swim slowest. Now suppose that the velocity is also dependent on the density of swimmers, $\rho$ \footnote{Note that the definition of `density' here involves considerable subtlety; see \cite{CatesReview2012,TailleurPRL}.}. If $v(\rho)$ is a decreasing function, then the swimmers slow down whenever they encounter a region of high particle density, which leads to positive feedback with the effect evidenced by Eq.~(\ref{eq:steady}). 

A key finding is this: if $v(\rho)$ and $\alpha(\rho)$ are {\it left-right symmetric} (i.e. the same for swimmers travelling in either direction), and depend only on the {\it local} density (i.e. they are not functionals of the density profile, e.g. $v \neq v[\rho(x)]$), then a mapping to an equilibrium system obeying detailed balance is possible, in this case, to a phase separating fluid described by a free energy density. This mapping then predicts a kind of `phase separation' in the active particle system, with coexisting regions of high-density low-motility and low-density high-motility particles. Interestingly, extension of this approach to 2D generates spatial patterns that are reminiscent of those observed in bacteria believed to have been driven to aggregation by chemotaxis, although chemotaxis does not feature in this theoretical model. Experimental confirmation of the possibility of `chemotactic patterns without chemotaxis' is not yet available.

\subsection{Non-stokeslet hydrodynamics} \label{subsec:nonstokes}

The flow field around a self-propelled colloid is very different from that around a particle dragged by an external force. The latter in the far field maps onto that of a force monopole, i.e. a stokeslet, which decays as $r^{-1}$. The far-field flow around a bacterium like {\it E. coli} is that generated by a force dipole, Section~\ref{subsubsec:propulsion}, and decays as $r^{-2}$. The far-field flow generated by phoresis decays even faster, as $r^{-3}$ \cite{AndersonFlow}. These flow fields give rise to unusual hydrodynamic interactions (HIs), which are a second main source of generic novelty in active colloids. The literature on HIs and the resulting collective effects in suspensions of swimming micro-organisms is large (for reviews, see \cite{Sriram,CatesReview2012,LaugaRev}). Here I can only introduce the basic physics and a conceptual framework for thinking about the subject. Much less is known about HI in suspensions of synthetic swimmers. Indeed, the subject can be expected to be intricate. While bacteria can swim using internal resources, all synthetic swimmers to date except one \cite{HerminghausActive} relies on external `fuel' in the suspending medium, so that HIs in suspensions of synthetic swimmers will couple to the reaction-diffusion of `fuel' in a highly non-trivial manner. Thus, collective behaviour driven by HIs in synthetic swimmers and how these differ from those in suspensions of micro-organisms should be a fruitful area for study.  

\begin{figure}
\begin{center}
\includegraphics[height=3cm]{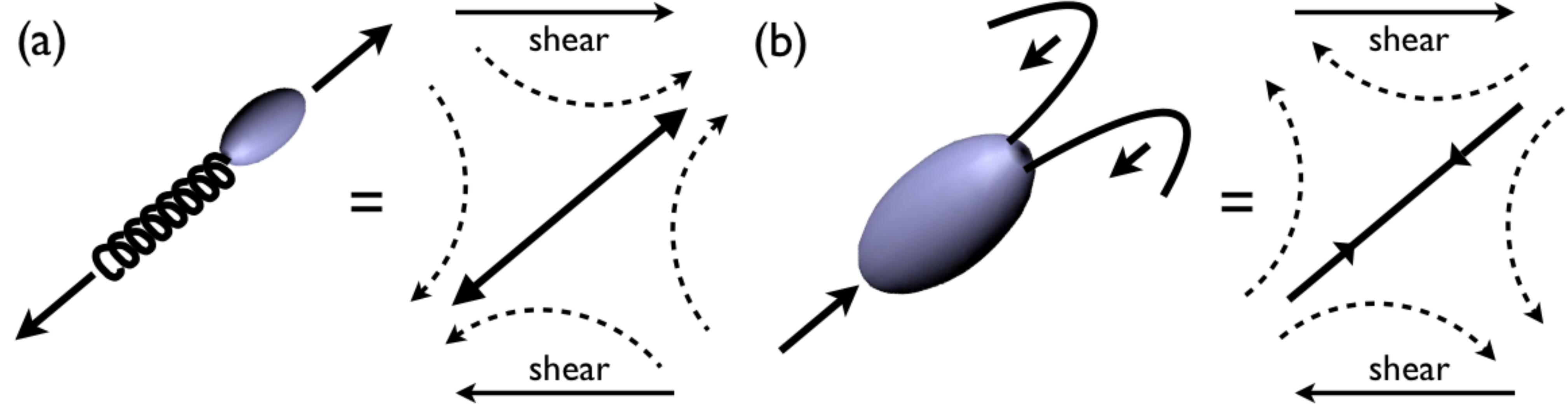}   
\end{center}  
\caption{Pushers and pullers, redrawn after \cite{Sriram}. (a) The forces exerted by a self-propelled {\it E. coli} bacterium is equivalent to an extensile dipole in the far field. (b) The forces exerted by a self-propelled {\it Chlamydomonas} is equivalent to a contractile dipole in the far field. In each case, the symmetry of the follow field is indicated by dashed arrows. Horizontal arrow pairs in each case represents an imposed shear, which aligns the swimmer in the direction shown. } \label{fig:pusherpuller}
\end{figure}

My analysis of forces acting on a motile {\it E. coli} bacterium, Fig.~\ref{fig:coliforces}, led to the conclusion that the bacterium exerted equal and opposite forces on the surrounding liquid; thus, the far-field flow must have dipolar symmetry. This kind of force dipole, whose flow field is shown schematically in Fig.~\ref{fig:pusherpuller}(a), is known as a `pusher' -- the cell body is pushed forward by the rotating flagellum. (The symmetry of the flow field is known as `extensile'.) I have also briefly introduce the swimming algae, {\it Chlamydomonas}. It is also a force dipole, but the symmetry of this dipole is fundamentally different from that of {\it E. coli}: the two beating flagella in {\it Chlamydomans} pulls the cell body forward, so that it is a `puller', Fig.~\ref{fig:pusherpuller}(b). (The symmetry of the flow field is known as `contractile'.) Interestingly, the bacterium {\it Caulobacter crescentus}, which has a single right-handed flagellum that can rotate in either direction, can change from being a pusher to being a puller \cite{LaugaRev}.  

\begin{figure}[t]
\begin{center}
\includegraphics[width=8cm]{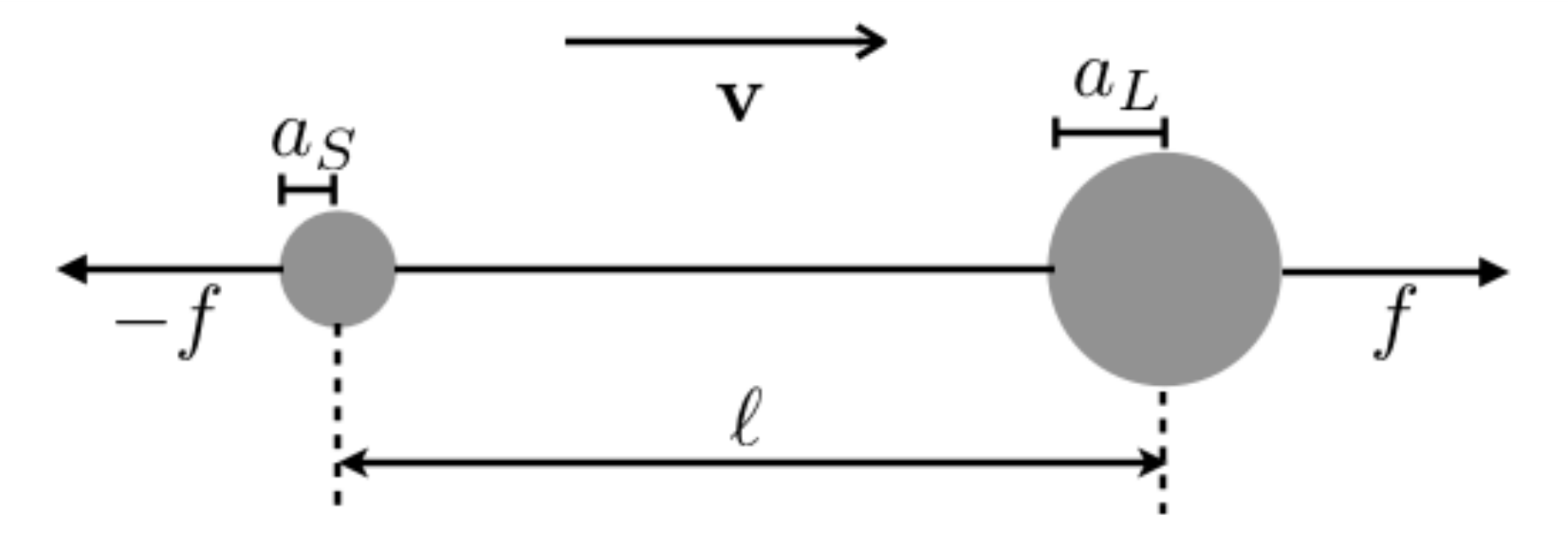}   
\end{center}  
\caption{An asymmetric dumbbell swimmer \cite{MarchettiDumbbell} consisting of large and small spheres joined by a zero-thickness rigid rod. The spheres exert equal and opposite forces, $\pm f$, on the surrounding liquid; $f > 0$ gives a pusher, $f < 0$ gives a puller, and a `shaker' results if $a_L = a_S$.} \label{fig:dumbbell}
\end{figure}

Whether a swimmer is of the pusher or puller type has profound consequences for all matters hydrodynamic. Before discussing some of these consequences, I therefore first introduce a simple model, the asymmetric dumbbell \cite{MarchettiDumbbell}, which is stripped of all system-specific details (flagella, etc.) to reveal the essential physics; this and other simple models are also useful for simulational purposes. The asymmetric dumbbell swimmer, Fig.~\ref{fig:dumbbell}, consists of large and small hard spheres (radii $a_L$ and $a_S$) joined by an infinitely thin rigid rod aligned along the $x$-axis separating their centres by length $\ell$; the spheres exert equal and opposite forces $\mathbf{f} = (\pm f, 0, 0)$ on the liquid. To find the self-propulsion velocity $\mathbf{v} = (v, 0, 0)$ of this model, we calculate the drag force on each sphere and impose the condition that the whole dumbbell must be force free. The flow pass each sphere is the sum of the self-propulsion velocity $\mathbf{v}$ and the velocity field induced at its position by the motion of the other sphere. The latter we find using the Oseen tensor, Eq.~(\ref{eq:oseen}) with $(x, y, z) = (\ell, 0, 0)$ and $r = x = \ell$, to be $(\pm f/4\pi\eta\ell, 0, 0)$, where the `+'/`$-$' signs are for the flow round the large/small sphere respectively. The force-free condition then reads:
\begin{equation}
6\pi\eta a_L\left(v + \frac{f}{4\pi\eta \ell}  \right) + 6\pi\eta a_S \left(v - \frac{f}{4\pi\eta \ell}  \right) = 0,
\end{equation}
which solves trivially to give
\begin{equation}
v = - \frac{f \Delta a}{8\pi\eta \ell \bar{a}},
\end{equation}
where $\Delta a = a_L - a_S$ and $\bar{a} = (a_L + a_S)/2$. In this model, $f >0$ gives a pusher: the dumbbell moves with the large sphere (the `payload') at the front, i.e. which is being pushed by the small sphere. On the other hand, $f < 0$ corresponds to a `puller' -- the big sphere lags, and is being pulled by the small sphere. Note that propulsion depends on asymmetry: $v\neq 0$ if and only if $\Delta a \neq 0$. The case of $\Delta a = 0$, however, is not uninteresting; it represents a class of active objects known as `shakers'. A shaker is not self-propelled, but nevertheless actively sets up a flow field around itself. 

I now turn to introduce some of the contrasting hydrodynamic behaviour of pushers and pullers. Consider first rheology. The viscosity of a suspension of passive particles at volume fraction $\phi$ is given by a power series expansion 
\begin{equation}
\eta_{\rm passive} = \eta_0\left(1 + c_1\phi + c_2\phi^2 + \ldots\right), \label{eq:viscosity}
\end{equation}
where $\eta_0$ is the viscosity of the solvent, and $c_1 = 5/2$ for hard spheres \cite{EinsteinBM,RusselBook,Wagner,Guyon}. Theory predicts that a suspension of pushers at the same $\phi$ has a viscosity {\it below} that given by Eq.~(\ref{eq:viscosity}), while the viscosity of a suspension of pullers is enhanced above $\eta_{\rm passive}$ \cite{SriramViscosity}. The qualitative reason for this is easy to understand: reference to the schematics in Fig.~\ref{fig:pusherpuller} shows that shear will seek to align a swimmer, irrespective of whether it is a pusher or a puller. Once aligned, however, a pusher enhances the aligning flow, while a puller works against the aligning flow, giving reduced and enhanced stress to shear rate ratios (i.e. viscosities) respectively. 

To date, there is a single set of experimental data supporting each of these predictions, although the data interpretation in either case is not yet wholly unequivocal. Conventional rheology was used to measure the viscosity of suspensions of motile {\it Chlamydomonas} (a puller) at a range of $\phi$, which was found to be higher than that of corresponding suspensions of dead cells \cite{Jibuti}. Experiments on dispersions of the flagellated bacterium {\it Bacillus subtilise} (a pusher) returned viscosities that are even below the value of water (i.e. $\eta < \eta_0$) \cite{AransonViscosity}.  

\begin{figure}
\begin{center}
\includegraphics[width=10cm]{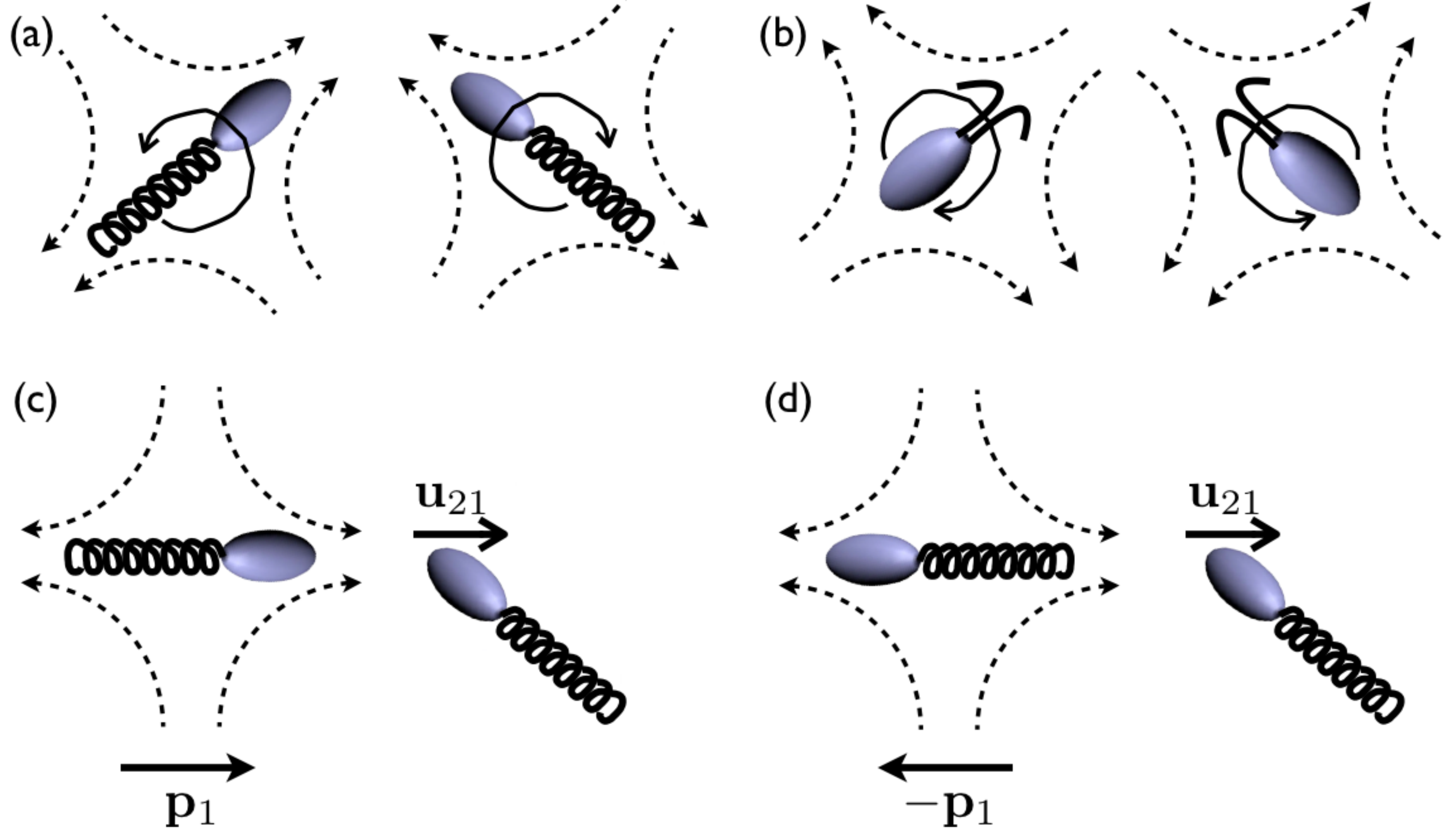}   
\end{center}  
\caption{Hydrodynamic interaction between two swimmers. (a) The flow fields of two convergent pushers orient them to swim parallel. (b) The flow fields of two convergent pullers orient them to approach head on. (c) Swimmer 1 with velocity $v\mathbf{p}_1$ sets up a flow of velocity $\mathbf{u}_{21}$ at the position of swimmer 2. (d) As in (c), only with the velocity of swimmer 1 reversed; the fore-aft symmetry of a dipole field leaves $\mathbf{u}_{21}$ is unchanged. Partly redrawn after \cite{LaugaRev}} \label{fig:reorient}
\end{figure}

Next I turn to the HI between two swimmers. The details are complex and the subject of ongoing research. However, certain gross features are intuitively understandable from simple arguments. One of the simplest examples is how two nearby swimmers reorient each other's trajectory. Figure~\ref{fig:reorient}(a) shows two convergent pushers; it is clear that their respective flow fields will seek to align each other to give parallel trajectories. On the other hand, Fig.~~\ref{fig:reorient}(b) shows two convergent pullers. The symmetry of the flow fields now tells us that they will reorient to approach each other head on. 

More quantitatively, consider the velocity pair correlation function between swimmers in a population, which is measured by the tensor
\begin{equation}
\uuline{\mbox{C}}(\mathbf{r}) = \frac{3\langle \mathbf{u}_1(\mathbf{r}_1)\mathbf{u}_2(\mathbf{r}_2) \rangle}{\langle u_1^2 \rangle} ,
\end{equation}
where $\mathbf{r} = \mathbf{r}_1 - \mathbf{r}_2$ is the vector separation between the two swimmers and the averaging is over all pairs of swimmers and time. This has been measured in a dilute suspension of {\it E. coli} using elegant image analysis \cite{LiaoPair}. In a dilute suspension, treating pair interactions suffice. In a pair of swimmers, the velocity of swimmer $1$ at position $\mathbf{r}_1$, $\mathbf{u}_1(\mathbf{r}_1)$, is the sum of its own propulsion speed $v$ (assumed the same for all swimmers) in direction $\mathbf{p}_1$ (this being a unit vector) and the convective flow produced at its position by swimmer $2$, which we write as
\begin{equation}
\mathbf{u}_1 = v\mathbf{p}_1 + \mathbf{u}_{12}, \label{eq:u12}
\end{equation}
where $\mathbf{u}_{12}$ will be a function of where swimmer $2$ is relative to swimmer 1, and the orientation of swimmer 2, $\mathbf{p}_2$, i.e. $\mathbf{u}_{12} = \mathbf{u}_{12}(\mathbf{r}, \mathbf{p}_2)$. Assuming that $u_{ij} \ll v$, we approximate the variance of swimmer speeds as $v^2$, so that
\begin{equation}
\uuline{C}(\mathbf{r}) = \frac{3}{v^2} \int \mathbf{u}_1(\mathbf{0}) \mathbf{u}_2(\mathbf{r}) P(\mathbf{p}_1,\mathbf{p}_2|\mathbf{r})\;d\mathbf{p}_1 d\mathbf{p}_2, \label{eq:correlation}
\end{equation}
where $P(\mathbf{p}_1,\mathbf{p}_2|\mathbf{r})$ is the probability distribution function of the orientation of two swimmers at relative position $\mathbf{r}$. If we neglect the reorientation of a swimmer due to the velocity of its neighbour, then this function is just the product of two isotropic distributions, i.e. $P(\mathbf{p}_1,\mathbf{p}_2|\mathbf{r}) = 1/(4\pi)^2$. Then, substituting Eq.~(\ref{eq:u12}) into Eq.~(\ref{eq:correlation}) and neglecting a term of order $u_{ij}^2$, we find (noting that $\int \mathbf{p}_i d\mathbf{p}_i = 0$)
\begin{eqnarray}
\uuline{\mbox{C}}(\mathbf{r}) & = & \frac{3}{(4\pi)^2 v^2} \int [\mathbf{p}_1 v\mathbf{u}_{21} + \mathbf{p}_2 v\mathbf{u}_{12}]\;d\mathbf{p}_1d\mathbf{p}_2 \nonumber \\
& = & \frac{3}{2\pi v}\int  \mathbf{p}_1 \mathbf{u}_{21}\;d\mathbf{p}_1. \label{eq:correlation2}
\end{eqnarray}
In other words, for a pair of swimmers 1 and 2, $\uuline{\mbox{C}}$ measures the average flow $\mathbf{u}_{21}$ generated at the position of swimmer 2 due to the swimming of its neighbour, swimmer 1, averaged over all orientations of swimmer 1. Two configurations in this averaging are shown in Fig.~\ref{fig:reorient}(c) and (d), with swimmer 1 in opposite orientations $\pm \mathbf{p}_1$. Due to the `fore-aft' symmetry of a dipolar flow field, $\mathbf{u}_{21}$ remains invariant when $\mathbf{p}_1$ is reversed, so that pairs of $\mathbf{p}_1 \mathbf{u}_{21}$ terms cancel in Eq.~(\ref{eq:correlation2}), giving $\uuline{\mbox{C}}(\mathbf{r})=0$. 

Interestingly, this contradicts experimental observations with {\it E. coli} \cite{LiaoPair}, which show finite components of the correlation tensor $\uuline{\mbox{C}}(\mathbf{r})$ at inter-cell distances $|\mathbf{r}| \lesssim 20 \mu$m. The resolution of this apparent contradiction lies in the realisation that the flow field around real pushers or pullers are only dipolar in the far field, where all higher-order contributions have decayed. However, the coarse-grained picture of a force dipole (giving $v \sim r^{-2}$) does not give an adequate account of the near-field flow, for which higher-order multipole terms must be taken into account. In fact, the observed pair correlation in {\it E. coli} can be account for by a quadrupolar field ($v \sim r^{-3}$) \cite{LiaoPair}.

\begin{figure}
\begin{center}
\includegraphics[width=10cm]{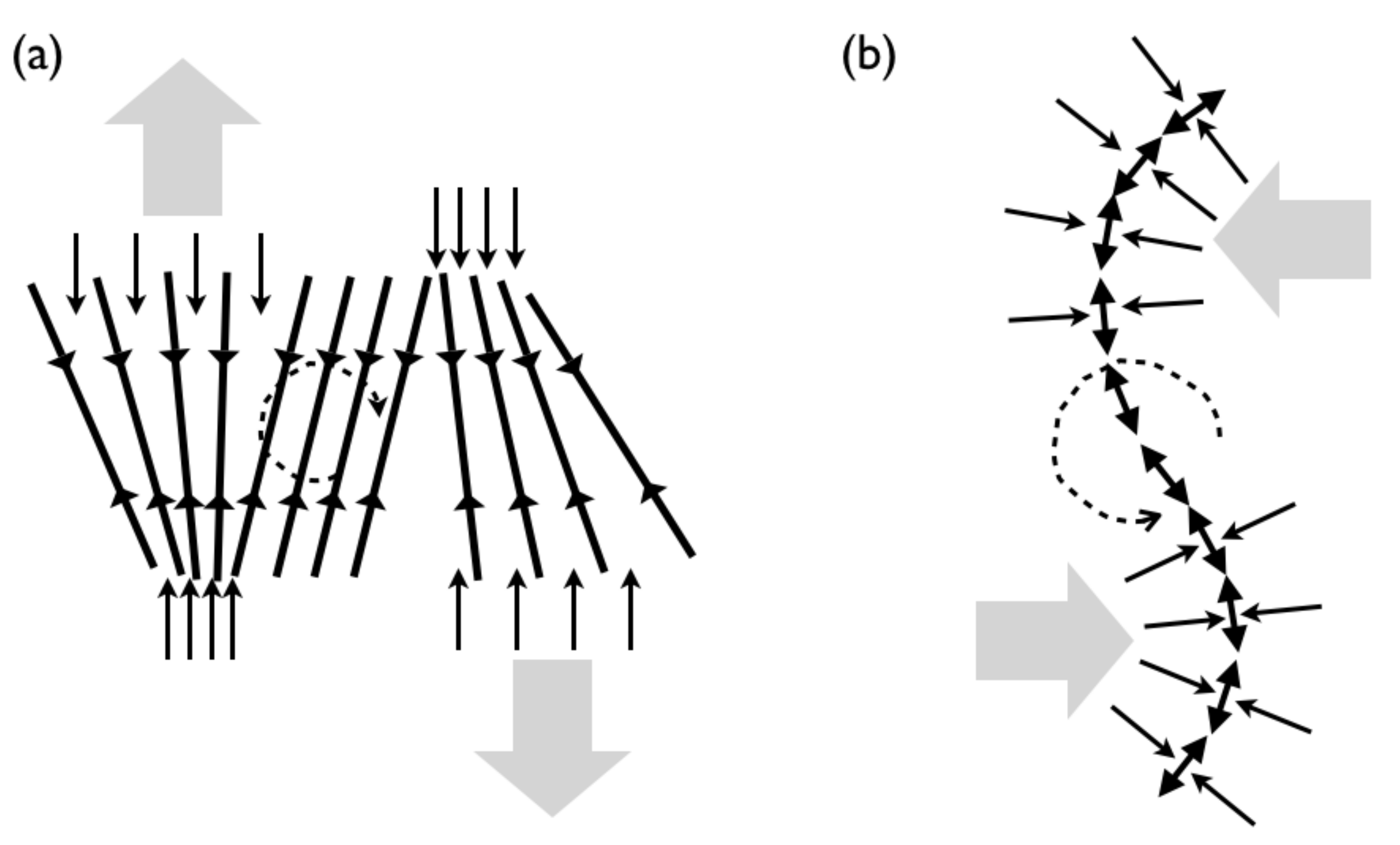}   
\end{center}  
\caption{Hydrodynamic instability in concentrated suspensions of swimmers: see text for a detailed description. (a) A puller pulls fluid in at both ends. A long-wavelength `splay' type perturbation is amplified in an initially ordered array of pullers. (b) A pusher suck in fluid at its `waist'. A long-wavelength `bend' type perturbation is amplified in an initially ordered array of pushers.  Redrawn after \cite{SriramFilaments}. } \label{fig:unstable}
\end{figure}

The HI between swimmers and surfaces also depend on whether they are pushers or pullers, as well as on the boundary condition at the wall or interface. Such effects can be treated using the method of images \cite{LaugaRev,diLeonardoInterface}: each of the stokeslets making up the dipole sets up a system of images that, together, satisfy (say) the no-slip boundary condition at a wall. Constructing images for the hydrodynamic problem is more complex than the corresponding case in electrostatics: in the latter, the boundary condition at (say) a metal surface is a scalar equation for the potential, $\phi = 0$, while in the former case, a vectorial boundary condition holds, e.g. $\mathbf{v} = 0$ for no slip. As a result, the image of a point charge at a conducting wall is simply a point charge of the opposite sign, but the image of a stokeslet at a no-slip boundary is a stokeslet of the opposite sign (this much is analogous to electrostatics) plus two high-order multipole terms (these are due to the vectorial boundary condition). Nevertheless, to leading order, the image of a force dipole (two stokeslets of opposite signs) is a dipole of the opposite sign. This suggests that a pusher will align with its image (cf. Fig.~\ref{fig:reorient}(a), and recall the fore-aft symmetry), i.e. it will be reoriented to swim parallel a no-slip wall. Detailed calculations using the full image system confirm this conclusion based on the leading-order image \cite{LaugaRev}. 

The HI between swimmers leads to a whole `zoo' of novel collective behaviour and instabilities: swirls, jets, and swarms of all kinds. Reviewing these phenomena in detail is beyond the scope of these lectures; readers who want to pursue this subject should consult one of the many reviews available \cite{Sriram,CatesReview2012,LaugaRev,PedleyKessler,PedleyHill}. Once again, however, arguments based on flow field symmetry can yield significant insight. Without HI, a suspension of more or less rod-like swimmers (such as {\it E. coli} with flagella bundle) may be expected to show orientational ordering. But HIs render an ordered state unstable. The form of the instability induced by the coupling between flow and orientation differs in pushers and pullers, as a pictorial argument makes clear \cite{SriramFilaments}. 

A single puller pulls in liquid at both ends and ejects liquid around its `waist', Fig.~\ref{fig:pusherpuller}(b). Consider a `splay' perturbation in an initially ordered array of pullers, Fig.~\ref{fig:unstable}(a). The pullers at the left half of the sketch pulls in liquid more strongly at the bottom than at the top (small arrows), so that is a net ejection of liquid at the top (large arrow). The opposite happens at the right half of the sketch, with a net ejection of liquid at the bottom. The resulting shear flow over the central portion of the sketch enhances the orientational perturbation (dotted arrow). There is therefore a long-wavelength splay instability.  
On the other hand, a single pusher ejects liquid at both ends and sucks in liquid around its `waist', Fig.~\ref{fig:pusherpuller}(a). Now consider a `bend' perturbation in an initially ordered array of pullers, Fig.~\ref{fig:unstable}(b). The pushers at the top of the sketch sucks in liquid more strongly on their right than on their left, so that there is a net liquid influx from the right. For the pushers at the bottom, the reverse is true. The resulting shear flow again enhances the original orientational perturbation (dotted arrow). There is now a long-wavelength bend instability.  
These two pictorial arguments are confirmed by detailed calculations \cite{Sriram,SriramFilaments}. 

Interestingly, a recent proposal seeks to unify many HI-induced instabilities under a single theoretical framework \cite{MarchettiDumbbell}. Many aspects of the predicted `universal phase diagram' of instabilities in active particle suspensions have been substantiated by simulations using the asymmetry dumbbell swimmer, Fig.~\ref{fig:dumbbell}.  

Before closing the discussion on collective effects, it is worthwhile noting a rather generic experimental issue. The study of collective effects requires the use of interacting, and therefore almost certainly concentrated, suspensions of swimmers. In all the systems I have reviewed in Sections~\ref{sec:bacteria} and \ref{sec:janus} except one \cite{HerminghausActive}, swimming depends on the consumption of external resources. Swimming cells in a suspension of {\it E. coli} bacteria slow down as a function of time, Fig.~\ref{fig:speedtime}, probably because dissolved oxygen is used up and cannot be replenished fast enough. In the case of Janus particles, they will eventually run out of `fuel' (such as H$_2$O$_2$). Thus, all but the briefest experiments will need to consider time dependence. Here, DDM really shows its advantage as a high-throughput method, allowing in principle the quasi-real-time monitoring of swimming behaviour so that the experimentalist can be alerted to resource exhaustion (cf. Fig.~\ref{fig:speedtime}). 

To end this section, I mention that collective effects can occur in active particle systems {\it without} hydrodynamics; a number of reviews of this topic exist \cite{Romanczuk,VicsekRev}.

\section{Summary and Conclusion} \label{sec:last} 

In these lectures, I have sought to give a pedagogical overview of the emergent field of active colloids from the point of view of an experimentalist. First, I have introduced two classes of active colloids available for experimentation: natural ones in the form of micro-organisms, and synthetic self-propelled particles. In the former case, I introduced the flagellated bacterium {\it E. coli} in detail, while in the latter case I paid most attention to synthetic spherical particles coated asymmetrically with catalyst enabling them to engage in self-phoretic motion. Secondly, I reviewed how motility information can be extracted from these suspensions using tracking, differential dynamic microscopy, and, potentially, dynamic light scattering. Finally, I ended with a survey of the kind of generic new physics that active colloids should display, relying here more on intuitive/pictorial arguments than formal mathematical derivations. After this concluding section, there is an appendix explaining to physicists aspects of practical microbiology. 

The introductory nature of these lectures means that I have focussed on the basics, and have only been able to offer a very cursory view of current research. However, even the small fraction of the recent research literature cited suffices to show that the subject of active colloids is in a phase of active growth at present. It is therefore interesting to conclude by commenting on possible avenues for future developments in each of the three areas I have surveyed: single particles, characterisation, and collective behaviour. 

First, the varieties of active colloids available will continue to grow. In the natural domain, physicists are moving away from an almost exclusive focus on a handful of model organisms (chief among which have been {\it E. coli}, {\it B. subtilis} and {\it Chlamydomonas}) and becomig more adventurous. As we do so, new modes of propulsion will need to be understood, and novel behavioural phenotypes, such as phototaxis, wait to be exploited. These developments combined with genetic manipulation will deliver an unprecedentedly rich collection of `smart' active particles for experimentation with tuneable swimming patterns and responses. The recently-engineered mutant of {\it E. coli} that only swims when illuminated \cite{Walter} is but one example. There is also wide scope for innovation in the area of synthetic self-propelled colloids. The mechanism of propulsion in even apparently simple systems such as Pt-coated polystyrene Janus particles dispersed in hydrogen peroxide does not appear yet fully understood, and the experimentalist's ability to control the motion of such particles is still at an elementary stage of development. Completely novel means of propulsion continue to appear on the scene: motile emulsion drops that carry their own `fuel' \cite{HerminghausActive} is again but one example. New ideas here can come from biomimesis (e.g. actin `rockets' \cite{CameronActin}) and/or the exploitation of new physics (e.g. the tuning of surface slip \cite{Ajdari} or Janus particles in binary liquids \cite{BechingerJanus}). Once again, significant opportunities exist in novel forms of control.

The detailed characterisation of active colloids is still in its infancy. It is somewhat embarrassing that the answers to many elementary questions remain illusive. Three examples are as follows. (1) Is the measured swimming speed distribution, $P(v)$, quenched or annealed? Put in other terms, is a slow particle always slow, or does each swimmer explore the whole range of speeds as time goes on (like molecules in a gas)? Deciding on this issue will presumably have fundamental implications for how the statistical mechanics of swimmers is formulated. (2) For swimmers that consume external sources of `fuel', what is the spatio-temporal distribution of fuel in its periphery? Quantitative information on this issue will be crucial for understanding, {\it inter alia}, the many-body physics of such swimmers, because at high concentrations, any `fuel depletion zone' surrounding each particle will start to overlap. (3) How is the swimming speed of swimmers dependent on their concentration? I have already reviewed theoretical predictions of novel phenomena predicated on a concentration-dependent speed in Section~\ref{subsec:detailed}. Finding answers to these and other similar questions will immensely increase our knowledge of precisely what physical systems we are dealing with, which is certainly a {\it sine qua non} for simulations and theory development. 

The structure of Section~\ref{sec:generic} already indicates a generic way to search for new collective physics: one could systematically perform experiments using active colloids in which there are passive counterparts (in which equilibrium statistical mechanics with detailed balanced applies), and look for similarities and differences. In particular, the self assembly of passive colloids, with or without the assistance of templates, is a mature research field with many possibilities for `knowledge transfer' to active particles. How monodisperse self-propelled spheres may crystallise is just one example among a multitude that can be given. Many (but by no means all) of the most interesting collective phenomena in passive colloids occurs at high particle concentrations. I have already mentioned one potential source of experimental difficulties for corresponding experiments using active particles: the rapid depletion of `fuel', so that any putative novel collective effects need to be carefully scrutinised to rule out fuel depletion as a causative factor. Of course, there are experiments in which active particles stand out as {\it sui generic}, e.g. how they may `run obstacle courses' \cite{BechingerJanus} -- here, no passive counterpart can really be imagined. 

Next, I mention two avenues for further work that cut across all three of the above areas. First, it will undoubtedly be fruitful to compare the behaviour of different kinds of active colloids, natural with synthetic, or different kinds of (say) motile organisms with differing modes of propulsion. Thus, experiments on polymer-induced phase separation in dilute suspensions of {\it E. coli} \cite{Jana} can be repeated using synthetic colloids. The depletion of fuel around the synthetic particles may cause aggregates to behave quite differently. 

Secondly, a potentially very fruitful area of research is the study of mixtures. One could envisage mixtures of different kinds of motile particles -- synthetic-synthetic, natural-natural, and natural-synthetic. The study of `natural-natural' mixtures in particular can be rather immediately biologically relevant -- there are predator-prey systems of this kind in the natural world of micro-organisms. Equally fascinating will be mixtures of passive and active colloids. The motivation here can be both fundamental and applied \footnote{Of course, these two motivations are `linearly independent but not orthogonal' -- progress in one direction will influence the other, but they remain conceptually distinct.}. Fundamentally, it is interesting to enquire, for example, how the motion of the passive particles are affected by the active ones. There is already some work on this topic, but mostly in 2D \cite{Gaston,WuLibchaber}; one exception is an experiment tracking polystyrene particles in a bulk suspension of swimming {\it Chlamydomonas} \cite{GoldsteinEnhanced}. An `enhanced diffusion' is observed in all cases reported to date, but the mechanism is still a matter of debate (e.g., direct collision vs. hydrodynamic interaction). Tracking of the passive particles can, of course, also give information on the fluctuating stresses in the system (microrheology \cite{WaighRev}), and has indeed been used in this manner to study the basic statistical physics of bacterial suspensions \cite{YodhBacteria}. Much more can be done on this front, especially since the equivalent of `active microrheology' (where the `probe' particle is dragged through the host suspension \cite{WilsonRev}) has not yet been implemented for motile particle suspensions. On the practical level, bacteria  \cite{Langlois} or self-propelled Janus particles \cite{DietrichCargo2} may be used to carry passive particles as `cargo', so that active-passive mixtures offer new avenues for the `assisted self assembly' of passive colloids. Theoretical analysis of synthetic `combo-particles' (active carrier + passive cargo) is yielding a number of interesting predictions \cite{DietrichCargo1}, e.g. a polystyrene particle {\it evenly coated} with Pt rigidly linked to a passive particle turns the combination into a self-propelled unit. Experimentally, the proper characterisation of any passive-active mixture will pose significant challenges. 

Finally, there is an area of investigation in `active colloids' that I have not touched on at all in these lectures. Motility is only one possible manifestation of the `active' nature of particles. Another manifestation, so far confined the natural world (but see \cite{ChaikinReproduce}), is growth and replication. A single cell of {\it E. coli} inoculated onto nutrient agar and kept at 37$^\circ$C will lengthen (at constant body diameter) and become two cells after about 20 minutes, and then the process will repeat \cite{Koch}. The shape of the bacterial colonies resulting from such growth and replication processes are characteristic of individual bacterial strains, and is used in classical microbiology for identification purposes. (See, e.g., the pictures in \cite{RaineyAdaptive} of `smooth', `wrinkly spreader' and `fuzzy spreader' phenotypes of {\it Psucomonas fluorescent}.) Understanding the spatio-temporal and mechanical properties of bacterial colonies is undoubtedly part of active colloid physics. Thus, for example, such colonies often display pronounced local liquid crystalline order, but manifest multiple defects on the global level. The active `nematohydrodynamics' of such colonies will be a fascinating area of study.   Progress here will also have significant implications for the study of biofilms \cite{Brenner} population dynamics \cite{NelsonSegregation}.

In conclusion, I hope I have demonstrated that active colloids is an active and growing sub-field of modern colloid physics, and that the concepts and tools used to study passive colloids can be applied to their study, but with challenging extensions or modifications. It is a nice irony that physics of passive colloids started with Robert Brown painstakingly proving that the incessant `jiggling' he observed in granules derived from pollens of {\it C. pulchella} were {\it not} biological in origin, but generic to all particles small enough to be suspended in a liquid against gravity; he announced this foundational discovery of colloid physics as `the general existence of {\it active} molecules in organic and inorganic solids' (italics mine) \cite{Brown}. Today, we have appropriated Brown's terminology of `activity' to describe particles that engage in self-propelled motion, some of which (bacteria, algae, etc.) {\it are} indeed alive. I think Brown would have been amused and pleased.

\appendix{} 

\section{`Microbiology Lab 101 for physicists', written in collaboration with Angela Dawson and Jana Schwarz-Linek}

We suggest that any physicist embarking on microbiological work should seek help and guidance from a friendly microbiologist. This Appendix aims to guide our readers through the kind of issues that they need to discuss with such a collaborator.

\subsection{Which bug?}

The first practical consideration is: `Which bacterium should I study?' The key point here is that one should seek to work with what biologists call `model organisms': organisms that have been studied in considerable detail because they are deemed exemplars of their kind. For motility work, {\it Escherichia coli},  {\it Salmonella typhiimurium} and {\it Bacillus subtilis} are good model organisms.  They are the subject of a huge body of pedagogical and research literature. In the case of the first two species, the definitive reference is the two-volume ``{\it Escherichia coli} and {\it Salmonella}: Cellular and Molecular Biology'' (ASM Press, 1996, editor in chief, F. C. Neidhardt) and its continuously updated electronic version \footnote{\url{http://ecosal.org/}, subscription required.}. For any model organism, thousands of documented mutants and genetic manipulation protocols are typically available. 

{\it Escherichia coli} is a work horse for much of modern biology. The bacterium is found in nature as a commensal in the human gut (i.e. it is a normal part of the gut flora).  There are, however, also pathogenic strains such as {\it E. coli} O157.  Most biology labs will have numerous strains of {\it E. coli} stored in their -80$^\circ$C freezers, because the bacterium is used for cloning genes from plants, animals, fungi or other bacteria, and as a host strain for over-expressing proteins in the first stage recombinant protein production. Many of these strains of {\it E. coli} would have been engineered for these specific purposes. Many will not be ideal for motility work, for which strains derived from K-12 or W are popular.

A second practical consideration is biological safety. Bacteria are classified according to their potential to behave as pathogens.  Are they pathogenic to a normal healthy individual?   Are they infectious?  Is there effective prophylaxis and treatment available?  On these bases bacteria (and other hazardous biological agents) are categorised into four biosafety containment levels (1 to 4, low to high risk).  In Great Britain, The Health and Safety Commission publishes this categorization in accordance with EU directives; in America, check with the Center for Disease Control.  The hazard group of a particular biological agent indicates the minimum level of containment under which it must be handled. Level 1 organisms are not harmful to human health; a level 1 containment laboratory nevertheless has to satisfy certain minimum requirements (see below). Strains of {\it E. coli} derived from K-12 or W are containment level 1 organisms. Containment level 2 organisms present about the same level of hazard as that typically encountered in domestic toilets. Containment levels 3 and 4 indicate the most dangerous pathogens. Before you start work, it is important that you know the categorisation of your target organism, and set up your laboratory and train personnel accordingly.

\subsection{Laboratory requirements}

We recommend that a physicist wanting to start bacterial work should visit a number of microbiology laboratories.   The minimum containment requirements to work with a bacterium depend on its hazard rating, for which consult your institution's biological safety officers.   Requirements for containment level 1 labs include the provision of benches impervious/resistant to chemicals, access to sterilisation equipment, hand washing facilities available but separate from general lab sinks, the wearing of lab coats, and disinfection and waste disposal procedures in place.  A microbiology work area should be light and draught free; windows need to remain shut to avoid air-borne contaminants.   

Many pieces of equipment in a microbiology lab are common to any wet lab, e.g. centrifuges, pH meters, balances and spectrophotometers.  Additional equipment is required to sterilise media and all glassware that is in contact with bacteria.  A biology department will frequently have dedicated washing up and sterilising facilities; these may be available to other scientists.  Alternatively small autoclaves can be purchased for sterilisation.   A convenient source of ultra-pure distilled water for making up solutions will be required.  Accessibility to a fume hood when working with certain chemicals may also be needed. In addition to sterilisation, microbiologists have dedicated equipment for growing bacteria under defined conditions.    Bacteria may be grown on solid media (usually agar in Petri dishes) or in liquid media.  Optimal conditions depend on the organism.  Thus, {\it E. coli} grows fastest around 37$^\circ$C with aeration, so in liquid culture it is grown in a temperature-controlled orbital shaker.    Sometimes the temperature is varied to enhance certain characteristics.   Microbiologists (ever wary of contamination) frequently set up dedicated growing areas so that the risk of cross-contamination is minimised. Fridges (or cold rooms), -20$^\circ$C freezers and access to a -80$^\circ$C freezer for long-term storage and preservation of strains, together with any labile chemicals, will be required.

Different species of bacteria are grown in/on different sterile media. Many biology departments have a central facility for making up media. Some media can be purchased as concentrates for dilution followed by sterilisation.  If this is not available for a specific organism then media will have to be made up from the individual chemical constituents.  Most species of bacteria are routinely grown on just one or two types of media. 

\subsection{Obtaining your bacteria}

The safest place to obtain a starting sample of bacteria from is a microbial culture collection, where strains are deposited, safeguarded and distributed.  Examples are the American Tissue Culture Collection (ATCC) and National Collection of Tissue Cultures (NCTC).   Obtaining a strain from such a collection should guarantee its genetic provenance, i.e. that it has the genotype it is supposed to have and hence the correct phenotype.  The cost is usually around GBP~50-200.  It may also be possible to obtain a culture from a nearby lab, but be beware that many samples in freezers are not as they should be.  It is common practice to write to the author of a paper who has isolated or prepared and characterised an interesting mutant and request a sample, offering to pay shipment costs.   It is an increasingly common procedure to be asked to sign a Materials Transfer Agreement (MTA), a legal document preventing the strain from being passed on to a third party.  It is good microbiological practise always to check the phenotype prior to starting work with a new strain \footnote{The right attitude is: `I won't believe it's motile until I see it move down a microscope!' \protect\label{fn:motile}}.

\subsection{Getting started}

To work safely and accurately with microorganisms requires training from an experienced microbiologist.  Most microbiologists work with only a handful of species during their whole career.  Years of handling and propagating these organisms give them the ability to recognise  barely perceptible changes that may indicate that there is contamination problem or that a mutation has arisen \footnote{The natural mutation rate, even without phage infection, is high enough that experimentalists must be alert to the possibility that the genotype of their strains may change with time.}. The cornerstone of being able to work well and reproducibly with bacteria is a good aseptic technique.  This is the technical expertise to propagate and work with your bacterial strain without cross-contamination arising.  This cannot be learnt from a book but has to be passed on from one individual to the next  \footnote{A useful (and free!) training website is available, \url{http://www.microeguide.com/}, although this does not replace being trained by a qualified person.}.  A microbiologist familiar with the relevant species will also be able to give instructions on how best to store the strain, recommend the best means of preservation and advise on how stable it is under different storage conditions.

\subsection{Preparing a motile culture}

Motility is not high on the agenda of most biology laboratories, so that after a physicist has learnt the basics from a biology colleague, there are extra steps to be optimised if the aim is to culture swimmers. Here footnote (\ref{fn:motile}) is particularly relevant --- it is unwise to assume that preparative techniques learnt from a general microbiology lab will automatically give a culture of `happy swimmers' even if the strain is a motility phenotype in principle: always look in a microscope. Phase contrast objectives at $\times 20$ or $\times 40$ work well for this purpose. 

Whether you develop your own protocol or use a published procedure, it is important to follow it consistently for every single preparation to maximise reproducibility. Growth conditions like temperature and medium composition will affect motility, therefore growth conditions for motile cultures are often sub-optimal compared to standard procedures (for instance incubation at 30$^\circ$C instead of 37$^\circ$C and the use of less-rich medium). After overnight incubation, a second growth step is performed to enable the cells to be harvested when they are in the so-called exponential growth phase, during which they tend to be more motile. Before bacteria can be used in experiments they need to be washed several times to remove any food and/or waste by either centrifugation or filtration. For this a buffer at specific pH and salt concentration is used, possibly with an added chelating agent like EDTA and an energy source (glucose or lactate). Care must be exercised when handling motile cell suspensions during washing and afterwards to prevent flagellar damage. Strong shaking and vortexing should be avoided at all costs \footnote{For those who have the relevant experience, this is a little like the care needed in transferring, shaking and pipetting high polymer solutions to avoid breaking the polymer coils.}. After washing, the cell concentration can then be adjusted according to experimental requirements (typically by optical density measurements in a spectrophotometer). 

\vspace{.3cm}

We end by mentioning a number of helpful texts. The general text by Schlegel \cite{Schlegel} is more manageable in size than most introductory microbiology texts, and offers a treatment that, perhaps because it has more on quantitative physiology than usual, may appeal more to the physics reader. Following on this track, two texts on bacterial physiology \cite{NeidhardtText,BartonBook} and a chemical engineering text on microbes \cite{Ollis} are good sources of numerical information of the kind that physicists may want \footnote{Readers may also wish to consult the (citable) web resource, The {\it CyberCell} Database, \url{http://ccdb.wishartlab.com/CCDB/}.}.

My work on active colloids was supported by grants EP/D071070/1, EP/E030173/1 and EP/J007404/1 from the EPSRC. I thank staff and students at the Faculty of Physics, University of Konstanz, who provided valuable feedback when they attended a draft version of these lectures, and Richard Blythe, Aidan Brown, Mike Cates, Daan Frenkel, Alastair Mailer, Alexander Morozov, Roberto Piazza and Ramin Golestanian for enlightening discussions during the preparation of the written version. 


\end{document}